\documentclass[11pt,a4paper]{article}

\usepackage{framed,multirow}

\usepackage{authblk}
\usepackage{graphicx}
\usepackage{amsmath}
\usepackage{amsfonts}
\usepackage{amssymb}
\usepackage{xcolor}
\usepackage{multirow}
\usepackage[left=2cm,right=2cm,top=2cm,bottom=2cm]{geometry}
\usepackage{url}
\usepackage{xcolor}

\usepackage{textcomp,calligra}

\newcommand{\ess}{\text{\Large \textcalligra{s}}\, }

\newcommand{\Reals}{\mathbb{R}}

\title{Kinetic theory of particle-in-cell simulation plasma and the ensemble averaging technique}%

\author[1]{Micha\"{e}l  Touati\footnote{michael.touati@marvelfusion.com}}
\author[2]{Romain Codur}
\author[2]{Frank Tsung}
\author[2]{Viktor K Decyk}
\author[2]{Warren B Mori}
\author[3]{Luis O Silva}

\affil[1]{Marvel Fusion Gmbh, Blumenstrasse 28, Munich, Germany}
\affil[2]{Department of Physics and Astronomy, University of California Los Angeles, Los Angeles, USA}
\affil[3]{GOLP/Instituto de Plasma e Fus\~{a}o Nuclear, Instituto Superior T\'{e}cnico, Universidade de Lisboa, Lisbon, Portugal}

\date{August 2022}

\begin{document}

\maketitle

\begin{abstract}
We derive the kinetic theory of fluctuations in physically and numerically stable particle-in-cell (PIC) simulations of electrostatic plasmas. The starting point is the single-time correlations at the simulation start between the statistical fluctuations of weighted densities of macroparticle centers in the plasma particle phase-space. The fluctuations are associated with different initial conditions, typically due to the random initial conditions (in velocity space) of the macroparticles/simulation plasma, assigned according to their initial distribution of probability. 
The single-time correlations at all time steps and in each spatial grid cell are then determined from the Laplace-Fourier transforms of the discretized Klimontovich-like equation for the macroparticles and Maxwell's equations for the fields as computed by modern PIC codes. 
We recover the expressions for the electrostatic field and the plasma particle density fluctuation autocorrelations spectra as well as the kinetic equations describing the average evolution of PIC-simulated plasma particles, first derived by \cite{Langdon1970a} using a macroparticle test approach perturbing a discretized Vlasovian plasma and then averaging the obtained physical quantity over the initial macroparticle velocity distribution. We generalize and extend these results to the modern algorithms in PIC codes and using arbitrary macroparticle weights.
Analytical estimates of statistical fluctuations single-time correlation amplitudes are derived as a function of the plasma simulation parameters, using the central limit theorem in the limit of a large number of macroparticles per cell.
The theory is then used to analyze the ensemble averaging technique of PIC simulations where statistical averages are performed over ensembles of PIC simulations, modeling the same plasma physics problem but using different statistical realizations of the initial distribution functions of the macroparticles. 
This method is illustrated with linear Landau damping uncovering (out of what is usually considered numerical, noise) the physical fluctuations driven by a single small amplitude electrostatic wave perturbing a PIC simulation plasma in equilibrium.\\
keywords: \textit{kinetic theory, electrostatic plasma, particle-in-cell codes, numerical noise, fluctuations correlations, ensemble averaging technique}
\end{abstract}

\clearpage

\tableofcontents

\clearpage

\section{Introduction}

A plasma is a set of charged particles consisting of electrons and ionized atoms that behave collectively through their self-consistent long-distance electromagnetic fields. In this paper, we consider (the, apparently, less complex) non-degenerate, non-relativistic, and non-coupled collisionless electrostatic plasmas without external forces. One possible approach to describe the plasma considers all particle equations of motion under the action of the plasma electrostatic fields, determined according to the self-consistent Maxwell equations. This fully deterministic approach allows for deducing the particle positions and velocities at every time (and thus the temporal evolution of the plasma) if the full microscopic details of the system are known at a given time. Differentiating the phase-space densities associated with this detailed description with respect to time, the Klimontovich equation for each plasma species \cite{Klimontovich1958} is obtained. The Klimontovich equation describes the evolution in time of all $N$ plasma particles in the 6-dimensional phase-space under the action of self-consistent microscopic electrostatic fields. The detailed knowledge of the positions and velocities of the particles is needed to evaluate the microscopic electrostatic fields, which renders it considerably difficult due to the challenge of computing simultaneously the particle dynamics and the microscopic self-consistent fields, even resorting to today's computer technology. The PIC method relaxes this constraint by considering particles with a finite spatial size, also called macroparticles \cite{Hockney1965} \cite{Yu1965} \cite{Hockney1966} \cite{Burger1965} \cite{Buneman1967} \cite{Birdsall1969} \cite{Dawson1983}. For dimensionless simulations mapped to a real physical system (or for simulations that use real physical units), each macroparticle usually represents several real particles -- the relation between real particles and their macroparticle representation depends on the compromise between the plasma density of the system under study and the available computer technology (for instance, restrictions on computer time and memory / represent the plasma particles). In the PIC method, macroparticle positions are interpolated onto a spatial grid used to compute the self-consistent discretized Maxwell equations. Therefore, for simulations of a plasma homogeneously distributed in the whole simulation box, the PIC algorithm is much more efficient than the direct Klimontovich model previously outlined, with a computational cost $C\sim \mathrm{O} \left ( N \right )$ compared to the computation of all binary Coulomb interactions between macroparticles that would need a computational cost $C'\sim \mathrm{O} \left ( N^2 \right )$. In terms of the physics being captured, the PIC method smooths the fields and thus underestimates close-encounter Coulomb collisions \cite{Okuda1970} \cite{Langdon1970b}.

Moreover, from classical thermodynamic theory, one can argue that the exact positions and velocities of all particles are never known/measured simultaneously. Instead, and for large N-body systems, knowledge of the system can be obtained via quantities averaged over time, infinitesimally small mesoscopic spatial volumes, or over different measurements performed under the same conditions. This naturally motivates a probabilistic approach, based on the temporal evolution of the probability distribution functions. Liouville theorem describes the temporal evolution of the full $N$-body probability distribution function of all particle position and velocity random variables \cite{Liouville1838}. As before, the information contained in the $N$-body distribution function is too large to allow for the computation of the Liouville equation coupled with the self-consistent microscopic Maxwell equations even with today's computer technology. Simplifications are also possible in this approach: smaller order $s$-body distribution functions can be obtained by integrating the Liouville equation over the position and velocity random variables of all other particles $s+1,\,...,N$. In the resulting equation, describing the temporal evolution of the $s$-body distribution functions in the 6$s$-dimensional phase-space, the presence of the "collision" integral\footnote{We observe this term is used, even if, sometimes, the "collision" integral does not strictly describe close-encounter Coulomb collisions}, also poses challenges, and some additional assumptions are needed to close the problem. The first order equation ($s=1$) in the BBGKY hierarchy of equations
% , -from the name of its founders Born, Bogoliubov, Green, Kirkwood and Yvon 
\cite{Bogoliubov1946,Born1946,Kirkwood1946,Yvon1935}, can be truncated to obtain the Vlasov equation by assuming an infinite number of electrons in a Debye sphere, thus neglecting 2-body correlations between 1-body distribution functions \cite{Vlasov1938}. However, the number of particles is necessarily finite in plasma, and particles are always correlated due to their long-range electrostatic interactions. The Vlasov equation remains valid only on time scales much smaller than this minimum interaction time. On larger time scales, the Bogoliubov hypothesis \cite{Boer1962} can be used to simplify the second order equation in the BBGKY hierarchy and estimate the 2-body correlations between plasma particles. The Bogoliubov hypothesis assumes the 2-body correlations relax on a time scale smaller than the time scale on which the 1-body distribution function relaxes. Further assumptions are still needed to get an analytical expression of correlations e.g considering only binary electrostatic Coulomb interaction between plasma particles and neglecting their 3-body correlations, it is possible to derive the Vlasov-Boltzmann equation or the Vlasov-Lenard-Balescu equation depending on the underlying assumptions \cite{Lenard1960} \cite{Balescu1960} \cite{Guernsey1960}. Both the Vlasov-Boltzmann equation or the Vlasov-Lenard-Balescu equation reduce to the Vlasov-Fokker-Planck-Landau equation when considering only non-screened and small angle Coulomb collisions between plasma particles \cite{Landau1936}. The numerical computation of these kinetic equations onto the resulting 7-dimensional phase-space-time grid is still challenging with modern computer technology and further approximations are usually used to reduce the number of variables \cite{Tzoufras2011,Touati2014,Joglekar2014}. 

The theory of fluctuations in collisionless plasmas is related to (and connects) both the deterministic (Klimontovich-based) and  probabilistic (Liouville-based) approaches. It is based on the study of single-time correlations between exact microscopic plasma quantities fluctuations, measuring the deviations of one statistical realization of the plasma consisting of a discrete number of self-interacting particles around the expected average, which is described by the continuous 1-body distribution function. \cite{Rostoker1961,Klimontovich1962,Dupree1963} have shown that the study of single-time correlations between fluctuations also leads to the Vlasov-Lenard-Balescu equation describing the 1-body distribution function in the case of homogeneous and stationary collisionless plasmas. This bears a striking resemblance with the fundamentals of PIC simulations: in PIC simulations, the initial macroparticle velocities are determined by random sampling of the (desired) probability distribution function. Therefore, this random sampling (via a random number generator) naturally produces statistical fluctuations around the initial (ensemble averaged) distribution function. 
The main motivation for this paper lies in this observation: we apply the theory of fluctuations to the discretized equations associated with electrostatic PIC code (taking into account both the discretized equations of motion and the discretized field equations), and we explore this theory to determine the statistical properties (and possible advantages) of ensemble averages over sets of PIC simulations. The paper is organized as follows. In section 2, we recapitulate the discretized equations of an electrostatic PIC code and we reformulate them within the mathematical framework of the kinetic theory of plasmas. In section 3, we derive the theory of statistical fluctuations in electrostatic PIC simulations, for stationary, homogeneous, infinite, and fully ionized plasmas. We discuss the assumptions and the key results of the theory in section 4. In section 5, we discuss the "ensemble averaging technique" \cite{Shanny1967} within the theoretical framework developed in the previous sections, and we determine the conditions under which this technique relaxes the discrete particle effects inherent to the use of a finite and relatively small number of macroparticles resulting from the limits of computer technology. In section 6, the main conclusions of the paper are presented.

%\tableofcontents

\section{Electrostatic PIC simulations}

\subsection{Discretized particle and field equations in an electrostatic PIC code}

In a PIC simulation, the phase-space of each plasma species $a$ is sampled according to $N_a $ macroparticles of mass $M_a$ and charge $Q_a$. When the equations computed by a PIC code are renormalized to physical units, each macroparticle represents $\delta N_a = M_a / m_a = Q_a / q_a$ real particles of mass $m_a$ and electrical charge $q_a$. $\delta N_a$ is therefore a number, a lineal density or an areal density of real particles from species $a$ in 3D, 2D or 1D PIC simulations, respectively. Macroparticle trajectories $\displaystyle \left ( \mathbf{r}_{a,\ell} \left ( t_n \right ),\, \mathbf{v}_{a,\ell}  \left ( t_n \right ) \right )$ in phase-space are computed at each time step $t_n = \left (n-1 \right ) \Delta_t$ according to their discretized equation of motion
\begin{equation}
\label{EoM}
\forall a,\,\forall \ell \in \left [1,\, N_a\right ],\, \forall n,\, \displaystyle \left \{ \begin{array}{lllr}
     {\displaystyle \left . \displaystyle \frac{d \mathbf{r}_{a,\ell} }{d t } \right |}^n  &=& \mathbf{v}_{a,\ell} \left ( t_n \right )                                                                                                  &, \mathbf{r}_{a,\ell} \left ( t_1 \right ) = \mathbf{R}_{a,\ell}
\cr {\displaystyle \left . \displaystyle \frac{d \mathbf{v}_{a,\ell} }{d t } \right |}^n &=& \displaystyle \frac{Q_a}{M_a} \mathbf{E}_{s} \left ( \mathbf{r}_{a,\ell} \left ( t_n \right ),\, t_n \right ) &, \mathbf{v}_{a,\ell} \left ( t_1  \right ) = \mathbf{V}_{a,\ell}
\end{array} \right .
\end{equation}
where
\begin{equation}
\label{field_interpolation}
\mathbf{E}_{s} \left ( \mathbf{r}_{a,\ell} \left ( t_n \right ),\, t_n \right ) = \displaystyle \sum_{i,j,k} \mathbf{E} \left ( \mathbf{r}_{i,j,k} ,\, t_n \right ) S \displaystyle \left ( \mathbf{r}_{a,\ell} \left ( t_n \right )  - \mathbf{r}_{i,j,k} \right ) \Delta_x \Delta_y \Delta_z
\end{equation}
is the electric field, interpolated at macroparticle center locations. The electrostatic field is deduced according to the discretized Maxwell equations
\begin{equation}
\label{Maxwell}
\displaystyle \left \{\begin{array}{lllll}
      {\displaystyle \left . \displaystyle \frac{\partial}{\partial \mathbf{r}} \,\, \cdot \,\mathbf{E} \hspace{0.5em}\right |}^{i,j,k,n} &=& 4 \pi \displaystyle \sum_a Q_a \displaystyle \sum_{\ell = 1}^{N_a} S \displaystyle \left ( \mathbf{r}_{i,j,k}  - \mathbf{r}_{a,\ell} \left ( t_n \right )  \right )
\cr  {\displaystyle \left . \displaystyle \frac{\partial}{\partial \mathbf{r}}  \times \mathbf{E} \hspace{0.5em} \right |} ^{i,j,k,n} &=& \mathbf{0}
\end{array} \right .
\end{equation}
where macroparticle centers $\mathbf{r}_{a,\ell} \left ( t_n \right )$ are interpolated at the grid points $\forall \left \{ i,\,j,\,k\right \} \in \left [1,N_x \right ] \times \left [1,N_y \right ] \times \left [1,N_z \right ],\, \mathbf{r}_{i,\,j,\,k} = \displaystyle \left (  \left ( i-1 \right ) \Delta_x,\,\displaystyle \left ( j-1 \right ) \Delta_y,\,\displaystyle \left ( k-1 \right ) \Delta_z \right )$. Here, $L_x = N_x \Delta_x$, $L_y= N_y \Delta_y$ and $L_z= N_z \Delta_z$ are the simulation box dimensions and $\Delta_x$, $\Delta_y$ and $\Delta_z$ are the spatial grid spacings that are usually constant. The index $k$ or indices $j$ and $k$ for 2D or 3D simulations, respectively, should be ommitted in the equations but we will keep writing them in all the following to be the more general as possible. A smoothed interpolating function \begin{equation}
\label{particle_shape}
S \displaystyle \left ( \mathbf{r}  \right ) = \displaystyle \left \{ \begin{array}{lll}
    S_x \displaystyle \left ( x  \right )                                                                             &\, \mathrm{in} \, 1\mathrm{D}
\cr S_x \displaystyle \left ( x  \right ) S_y \displaystyle \left ( y  \right )                                       &\, \mathrm{in} \, 2\mathrm{D}
\cr S_x \displaystyle \left ( x  \right ) S_y \displaystyle \left ( y  \right ) S_z \displaystyle \left ( z  \right ) &\, \mathrm{in} \, 3\mathrm{D}
\end{array} \right . \,\mathrm{with}\,\forall \xi \in[x,\,y,\,z],\, S_\xi(\xi) = \displaystyle \left [ \Pi_{\Delta_\xi}^{(n)} * S_{0_\xi} \right ] \displaystyle \left ( \xi \right )
\end{equation}
is used where usually $\Pi_{\Delta}^{(n)} = \Pi_{\Delta} * ... * \Pi_{\Delta}$ is the $n$th-order B-spline, i.e. the  centered square distribution of support $\Delta$, $ \Pi_{\Delta} \left ( \xi \right ) = 1 / \Delta$ if $\left | \xi\right | \leq \Delta / 2$ and $0$ else, convoluted $n$ times with itself. $S_{0_\xi}$ are optional filters or smoothing functions that mitigate spatial aliasing effects such as, for example, a Gaussian filter
\begin{equation}
\label{smoothing_function}
S_{0_\xi} \displaystyle \left ( \xi \right ) = \displaystyle \frac{1}{ \sqrt{2 \pi {a_\xi}^2 } } \exp{ \displaystyle \left [ - \displaystyle \frac{ \xi^2 }{ 2 {a_\xi}^2 } \right ] }
\end{equation}
for spectral Maxwell solvers or such as a compensated $N$th-order binomial filter for finite-difference time-domain (FDTD) Maxwell solvers. Interpolating functions and filters vary a lot across codes. 
For example, in PIC codes using a FDTD scheme, the smoothing function is sometimes only applied to the electrostatic field after computing it according to Maxwell equations (\ref{smoothing_function}). In all the following, we will consider the same interpolating function in both (\ref{field_interpolation}) and (\ref{Maxwell}) so that the smoothed interpolating function (\ref{particle_shape}) can be interpreted physically as the spatial shape of macroparticles. The generalization of our results to the case where different interpolating functions are used for (\ref{field_interpolation}) and (\ref{Maxwell}) is straightforward. In general, the interpolating and smoothing functions have the following properties
\begin{equation}
\label{particle_shape_property}
\displaystyle \left \{ \begin{array}{ll}
     \forall \mathbf{r},& \displaystyle \sum_{i,j,k} \Pi_{\mathbf{\boldsymbol{\Delta}}}^{(n)} \displaystyle \left ( \mathbf{r}_{i,j,k} - \mathbf{r} \right )\Delta_x \Delta_y \Delta_z = 1,\\
\cr \forall \mathbf{r},&\displaystyle \int_{\Reals^3} \Pi_{\mathbf{\boldsymbol{\Delta}}}^{(n)} \displaystyle \left ( \mathbf{r} \right ) d^3 \mathbf{r} = 1\\
\cr    \mathrm{and}  &\displaystyle \left \{ \begin{array}{lll}
                                      \displaystyle \int_{\Reals^3} S_0 \displaystyle \left ( \mathbf{r} \right ) d^3 \mathbf{r} = 1&\mathrm{for\,spectral\,Maxwell\,solver\,or}\\
                                 \cr \displaystyle \sum_{i,j,k} S_0 \displaystyle \left ( \mathbf{r}_{i,j,k}  \right )\Delta_x \Delta_y \Delta_z = 1&\mathrm{for\,FDTD\,Maxwell\,solvers}\\
                                 \end{array} \right .
\end{array} \right .
\end{equation}
Here, we have noted $ \mathbf{\boldsymbol{\Delta}} = {\displaystyle \left ( \Delta_{x} ,\,   \Delta_{y} ,\, \Delta_{z} \right )}^t $ with the uppercase $^t$ meaning the vector transposition, $\Pi_{\mathbf{\boldsymbol{\Delta}}}^{(n)} \displaystyle \left ( \mathbf{r} \right ) = \Pi_{\Delta_x}^{(n)} \displaystyle \left ( x \right )\Pi_{\Delta_y}^{(n)} \displaystyle \left ( y \right )\Pi_{\Delta_z}^{(n)} \displaystyle \left ( z \right )$ and $S_0 \displaystyle \left ( \mathbf{r} \right ) = S_{0_x} \displaystyle \left ( x \right ) S_{0_y} \displaystyle \left ( y \right ) S_{0_z} \displaystyle \left ( z \right )$ to simplify the notations. Let us finally stress here that even in the case where the PIC codes computes Maxwell equations by using the second order FDTD scheme proposed by \cite{Yee1966} coupled with the charge conserving scheme proposed by \cite{Villasenor1992} for first order B-spline macroparticle shapes or by \cite{Esirkepov2001} for higher orders, the resulting numerical scheme can be written with the form (\ref{Maxwell}) if the discretized Maxwell-Poisson equation is verified at the simulation start.

\subsection{Initialization of a fully ionized, infinite and homogeneous plasma in PIC simulations}

Let us consider the electrostatic PIC simulation of a fully ionized, infinite and homogeneous plasma of atomic number $Z$. We note $\bar{n}_{i}$ the ion density and $\bar{n}_{e} = Z \bar{n}_{i}$ the electron density. We use $L_x$, $L_y$ and $L_z$-periodic boundary conditions so that
\begin{equation}
\label{density}
\forall a \in \{e,\,i\},\,\bar{n}_a = \displaystyle \frac{ N_a \delta N_a }{ L_x L_y L_z }
\end{equation}
(In 1D simulations, $\bar{n}_a = N_a \delta N_a / L_x$ and in 2D simulations, $\bar{n}_a = N_a \delta N_a / L_x L_y$). We choose a number $N_i$ and a number $N_e = Z N_i$ of macroions and macroelectrons to sample the plasma particle phase-spaces such that $\delta N_e = \delta N_i = \delta N$. At the simulation start, we use a random number generator to initialize macroparticle velocities $\mathbf{V}_{a,\ell}$ according to the desired probability distribution $F_{a0} \left ( \mathbf{V}_a \right )$. For example, we have for the non-drifting Maxwell-Boltzmann equilibrium probability distribution of velocities that maximizes entropy:
\begin{equation}
\label{Maxwellian}
F_{a0} \left ( \mathbf{V}_a \right ) = \displaystyle \frac{1}{ { \left ( 2 \pi {v_{T_a}}^2 \right ) }^{3/2} } \exp{ \displaystyle \left ( - \displaystyle \frac{ {\mathbf{V}_a}^2 }{ 2 {v_{Ta}}^2 } \right ) }
\end{equation}
where $\mathbf{V}_{a}$ is the velocity random variable of macroparticles from species $a$, $v_{T_a} = \sqrt{ k_B T_a / m_a }$ is the thermal velocity of particles, $k_B$ is the Boltzmann constant and $T_a$ are the species temperatures.
However, in order to ensure homogeneous electrical charge neutrality in the whole PIC simulation plasma, macroparticle locations at the simulation start $\mathbf{R}_{a,\ell}$ are not randomly chosen according to a uniform probability law. Instead, we deposit each macroparticle $\ell \in \left [ 1,\,N_a\right ] $ homogeneously inside all spatial grid cells. If we note $N_{a_x}$, $N_{a_y}$ and $N_{a_z}$ the number of macroparticles distributed along the $x$, $y$ and $z$-axis respectively such that $N_a = N_{a_x} N_{a_y} N_{a_z}$, the macroparticle locations read $\forall \left ( \ell_x,\,\ell_y,\,\ell_z\right ) \in \left [ 1,\,N_{a_x}\right ] \times \left [ 1,\,N_{a_y}\right ] \times \left [ 1,\,N_{a_z}\right ],$
\begin{equation}
\label{initial_macroparticle_location}
\mathbf{R}_{a,\ell} = \displaystyle \left ( \displaystyle \begin{array}{c} 
     X_{a,\ell_x} 
\cr Y_{a,\ell_y} 
\cr Z_{a,\ell_z} 
\end{array} \right ) = \displaystyle \left ( \displaystyle \begin{array}{c} 
     \left ( \ell_x - \displaystyle \frac{1}{2}\right ) \displaystyle\frac{L_x}{ N_{a_x} }
\cr \left ( \ell_y - \displaystyle \frac{1}{2}\right ) \displaystyle \frac{L_y}{ N_{a_y} }
\cr \left ( \ell_z - \displaystyle \frac{1}{2}\right ) \displaystyle \frac{L_z}{ N_{a_z} }
\end{array} \right )
\end{equation}
where $\ell = \ell_x + \left ( \ell_y - 1 \right ) N_{a_x} + \left ( \ell_z - 1 \right ) N_{a_x} N_{a_y}$ and where $-1/2$ is used to avoid having half the box have more charge than the other half, which gives rise to non-physical larger amplitude oscillations due to initial charge imbalances. The numbers of macroparticles are related to the number of macroparticles per cell according to
\begin{equation}
\label{NMPC}
N_{a,\mathrm{mpc}} = \displaystyle \frac{N_a}{N_x N_y N_z } = \displaystyle \frac{N_{a_x}}{N_x } \displaystyle \frac{N_{a_y}}{N_y } \displaystyle \frac{N_{a_z}}{N_z }
\end{equation}
(In 1D simulations, $N_{a,\mathrm{mpc}} = N_a / N_x = N_{a_x} / N_x$ and in 2D simulations, $N_{a,\mathrm{mpc}} = N_a / N_x N_y = N_{a_x} N_{a_y} / N_x N_y$). The latter is a key parameter of PIC simulations as we are going to see in the next sections. 

\subsection{Kinetic theory of PIC simulation electrostatic plasmas}

One can reformulate equations (\ref{EoM}) and (\ref{Maxwell}) computed by an electrostatic PIC code by defining the phase-space densities of macroparticle centers multiplied by their weights
\begin{equation}
\label{MCPSD}
f_{a_c} \displaystyle \left ( \mathbf{r},\,\mathbf{v},\,t \right ) = \delta N_a \displaystyle \sum_{\ell = 1}^{N_a} \delta \left ( \mathbf{r} - \mathbf{r}_{a,\ell} \left ( t \right ) \right ) \delta \left ( \mathbf{v} - \mathbf{v}_{a,\ell} \left ( t \right ) \right )
\end{equation}
where we have noted $\delta \left ( \mathbf{r}\right ) = \delta \left ( x \right )\delta \left ( y \right )\delta \left ( z \right )$ ($\delta \left ( \mathbf{r}\right ) = \delta \left ( x \right )\delta \left ( y \right )$ in 2D and $\delta \left ( \mathbf{r}\right ) = \delta \left ( x \right )$ in 1D) the Dirac distribution. Firstly, by differentiating the latter with time and then discretizing it in space and time, we obtain
\begin{equation}
\label{Klimontovich}
 {\displaystyle \left . \displaystyle \frac{ \partial f_{a_c} }{\partial t} \right |}^{i,j,k,n}  + {\displaystyle \left . \displaystyle \frac{ \partial  }{\partial \mathbf{r}} \cdot \displaystyle \left ( \mathbf{v} f_{a_c} \right ) \right |}^{i,j,k,n}   + \displaystyle \frac{ \partial  }{\partial \mathbf{v}} \cdot \displaystyle \left ( \displaystyle \frac{q_a}{m_a} \mathbf{E}_s \left ( \mathbf{r}_{i,j,k},\,t_n \right )  f_{a_c} \left ( \mathbf{r}_{i,j,k},\, \mathbf{v},\,t_n \right )  \right ) = 0
\end{equation}
at the grid points $\mathbf{r}_{i,j,k}$ and time steps $t_n$ according to macroparticle discretized equations of motion (\ref{EoM}). Secondly, one may express Maxwell equations (\ref{Maxwell}) as a function of weighted macroparticle center phase-space densities (\ref{MCPSD}) according to
\begin{equation}
\label{Maxwell2}
\displaystyle \left \{\begin{array}{lllll}
      {\displaystyle \left . \displaystyle \frac{\partial}{\partial \mathbf{r}} \,\, \cdot \,\mathbf{E} \hspace{0.5em}\right |}^{i,j,k,n} &=& 4 \pi \displaystyle \sum_a q_a \displaystyle \int_{\Reals^3} \displaystyle \int_{\Reals^3} f_{a_c} \displaystyle \left ( \mathbf{r}',\,\mathbf{v},\,t_n \right ) S \left ( \mathbf{r}_{i,j,k} - \mathbf{r}' \right ) d^3 \mathbf{r}' d^3 \mathbf{v}
\cr{\displaystyle \left . \displaystyle \frac{\partial}{\partial \mathbf{r}}  \times \mathbf{E} \hspace{0.5em} \right |}^{i,j,k,n} &=& \mathbf{0}
\end{array} \right . .
\end{equation}
Here, we recognize the phase-space densities of real particles
\begin{equation}
\label{RPPSD}
\begin{array}{lll}
f_a \displaystyle \left ( \mathbf{r},\,\mathbf{v},\,t\right ) &=& \displaystyle \int_{\Reals^3} f_{a_c} \displaystyle \left ( \mathbf{r}',\,\mathbf{v},\,t \right ) S \left ( \mathbf{r} - \mathbf{r}' \right ) d^3 \mathbf{r}'
\cr &=& \delta N_a  \displaystyle \sum_{\ell = 1}^{N_a}  S \displaystyle \left ( \mathbf{r} - \mathbf{r}_{a,\ell} \left ( t \right ) \right ) \delta \left ( \mathbf{v} - \mathbf{v}_{a,\ell} \left ( t \right ) \right )
\end{array}
\end{equation}
as it is sampled according to the PIC method. By differentiating the latter with time in order to get the deterministic description of electrostatic PIC simulations, we obtain the discretized kinetic equation
\begin{equation}
\label{Klimontovich_like_equation}
 { \begin{array}{l}
{\displaystyle \left . \displaystyle \frac{ \partial f_{a} }{\partial t} \right |}^{i,j,k,n} + {\displaystyle \left . \displaystyle \frac{ \partial  }{\partial \mathbf{r}} \cdot \displaystyle \left ( \mathbf{v} f_{a} \right ) \right |}^{i,j,k,n} + \displaystyle \frac{ \partial  }{\partial \mathbf{v}} \cdot \displaystyle \left ( \displaystyle \frac{q_a}{m_a} \mathbf{E}_s \left ( \mathbf{r}_{i,j,k},\,t_n\right ) f_{a} \left ( \mathbf{r}_{i,j,k},\,\mathbf{v},\, t_n \right ) \right )  
= C_a \displaystyle \left ( \mathbf{r}_{i,j,k},\,\mathbf{v},\,t_n \right ).
\end{array}
}
\end{equation}
It is self-consistently coupled with the discretized Maxwell equations
\begin{equation}
\label{PIC_Maxwell}
\displaystyle \left \{\begin{array}{lllll}
       {\displaystyle \left . \displaystyle \frac{\partial}{\partial \mathbf{r}} \,\, \cdot \,\mathbf{E} \hspace{0.5em}\right |}^{i,j,k,n} &=& 4 \pi \displaystyle \sum_a q_a \displaystyle \int_{\Reals^3} f_{a} \displaystyle \left ( \mathbf{r}_{i,j,k},\,\mathbf{v},\,t_n \right ) d^3 \mathbf{v}
\cr {\displaystyle \left . \displaystyle \frac{\partial}{\partial \mathbf{r}}  \times \mathbf{E} \hspace{0.5em} \right |}^{i,j,k,n} &=& \mathbf{0}
\end{array} \right . .
\end{equation}
The source term 
\begin{equation}
\label{non_physical_term}
C_a  \displaystyle \left ( \mathbf{r}_{i,j,k},\,\mathbf{v},\,t_n \right ) 
=  -  \displaystyle \frac{ \partial  }{\partial \mathbf{v}} \cdot \displaystyle \left ( \displaystyle \frac{q_a}{m_a} \delta N_a \displaystyle \sum_{\ell=1}^{N_a} \displaystyle \left [ \mathbf{E}_{s} \displaystyle \left ( \mathbf{r}_{a,\ell} \left ( t_n \right ),\, t_n \right ) - \mathbf{E}_s \displaystyle \left ( \mathbf{r}_{i,j,k},\,t_n\right ) \right ] S \displaystyle \left ( \mathbf{r}_{i,j,k} - \mathbf{r}_{a,\,\ell} \displaystyle \left ( t_n \right ) \right ) \delta \displaystyle \left ( \mathbf{v} - \mathbf{v}_{a,\,\ell} \displaystyle \left ( t_n \right ) \right ) \right)
\end{equation}
is due to the sum of forces $(q_a / m_a) \left [ \mathbf{E}_{s} \displaystyle \left ( \mathbf{r}_{a,\ell} \left ( t_n \right ),\, t_n \right ) - \mathbf{E}_s \displaystyle \left ( \mathbf{r}_{i,j,k},\,t_n\right ) \right ]$ acting on each macroparticle $a,\ell$ because of the internal tension force which is constraining the macroparticle charge and inertia as a unit, when different parts of the macroparticle feel a different electrostatic field \cite{Decyk1982}. According to its mathematical expression (\ref{non_physical_term}), these non-physical forces  decreases with decreasing spatial grid spacings and/or with increasing order of interpolation and smoothing functions. One must always check 
\begin{equation}
\begin{array}{lll}
&&\displaystyle \sum_{i,j,k} \displaystyle \int_{\Reals^3} \displaystyle \frac{m_a \mathbf{v}^2}{2} C_a  \displaystyle \left ( \mathbf{r}_{i,j,k},\,\mathbf{v},\,t_n \right ) d^3 \mathbf{v} \Delta_x \Delta_y \Delta_z
\cr &=& q_a \delta N_a \displaystyle \sum_{\ell=1}^{N_a} \displaystyle \left [ \mathbf{E}_{s} \displaystyle \left ( \mathbf{r}_{a,\ell} \left ( t_n \right ),\, t_n \right ) - \displaystyle \sum_{i,j,k} \mathbf{E}_s \displaystyle \left ( \mathbf{r}_{i,j,k},\,t_n\right ) S \displaystyle \left ( \mathbf{r}_{i,j,k} - \mathbf{r}_{a,\,\ell} \displaystyle \left ( t_n \right ) \right ) \Delta_x \Delta_y \Delta_z\right ] \cdot \mathbf{v}_{a,\,\ell} \displaystyle \left ( t_n \right )
\end{array}
\end{equation}
remains sufficiently small in case of electrostatic oscillations on the macroparticle size space scale by checking this energy flow associated with this internal tension force at each time step $t_n$ during the whole simulation.

The random number generator used to initialize macroparticle velocities $\mathbf{V}_{a,\ell}$  introduces statistical fluctuations 
\begin{equation}
\label{density_fluctuations}
\delta f_{a_c} \displaystyle \left ( \mathbf{r},\,\mathbf{v},\,t \right ) = f_{a_c} \displaystyle \left ( \mathbf{r},\,\mathbf{v},\,t \right ) - \bar{f}_{a_c} \displaystyle \left ( \mathbf{r},\,\mathbf{v},\,t \right )
\end{equation}
between the weighted macroparticle center phase-space densities (\ref{MCPSD}) from the simulation, seen here as a statistical realization of the plasma that one wants to simulate, and the corresponding weighted macroparticle center distribution functions
\begin{equation}
\label{average_density}
\bar{f}_{a_c} \displaystyle \left ( \mathbf{r},\,\mathbf{v},\,t \right ) = \mathbb{E} \displaystyle \left \{ f_{a_c} \displaystyle \left ( \mathbf{r},\,\mathbf{v},\,t \right )\right \}
\end{equation}
that ones would obtain on average over an infinite number of simulations and where we have thus noted $\mathbb{E} \displaystyle \left \{ X \right \}$ the expectation of the random variable $X$ from the probability theory. According to the previous section 2.2, the latter reads
\begin{equation}
\bar{f}_{a_c} \displaystyle \left ( \mathbf{r},\,\mathbf{v},\,t_1 \right ) = n_{a_c} \displaystyle \left ( \mathbf{r},\,t_1 \right ) F_{a_0} \displaystyle \left ( \mathbf{v}\right )
\end{equation}
at the simulation start $t_1$ where
\begin{equation}
\label{weighted_center_spatial_density}
n_{a_c} \displaystyle \left ( \mathbf{r},\,t_1 \right ) = \delta N_a \displaystyle \sum_{\ell= 1}^{N_a} \delta \displaystyle \left ( \mathbf{r} - \mathbf{R}_{a,\ell} \right )
\end{equation}
is the weighted macroparticle center spatial density. They lead to electrostatic field statistical fluctuations 
\begin{equation}
\label{field_fluctuations}
\delta \mathbf{E} \displaystyle \left ( \mathbf{r},\,t \right ) = \mathbf{E} \displaystyle \left ( \mathbf{r},\,t \right ) - \bar{\mathbf{E}} \displaystyle \left ( \mathbf{r},\,t \right )
\end{equation}
between the electrostatic field from the simulation and its expected value
\begin{equation}
\label{average_field}
\bar{\mathbf{E}} \displaystyle \left ( \mathbf{r},\,t \right ) =  \mathbb{E} \displaystyle \left \{  \mathbf{E} \displaystyle \left ( \mathbf{r},\,t \right ) \right \}
\end{equation}
corresponding to the plasma macroscopic state that one wants to simulate. We remind the reader that we're dealing here with the statistical fluctuations of a statistical realization of the plasma that one wants to simulate around its mean behaviour described by its distribution functions and not with spatial nor temporal fluctuations. In this sense, a PIC simulation is seen here as a numerical experiment. Taking the expected value of (\ref{Klimontovich}) and (\ref{Maxwell2}) at grid points $\mathbf{r}_{i,j,k}$ and time steps $t_n$, we obtain without any approximation
\begin{equation}
\label{averaged_Klimontovich}
\begin{array}{lll}
&&{\displaystyle \left .  \displaystyle \frac{ \partial \bar{f}_{a_c} }{\partial t} \right |}^{i,j,k,n} + {\displaystyle \left .  \displaystyle \frac{ \partial  }{\partial \mathbf{r}} \cdot \displaystyle \left ( \mathbf{v} \bar{f}_{a_c} \right ) \right |}^{i,j,k,n}+ \displaystyle \frac{ \partial  }{\partial \mathbf{v}} \cdot \displaystyle \left ( \displaystyle \frac{q_a}{m_a} \bar{\mathbf{E}}_s \left ( \mathbf{r}_{i,j,k},\,t_n\right ) \bar{f}_{a_c} \left ( \mathbf{r}_{i,j,k},\, \mathbf{v},\,t_n \right ) \right ) 
\cr &=& - \displaystyle \frac{q_a}{m_a}  \displaystyle \frac{ \partial }{\partial \mathbf{v}} \cdot \mathbb{E} \displaystyle \left \{ {\delta \mathbf{E}_s } \left ( \mathbf{r}_{i,j,k},\,t_n\right ) \delta f_{a_c}  \left ( \mathbf{r}_{i,j,k},\, \mathbf{v},\,t_n  \right )\right \} 
\end{array}
\end{equation} 
self-consistently coupled with the discretized macroscopic Maxwell equations
\begin{equation}
\label{averaged_Maxwell}
\displaystyle \left \{\begin{array}{lllll}
      {\displaystyle \left . \displaystyle \frac{\partial}{\partial \mathbf{r}} \,\, \cdot \,\bar{\mathbf{E}} \hspace{0.5em}\right |}^{i,j,k,n} &=& 4 \pi \displaystyle \sum_a q_a \displaystyle \int_{\Reals^3} \displaystyle \int_{\Reals^3} \bar{f}_{a_c} \displaystyle \left ( \mathbf{r}',\,\mathbf{v},\,t_n \right ) S \left ( \mathbf{r}_{i,j,k} - \mathbf{r}' \right ) d^3 \mathbf{r}' d^3 \mathbf{v}
\cr {\displaystyle \left . \displaystyle \frac{\partial}{\partial \mathbf{r}}  \times \bar{\mathbf{E}} \hspace{0.5em} \right |}^{i,j,k,n} &=& \mathbf{0}
\end{array} \right . .
\end{equation}
Here,
\begin{equation}
\bar{\mathbf{E}}_s \displaystyle \left ( \mathbf{r}_{a,\ell} \left ( t_n \right ),\,t_n \right ) =  \displaystyle \sum_{i,j,k} \bar{\mathbf{E}} \left ( \mathbf{r}_{i,j,k},\, t_n \right ) S \displaystyle \left ( \mathbf{r}_{a,\ell} \left ( t_n \right ) - \mathbf{r}_{i,j,k} \right ) \Delta_x \Delta_y \Delta_z
\end{equation}
is the expected electrostatic field interpolated at macroparticle center locations and
\begin{equation}
\delta \mathbf{E}_s \displaystyle \left ( \mathbf{r}_{i,j,k},\,t_n \right ) = \displaystyle \sum_{i,j,k} \delta {\mathbf{E}} \left ( \mathbf{r}_{i,j,k},\, t_n \right ) S \displaystyle \left ( \mathbf{r}_{a,\ell} \left ( t_n \right ) - \mathbf{r}_{i,j,k} \right ) \Delta_x \Delta_y \Delta_z
\end{equation}
the interpolated electrostatic field fluctuating component from the simulation.
Subtracting (\ref{averaged_Klimontovich}) from (\ref{Klimontovich}) and (\ref{Maxwell2}) from (\ref{averaged_Maxwell}), we obtain the self-consistently coupled equations
\begin{equation}
\label{linearized_Klimontovich}
\begin{array}{lll}
& &{\displaystyle \left . \displaystyle \frac{ \partial \delta f_{a_c} }{\partial t} \right |}^{i,j,k,n}  + {\displaystyle \left . \displaystyle \frac{ \partial  }{\partial \mathbf{r}} \cdot \displaystyle \left ( \mathbf{v} \delta f_{a_c} \right ) \right |}^{i,j,k,n}  +  \displaystyle \frac{ \partial  }{\partial \mathbf{v}} \cdot \displaystyle \left ( \displaystyle \frac{q_a}{m_a} \delta \mathbf{E}_s  \left ( \mathbf{r}_{i,j,k},\,t_n \right ) \bar{f}_{a_c}  \left ( \mathbf{r}_{i,j,k},\, \mathbf{v},\,t_n \right )\right ) 
 \cr &=& -\displaystyle \frac{q_a}{m_a} \displaystyle \frac{ \partial  }{\partial \mathbf{v}} \cdot \displaystyle \left (  \delta \mathbf{E}_s \left ( \mathbf{r}_{i,j,k},\,t_n \right ) \delta f_{a_c} \left ( \mathbf{r}_{i,j,k},\, \mathbf{v},\,t_n \right ) - \mathbb{E}{ \displaystyle \left \{ \delta \mathbf{E}_s \left ( \mathbf{r}_{i,j,k},\,t_n \right ) \delta f_{a_c} \left ( \mathbf{r}_{i,j,k},\, \mathbf{v},\,t_n \right ) \right \} }\right )
 \cr && - \displaystyle \frac{q_a}{m_a} \displaystyle \frac{ \partial  }{\partial \mathbf{v}} \cdot \displaystyle \left (  \bar{\mathbf{E}}_s \left ( \mathbf{r}_{i,j,k},\,t_n \right ) \delta f_{a_c} \left ( \mathbf{r}_{i,j,k},\, \mathbf{v},\,t_n \right ) \right )
\end{array}
\end{equation}
and
\begin{equation}
\label{Maxwell3}
\displaystyle \left \{\begin{array}{lllll}
      {\displaystyle \left . \displaystyle \frac{\partial}{\partial \mathbf{r}} \,\, \cdot \,\delta \mathbf{E} \hspace{0.5em}\right |}^{i,j,k,n}  &=& 4 \pi \displaystyle \sum_a q_a \displaystyle \int_{\Reals^3} \displaystyle \int_{\Reals^3} \delta f_{a_c} \displaystyle \left ( \mathbf{r}',\,\mathbf{v},\,t_n \right ) S \left ( \mathbf{r}_{i,j,k} - \mathbf{r}' \right ) d^3 \mathbf{r}' d^3 \mathbf{v}
\cr {\displaystyle \left . \displaystyle \frac{\partial}{\partial \mathbf{r}}  \times \delta \mathbf{E} \hspace{0.5em} \right |}^{i,j,k,n}  &=& \mathbf{0}
\end{array} \right . .
\end{equation}
They describe the evolution with time of statistical fluctuations introduced by different initial conditions implemented by random numbers. Knowing the statistical fluctuations $\delta \mathbf{E} \left ( \mathbf{r}_{i,j,k},\,t_1\right ) $ and $ \delta f_{a_c} \left ( \mathbf{r}_{i,j,k},\, \mathbf{v},\,t_1 \right )$ in all spatial grid cells $\mathbf{r}_{i,j,k}$ at the simulation start $t_1$, they allow to deduce their values at any time steps $t_n$.  

Let us emphasize here that the weighted macroparticle center phase-space density fluctuations (\ref{density_fluctuations}) are not the statistical fluctuations 
\begin{equation}
\delta f_{a} \displaystyle \left ( \mathbf{r},\,\mathbf{v},\,t \right )  = f_{a} \displaystyle \left ( \mathbf{r},\,\mathbf{v},\,t \right ) - \bar{f}_{a} \displaystyle \left ( \mathbf{r},\,\mathbf{v},\,t \right )
\end{equation}
between the expected plasma distribution functions
\begin{equation}
\label{distribution_function}
\bar{f}_{a} \displaystyle \left ( \mathbf{r},\,\mathbf{v},\,t \right ) = \mathbb{E} \displaystyle \left \{ f_{a} \displaystyle \left ( \mathbf{r},\,\mathbf{v},\,t \right )\right \}
\end{equation}
that one wants to simulate and the particle phase-space densities (\ref{RPPSD}) from the PIC simulation. Therefore, one has to take the expected value of (\ref{Klimontovich_like_equation}) and (\ref{PIC_Maxwell}) at grid points $\mathbf{r}_{i,j,k}$ and time steps $t_n$ in order to get the probabilistic description of electrostatic PIC simulations. One thus obtains without any approximation the discretized kinetic equation
\begin{equation}
\label{averaged_PIC_Klimontovich}
\begin{array}{lll}
&&{\displaystyle \left .  \displaystyle \frac{ \partial \bar{f}_{a} }{\partial t} \right |}^{i,j,k,n} + {\displaystyle \left .  \displaystyle \frac{ \partial  }{\partial \mathbf{r}} \cdot \displaystyle \left ( \mathbf{v} \bar{f}_{a} \right ) \right |}^{i,j,k,n} + \displaystyle \frac{ \partial  }{\partial \mathbf{v}} \cdot \displaystyle \left ( \displaystyle \frac{q_a}{m_a} \bar{\mathbf{E}}_s \left ( \mathbf{r}_{i,j,k},\,t_n\right ) \bar{f}_{a} \left ( \mathbf{r}_{i,j,k},\, \mathbf{v},\,t_n \right ) \right ) 
\cr &=& - \displaystyle \frac{q_a}{m_a}  \displaystyle \frac{ \partial }{\partial \mathbf{v}} \cdot \mathbb{E} \displaystyle \left \{ {\delta \mathbf{E}_s } \left ( \mathbf{r}_{i,j,k},\,t_n\right ) \delta f_{a}  \left ( \mathbf{r}_{i,j,k},\, \mathbf{v},\,t_n  \right )\right \} + \mathbb{E}{ \displaystyle \left \{ C_a \left ( \mathbf{r}_{i,j,k},\, \mathbf{v},\,t_n \right )  \right \} }
\end{array}
\end{equation} 
that is self-consistently coupled with the discretized macroscopic Maxwell equations
\begin{equation}
\label{averaged_PIC_Maxwell}
\displaystyle \left \{\begin{array}{lllll}
      {\displaystyle \left . \displaystyle \frac{\partial}{\partial \mathbf{r}} \,\, \cdot \,\bar{\mathbf{E}} \hspace{0.5em}\right |}^{i,j,k,n}&=& 4 \pi \displaystyle \sum_a q_a \displaystyle \int_{\Reals^3} \bar{f}_{a} \displaystyle \left ( \mathbf{r}_{i,j,k},\,\mathbf{v},\,t_n \right )  d^3 \mathbf{v}
\cr {\displaystyle \left . \displaystyle \frac{\partial}{\partial \mathbf{r}}  \times \bar{\mathbf{E}} \hspace{0.5em} \right |}^{i,j,k,n} &=& \mathbf{0}
\end{array} \right . .
\end{equation}

\section{Statistical fluctuations in electrostatic PIC simulations}

\subsection{Assumptions and notations}

We consider a stationary, homogeneous, infinite and fully ionized plasma consisting of a stable PIC simulation plasma that is initialized as detailed in the previous section 2.2 and that is reproduced periodically an infinite number of times in all directions and that evolves perpetually from the simulation start $t_1$ to infinity. Finite simulation box $L_x \times L_y \times L_z$ and time duration $L_t$ would only modify the spectral quantities by making them discrete with a resolution $\mathbf{\Delta}_k = \left ( 2 \pi / L_x,\, 2 \pi / L_y,\,2\pi / L_z \right )^t$ and $\Delta_\omega = 2 \pi / L_t$ due to the use of discrete inverse Laplace-Fourier transforms instead of the continuous ones (\ref{inverse_Fourier_transform}) and (\ref{inverse_Laplace_Fourier}). In addition, we assume
\begin{enumerate}
\item \label{Assumption_consistency_errors}Numerical consistency errors of numerical schemes used to compute the equations are small compared to statistical fluctuations introduced by different initial conditions implemented by random numbers.
\item \label{Assumption_homogeneous}The weighted macroparticle centers spatial densities are homogeneous at the simulation start
\begin{equation}
\forall i,\,j,\,k,\, n_{a_c} \left ( \mathbf{r}_{i,j,k},\,t_1 \right ) = \bar{n}_a
\end{equation}
so that
\begin{equation}
\forall i,\,j,\,k,\, f_{a_c} \left ( \mathbf{r}_{i,j,k},\,\mathbf{v},\,t_1 \right ) = \displaystyle \frac{ \bar{n}_a }{ N_{a,\mathrm{mpc}} } \displaystyle \sum_{\ell=1}^{N_{a,\mathrm{mpc}}} \delta \left ( \mathbf{v} - \mathbf{V}_{a,\ell_{i,j,k}} \right )
\end{equation}
where $\ell_{i,j,k} = \ell + \left [ \left ( i - 1 \right )  + \left ( j - 1 \right ) N_{x} + \left ( k - 1 \right ) N_{x} N_{y} \right ] N_{a,\mathrm{mpc}} $ and
\begin{equation}
\bar{f}_{a_c} \displaystyle \left ( \mathbf{r}_{i,j,k},\,\mathbf{v},\,t_1\right ) = \bar{n}_a F_{a_0} \left ( \mathbf{v} \right ).
\end{equation}
Therefore, we have also in this case 
\begin{equation}
\forall i,\,j,\,k,\, \bar{\mathbf{E}}  \left ( \mathbf{r}_{i,j,k},\,t_1 \right )  = \mathbf{0}
\end{equation} 
according to macroscopic Maxwell equations (\ref{averaged_Maxwell}), the macroparticle shape properties (\ref{particle_shape_property}) and Assumption \ref{Assumption_consistency_errors}.
\item \label{Assumption_collisions}We neglect the quadratic terms
\begin{equation}
 \displaystyle \frac{q_a}{m_a} \displaystyle \frac{ \partial  }{\partial \mathbf{v}} \cdot \displaystyle \left ( \delta \mathbf{E}_s \displaystyle \left ( \mathbf{r}_{i,j,k},\,t_n\right ) \delta f_{a_c} \displaystyle \left ( \mathbf{r}_{i,j,k},\,\mathbf{v},\,t_n\right ) - \mathbb{E}{ \displaystyle \left \{ \delta \mathbf{E}_s \displaystyle \left ( \mathbf{r}_{i,j,k},\,t_n\right )  \delta f_{a_c} \displaystyle \left ( \mathbf{r}_{i,j,k},\,\mathbf{v},\,t_n\right ) \right \} }\right ) 
\end{equation}
describing the influence of macroparticle close-encounter Coulomb collisions on the weighted macroparticle center discretized phase-space density statistical fluctuations (\ref{linearized_Klimontovich}).
\item \label{Assumption_stationary}The weighted macroparticle center distribution functions and macroscopic electrostatic fields are stationary
\begin{equation}
\forall n,\,  \bar{f}_{a_c} \displaystyle \left ( \mathbf{r}_{i,j,k},\,\mathbf{v},\,t_n\right ) = \bar{f}_{a_c} \displaystyle \left ( \mathbf{r}_{i,j,k},\,\mathbf{v},\,t_1\right ) \,\mathrm{ and }\, \bar{\mathbf{E}} \displaystyle \left ( \mathbf{r}_{i,j,k},\,t_n\right ) = \bar{\mathbf{E}} \displaystyle \left ( \mathbf{r}_{i,j,k},\,t_1 \right ).
\end{equation}
\end{enumerate}
In order to check Assumption \ref{Assumption_homogeneous}, let us introduce the Fourier transform in space of discrete functions
\begin{equation}
\label{Fourier_transform}
\widehat{\mathrm{F}} \left ( \mathbf{k} \right ) = \Delta_x \displaystyle \sum_{i=-\infty}^{\infty} \Delta_y \sum_{j=-\infty}^{\infty} \Delta_z \sum_{k=-\infty}^{\infty}  F \displaystyle \left ( \mathbf{r}_{i,j,k} \right ) \exp{\displaystyle \left ( - \iota \mathbf{k} \cdot \mathbf{r}_{i,j,k} \right ) }.
\end{equation}
such that 
\begin{equation}
\label{inverse_Fourier_transform}
F \displaystyle \left ( \mathbf{r}_{i,j,k} \right ) =  \displaystyle \int_{V_{\mathbf{k}_g}} \displaystyle \frac{d^3 \mathbf{k}}{ {\left ( 2 \pi \right )}^3 } \widehat{\mathrm{F}} \left ( \mathbf{k} \right )  \exp{ \displaystyle \left (  \iota  \mathbf{k} \cdot \mathbf{r}_{i,j,k} \right ) } 
\end{equation}
where the integration domain is the first Brillouin zone ${V_{\mathbf{k}_g}} = \left [ - k_{g_x} / 2, k_{g_x} / 2\right ] \times \left [ - k_{g_y} / 2, k_{g_y} / 2\right ] \times \left [ - k_{g_z} / 2, k_{g_z} / 2\right ]  $. The inverse Fourier transforms (\ref{inverse_Fourier_transform}) can be seen as Fourier series components of (\ref{Fourier_transform}). Indeed, according to its mathematical expression, the latter is necessarily $\mathbf{k}_g$-periodic where
\begin{equation}
\label{Nyquist_frequency}
\mathbf{k}_g = \displaystyle \left ( \begin{array}{c}
     2 \pi / \Delta_x
\cr 2 \pi / \Delta_y
\cr 2 \pi / \Delta_z
\end{array} \right )
\end{equation}
is the Nyquist frequency. It is defined as the highest spatial frequency that one can resolve in a PIC simulation due to discretization of space.
According to (\ref{density}), (\ref{initial_macroparticle_location}), (\ref{weighted_center_spatial_density}), (\ref{Fourier_transform}) and the Poisson summation formula applied first to the infinite sums over macroparticle indices $\ell_x$, $\ell_y$ and $\ell_z$ and then to the infinite sums over spatial indices $i$, $j$ and $k$, one finds
$$
\widehat{n}_{a_c} \displaystyle \left ( \mathbf{k},\, t_1 \right ) = {\left ( 2 \pi \right )}^3 \bar{n}_a \displaystyle \sum_{p,q,r} {\left ( - 1 \right )}^{p+q+r} \delta \left ( \mathbf{k} - \mathbf{I}_{p,q,r} \mathbf{k}_d,\,\mathbf{k}_g \right )
$$
where we have noted
$$
\delta \left ( \mathbf{k},\,\mathbf{k}_g \right ) = \displaystyle \sum_{p,q,r} \delta \displaystyle \left ( \mathbf{k} - \mathbf{I}_{p,q,r} \cdot \mathbf{k}_g\right ) = \delta \displaystyle \left ( k_x , k_{g_x} \right ) \delta \displaystyle \left ( k_y , k_{g_y} \right ) \delta \displaystyle \left ( k_z  , k_{g_z} \right )
$$
the 3D Dirac comb distribution,
$$
\delta \displaystyle \left ( k_x , k_{g_x} \right ) = \displaystyle \sum_{p=-\infty}^{\infty} \delta \displaystyle \left ( k_x - p k_{g_x} \right ),\,
\mathbf{I}_{p,q,r} = \displaystyle \left ( \begin{array}{ccc}
     p & 0 & 0
\cr 0 & q & 0
\cr 0 & 0 & r
\end{array} \right )
\, \mathrm{ and } \,
\mathbf{k}_d = \displaystyle \left ( \begin{array}{c}
     2 \pi N_{a_x} / L_x
\cr 2 \pi N_{a_y} / L_y
\cr 2 \pi N_{a_z} / L_z
\end{array} \right )
$$
to simplify the notations. Considering only the first Brillouin zone due to the discretization of space to compute Maxwell equations and a number of macroparticles per cell distributed in all directions greater than one
\begin{equation}
\label{Assumption2_condition}
\forall \xi \in \displaystyle \left \{ x,\,y,\,z\right \},\, \displaystyle \frac{N_{a_\xi}}{N_\xi } > 1,
\end{equation}
the Nyquist frequencies $ \mathbf{k}_{g_\xi}$ are necessarily lower than $\mathbf{k}_{d_\xi} $ and we thus have necessarily
\begin{equation}
\label{macroparticle_center_density_Fourier_transform}
\forall \mathbf{k} \in V_{\mathbf{k}_g},\,  \widehat{n}_{a_c} \displaystyle \left ( \mathbf{k},\, t_1 \right ) = {\left ( 2 \pi \right )}^3 \bar{n}_a \delta \displaystyle \left ( \mathbf{k}, \mathbf{k}_g\right ).
\end{equation}
The inverse Fourier transform  (\ref{inverse_Fourier_transform}) of the latter validates consequently Assumption \ref{Assumption_homogeneous} if the condition (\ref{Assumption2_condition}) is respected. A consequence of the condition (\ref{Assumption2_condition}) is that the number of macroparticles per cell must necessarily be $N_{a,\mathrm{mpc}} \geq 2$ in 1D, $\geq 4$ in 2D and $\geq 8$ in 3D according to its definition (\ref{NMPC}) in such a way that, by depositing the same amount strictly greater than 1 of macroparticles per cell in each directions, the code necessarily sees homogeneous weighted macroparticle center spatial densities. 

\begin{figure}
\centering
\includegraphics[height=6.cm]{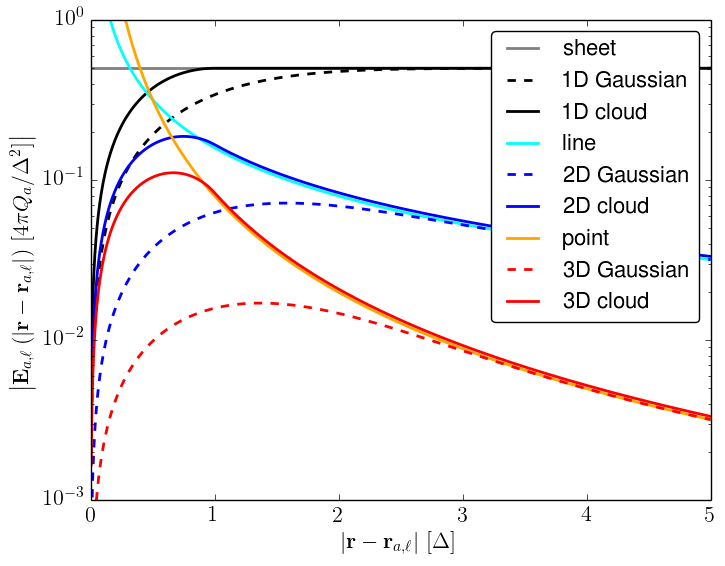}
\caption{Coulomb electrostatic field generated by one macroparticle $a,\ell$ compared to one "point" particle of same electrical charge in 1D (sheet), 2D (line) and 3D (point).}
\label{Figure1}
\end{figure}
In Figure \ref{Figure1}, we have plotted the Coulomb electrostatic field generated by one macroparticle $a,\ell$ in the cases it has a Gaussian shape function (\ref{smoothing_function}) with $a_x = a_y =a_z = \Delta$ or a cloud shape function (linear interpolating function $\Pi_{\mathbf{\boldsymbol{\Delta}}}^{(1)}$) with $\Delta = \Delta_x = \Delta_y = \Delta_z$. One can see that it is zero at the macroparticle location. Then, it increases until the distance $\left | \mathbf{r} - \mathbf{r}_{a,\ell} \right | \sim \Delta $ and it finally tends to the classical Coulomb electrostatic field generated by one point particle of equivalent electrical charge $Q_a$ when $\left | \mathbf{r} - \mathbf{r}_{a,\ell} \right | \gtrapprox \Delta $. Therefore, long range electrostatic interactions between macroparticles are correctly taken into account in electrostatic PIC simulations while short range electrostatic interaction between macroparticles when $\left | \mathbf{r} - \mathbf{r}_{a,\ell} \right | < \Delta $ are underestimated \cite{Okuda1970} \cite{Langdon1970c}. We can thus neglect Coulomb collisions between macroparticles and consider Assumption \ref{Assumption_collisions} to be valid.  We will check Assumptions \ref{Assumption_consistency_errors} and \ref{Assumption_stationary} later in section 4.1 and section 4.2, respectively. The following derivations are based on the theory of fluctuations \cite{Rostoker1961} \cite{Klimontovich1962} \cite{Dupree1963} well summarized in the Landau and Lifschitz textbook \cite{Landau1981}. Notations used are recapitulated in Table \ref{Table1}.

\begin{table}
\centering
\begin{tabular}{|c|l|}
    \hline $a$                                                             & index of plasma species
\cr $m_a$                                                                  & particle mass
\cr $q_a$                                                                  & particle charge
\cr $M_a$                                                                  & macroparticle mass
\cr $Q_a$                                                                  & macroparticle charge
\cr $\delta N_a$                                                           & macroparticle  weight
\cr $S$                                                                    & macroparticle shape function
\cr $N_a$                                                                  & total number of macroparticles
\cr $\ell$                                                                 & index of macroparticles
\cr $\mathbf{r}_{a,\ell}(t)$                                               & macroparticle center location
\cr $\mathbf{R}_{a,\ell}$                                                  & macroparticle center location at the simulation start
\cr $\mathbf{v}_{a,\ell}(t)$                                               & macroparticle velocity
\cr $\mathbf{V}_{a,\ell}$                                                  & macroparticle velocity at the simulation start
\cr $F_{a_0} (\mathbf{V}_{a})$                                             & probability distribution of macroparticle velocities at the simulation start
\cr \hline $\mathbf{r} $                                                   & space variable
\cr $\mathbf{v} $                                                        & velocity variable
\cr $t$                                                                    & time variable
\cr $\mathbf{E} (\mathbf{r},\,t) $                                         & self-consistent electrostatic field
\cr $\mathbf{E}_s(\mathbf{r}_{a,\ell}(t),\,t)$                             & self-consistent electrostatic field interpolated at macroparticle center locations
\cr $f_{a}(\mathbf{r},\,\mathbf{v}_a,\,t)$                                 & plasma species phase-space density sampled according to the PIC method
\cr $n_{a}(\mathbf{r},\,t)$                                                & plasma species density
\cr $f_{a_c}(\mathbf{r},\,\mathbf{v}_a,\,t)$                               & weighted phase-space density of macroparticle centers
\cr $n_{a_c}(\mathbf{r},\,t)$                                              & weighted density of macroparticle centers
\cr \hline $N_x,\,N_y,\,N_z$                                               & number of spatial grid points along $x$, $y$ and $z$-axis
\cr $i,\,j,\,k$                                                            & indices of spatial grid points along $x$, $y$ and $z$-axis
\cr $\Delta_x,\,\Delta_y,\,\Delta_z$                                       & spatial cell size  along $x$, $y$ and $z$-axis
\cr $\mathbf{r}_{i,j,k} $                                                  & spatial grid point location
\cr $N_t$                                                                  & number of time iterations
\cr $n$                                                                    & index of time iterations
\cr $\Delta_t$                                                             & time step
\cr $t_n $                                                                 & iteration time
\cr $N_{a,\mathrm{mpc}} $                                                  & number of macroparticles per cell
\cr \hline $\widehat{\mathrm{F}} \left ( \mathbf{k} \right ) $             & Fourier transform of discrete function $F \left ( \mathbf{r}_{i,j,k} \right )$
\cr $\overset{\frown}{\mathrm{F}} \left ( \mathbf{k} \right )$             & Fourier transform of continuous function $F \left ( \mathbf{r} \right )$
\cr $\widehat{\widehat{\mathrm{F}}} \left ( \omega,\,\mathbf{k} \right ) $ & Laplace-Fourier transform of discrete function $F \left ( \mathbf{r}_{i,j,k},\,t_n \right )$
\cr $k_{g,x},\,k_{g,y},\,k_{g,z}$                                          & Nyquist spatial angular frequency along $k_x$, $k_y$ and $k_z$-axis
\cr $p,\,q,\,r $                                                           & indices of spatial angular frequency aliases along $k_x$, $k_y$ and $k_z$-axis
\cr $\mathbf{k}_{p,q,r} $                                                  & spatial angular frequency vector alias
\cr $\displaystyle \left \{ \mathbf{k} \right \}_\mathbf{r} $              & $\mathbf{k}$-bracket depending on the Maxwell solver
\cr $\omega_g$                                                             & Nyquist temporal  angular frequency
\cr $m $                                                                   & index of temporal angular frequency aliases 
\cr $\omega_{m} $                                                          & temporal angular frequency aliases
\cr $\displaystyle \left \{ \omega \right \}_t $                           & $\omega$-bracket depending on the macroparticle pusher
\cr \hline $\bar{F} = \mathbb{E} \displaystyle \left \{ F \right \}$       & expectation of stochastic process $F$ (sometimes called ensemble average)
\cr $\delta{F} = F - \bar{F}$                                              & fluctuations of stochastic process $F$
\cr $ \mathbb{E} \displaystyle \left \{ \delta F(t)^2 \right \}$           & single-time autocorrelation of fluctuations $\delta F$
\cr $ \mathbb{E} \displaystyle \left \{ \delta F(t)\delta G(t) \right \}$  & single-time correlation between fluctuations of stochastic processes $F$ and $G$
\cr $N_{\ess}$                                                             & total number of  statistical realizations
\cr $\ess$                                                                 & index of statistical realizations
\cr $F^{(\ess)}$                                                           & statistical realization of $F$
\cr $\langle F \rangle $                                                   & finite ensemble average over $N_{\ess}$ statistical realizations
\cr \hline
\end{tabular}
\caption{Recapitulation of main notations used in the text.}
\label{Table1}
\end{table}

\subsection{Estimate of statistical fluctuations in a PIC simulation electrostatic plasma}

According to their definitions (\ref{density_fluctuations}), (\ref{average_density}), (\ref{field_fluctuations}) and (\ref{average_field}), there is no electrostatic field statistical fluctuations 
\begin{equation}
\label{mean0}
\mathbb{E} \displaystyle \left \{ \delta \mathbf{E} \displaystyle \left ( \mathbf{r}_{i,j,k},\, t_n \right ) \right \} = 0
\end{equation}
nor weighted macroparticle center phase-space density statistical fluctuations
\begin{equation}
\mathbb{E} \displaystyle \left \{ \delta f_{a_c} \displaystyle \left ( \mathbf{r}_{i,j,k},\,\mathbf{v},\, t_n \right ) \right \}= 0
\end{equation}
on average. Therefore, in order to get an estimate of the non-zero electrostatic field -usually called "numerical noise"- that one observes in PIC simulations  due to the macroparticle initial conditions implemented by random numbers, we propose here to estimate it according to the square root of its statistical fluctuation single-time autocorrelation 
\begin{equation}
\label{variance0}
{\displaystyle \left | \mathbf{E}_\mathrm{fluct} \right |} = \displaystyle \sqrt{ \mathbb{E} \displaystyle \left \{ \delta \mathbf{E} \displaystyle \left ( \mathbf{r}_{i,j,k},\, t_n \right )^2 \right \} }.
\end{equation}
Indeed, as it will be justified later in section 5.1, if the number of macroparticles per cell is sufficiently large, the electrostatic field statistical fluctuation $ \delta \mathbf{E} \displaystyle \left ( \mathbf{r}_{i,j,k},\,,t_n \right ) $ at grid points $\mathbf{r}_{i,j,k}$ and time steps $t_n$ follows a normal distribution law with mean (\ref{mean0}) and variance (\ref{variance0}) according to the central limit theorem. As a consequence,
\begin{equation}
\label{statistical_fluctuation_amplitude_0}
 \displaystyle \left | \displaystyle \delta \mathbf{E} \displaystyle \left ( \mathbf{r}_{i,j,k},\,t_n \right ) \right | \underset{N_{e,\mathrm{mpc}} \gg 1}{\leq} \displaystyle \sqrt{ 2 }\,\mathrm{erf}^{-1} \displaystyle \left ( \displaystyle \frac{p}{100} \right )\displaystyle \left | \mathbf{E}_\mathrm{fluct} \right |
\end{equation}
with a $p \%$-confidence and the square root of electrostatic field statistical fluctuation autocorrelation (\ref{variance0}) thus provides a good estimate of statistical fluctuations.

To carry out such a program, we use the Laplace-Fourier transforms of discrete functions
\begin{equation}
\label{discrete_Laplace_Fourier}
\begin{array}{lll}
\widehat{\widehat{\mathrm{F}}} \left ( \omega,\,\mathbf{k} \right ) &=&  \Delta_t \displaystyle \sum_{n=1}^{\infty} \Delta_x \displaystyle \sum_{i=-\infty}^{\infty} \Delta_y \sum_{j=-\infty}^{\infty} \Delta_z \sum_{k=-\infty}^{\infty} \displaystyle F \displaystyle \left ( \mathbf{r}_{i,j,k},\,t_n \right ) \exp{\displaystyle \left [ \iota \displaystyle \left ( \omega t_n - \mathbf{k} \cdot \mathbf{r}_{i,j,k} \right ) \right ]}
\cr  &=&\Delta_t \displaystyle \sum_{n=1}^{\infty}  \,\, \widehat{\mathrm{F}} \displaystyle \left ( \mathbf{k},\,t_n \right ) \exp{\displaystyle \left ( \iota \omega t_n \right )}
\end{array}
\end{equation}
such that
\begin{equation}
\label{inverse_Laplace_Fourier}
F \displaystyle \left ( \mathbf{r}_{i,j,k},\,t_n \right ) =  \displaystyle \int_{\iota \nu - \omega_g / 2}^{\iota \nu + \omega_g/2} \displaystyle \frac{d \omega}{ 2 \pi } \displaystyle \int_{V_{\mathbf{k}_g}}  \displaystyle \frac{d^3 \mathbf{k} }{ {\left ( 2 \pi \right )}^3 }  \,\, \widehat{\widehat{\mathrm{F}}} \left ( \omega,\,\mathbf{k} \right )  \exp{\displaystyle \left [ - \iota \displaystyle \left ( \omega t_n - \mathbf{k} \cdot \mathbf{r}_{i,j,k} \right ) \right ]}
\end{equation}
where $\nu \in \Reals$ such that the contour path of integration is in the region of convergence. Again, according to their definitions (\ref{discrete_Laplace_Fourier}), one can see that they are necessarily $\left ( \omega_g,\,\mathbf{k}_g \right ) $-periodic where the Nyquist frequencies are (\ref{Nyquist_frequency}) and 
\begin{equation}
\omega_g = 2 \pi / \Delta_t.
\end{equation} 
This is not the case for the macroparticle shape function (\ref{particle_shape}) whose support is necessarily continuous since macroparticle positions $\mathbf{r}_{a,\ell} \left ( t_n \right )$ may have spatial variations less than the grid spacing. We will therefore note
\begin{equation}
\overset{\frown}{\mathrm{S}} \left ( \mathbf{k} \right ) = \displaystyle \int_{\Reals^3} d^3 \mathbf{r} \displaystyle S \displaystyle \left ( \mathbf{r} \right ) \exp{\displaystyle \left (-  \iota \mathbf{k} \cdot \mathbf{r} \right ) }
\end{equation}
its continuous Fourier transform such that
\begin{equation}
\displaystyle S \displaystyle \left ( \mathbf{r} \right ) = \displaystyle \int_{\Reals^3} \displaystyle \frac{d^3 \mathbf{k}}{ { \left ( 2 \pi \right )}^3 } \overset{\frown}{\mathrm{S}} \left ( \mathbf{k} \right ) \exp{\displaystyle \left (  \iota \mathbf{k} \cdot \mathbf{r} \right ) }
\end{equation}
to avoid misunderstandings. For example, in a spectral simulation using a $n$th-order B-spline interpolating function and a Gaussian filter, it simply reads
$$ 
\overset{\frown}{\mathrm{S}} \left ( \mathbf{k} \right ) = \prod_{\xi=x,y,z} { \displaystyle \left [ \displaystyle \frac{ \sin{ \displaystyle \left ( k_{\xi} \Delta_\xi / 2 \right ) } }{  k_{\xi} \Delta_\xi / 2 } \right ]}^{n+1} \exp{ \displaystyle \left ( - \displaystyle  \frac{ {a_\xi}^2 {k_\xi}^2 }{2}\right ) }.
$$

By linearizing the leap-frog macroparticle pusher algorithm 
\begin{equation}
\label{discretized_EoM}
\displaystyle \left \{ \begin{array}{clllr}
     \displaystyle \frac{\mathbf{r}_{a,\ell}\displaystyle \left ( t_{n+1} \right )-  \mathbf{r}_{a,\ell}\displaystyle \left ( t_{n} \right ) }{\Delta_t } &=& \mathbf{v}_{a,\ell}\displaystyle \left ( t_{n+1/2}  \right )  ,& \mathbf{r}_{a,\ell}\displaystyle \left ( t_{1} \right ) &= \mathbf{R}_{a,\ell}
\cr \displaystyle \frac{\mathbf{v}_{a,\ell}\displaystyle \left ( t_{n+1/2} \right )- \mathbf{v}_{a,\ell}\displaystyle \left ( t_{n-1/2} \right ) }{\Delta_t } &=& \displaystyle \frac{Q_a}{M_a} \mathbf{E}_{s} \left ( \mathbf{r}_{a,\ell}^{n},\, t_n \right ) ,& \mathbf{v}_{a,\ell}\displaystyle \left ( t_{1/2} \right ) &= \mathbf{V}_{a,\ell}
\end{array} \right ., 
\end{equation}
that is commonly used to compute the discretized macroparticle equations of motions (\ref{EoM}), and by associating the non-perturbed straight line orbits $( \mathbf{r}_{a,\ell}^{(0)} \left ( t_n \right ),\, \mathbf{v}_{a,\ell}^{(0)} \left ( t_n \right ) ) = \left ( \mathbf{R}_{a,\ell} + \mathbf{V}_{a,\ell} t,\,\mathbf{V}_{a,\ell} \right )$ with $\bar{f}_{a_c} $ and the first order perturbed orbits $\left ( \delta \mathbf{r}_{a,\ell} \left ( t_n \right ),\, \delta \mathbf{v}_{a,\ell} \left ( t_n \right ) \right )$ with $\delta f_{a_c}$ according to Assumptions \ref{Assumption_consistency_errors},  \ref{Assumption_homogeneous},  \ref{Assumption_collisions} and  \ref{Assumption_stationary} \cite{Lindman1970},  the discrete Laplace-Fourier transform (\ref{discrete_Laplace_Fourier}) of the resulting discretized equation (\ref{linearized_Klimontovich}) gives 
\begin{equation}
\label{discretized_linearized_Klimontovich}
\iota \displaystyle \left \{  \mathbf{k} \cdot \mathbf{v} - \omega \right \}_{t}  \widehat{\widehat{\delta f}}_{a_c} \displaystyle \left ( \omega,\, \mathbf{k},\, \mathbf{v} \right ) + \bar{n}_a \displaystyle \frac{q_a}{m_a} \displaystyle \sum_{p,q,r} \overset{\frown}{\mathrm{S}} \left ( \mathbf{k}_{p,q,r} \right )   \widehat{\widehat{\delta\mathbf{E}}} \displaystyle \left ( \omega,\, \mathbf{k}_{p,q,r} \right ) \cdot \displaystyle \frac{d F_{a_0}}{ d \mathbf{v} } = \widehat{\delta f}_{a_c} \displaystyle \left ( \mathbf{k},\, \mathbf{v},\, t_1 \right )
\end{equation}
where
\begin{equation}
\displaystyle \frac{1}{\displaystyle \left \{  \mathbf{k} \cdot \mathbf{v} - \omega \right \}_{t}} = \displaystyle \frac{ \Delta_t }{ 2 } \cot{ \displaystyle \left [ \displaystyle \left ( \mathbf{k} \cdot \mathbf{v} - \omega \right ) \displaystyle \frac{ \Delta_t }{ 2 }  \right ]} = \displaystyle \sum_{ m } \displaystyle \frac{1}{  \mathbf{k} \cdot \mathbf{v} - \omega_m }
\end{equation}
\cite{Langdon1970a} \cite{Birdsall1991} \cite{Abramowitz}. Here, we have noted
\begin{equation}
\omega_m = \omega - m \omega_g
\end{equation}
and
\begin{equation}
\mathbf{k}_{p,q,r} = \mathbf{k} - \mathbf{I}_{p,q,r}  \cdot \mathbf{k}_g
\end{equation}
the different temporal and spatial angular frequency aliases, following the same notation as in \cite{Langdon1970b} and \cite{Birdsall1991}. Maxwell equations (\ref{Maxwell2}) are usually computed by using a spectral solver or the second order FDTD method proposed by \cite{Yee1966} coupled with the charge conserving scheme proposed by \cite{Villasenor1992} or \cite{Esirkepov2001} rather than the Poisson equation solver considered in \cite{Langdon1970a}. In all cases, the Laplace-Fourier transform (\ref{discrete_Laplace_Fourier}) of the resulting Maxwell equations (\ref{Maxwell3}) linearized according to Assumptions \ref{Assumption_consistency_errors},  \ref{Assumption_homogeneous} and  \ref{Assumption_stationary} gives
\begin{equation}
\label{discretized_Maxwell3}
\displaystyle \left \{\begin{array}{lllll}
      \iota \displaystyle \left \{ \mathbf{k} \right \}_\mathbf{r} \,\, \cdot \, \widehat{\widehat{\delta\mathbf{E}}} \left ( \omega,\,\mathbf{k} \right ) &=& 4 \pi \displaystyle \sum_a q_a \displaystyle \sum_{p',q',r'} \overset{\frown}{\mathrm{S}} \left ( \mathbf{k}_{p',q',r'}  \right ) \displaystyle \int_{\Reals^3} \widehat{\widehat{\delta f}}_{a_c} \displaystyle \left ( \omega,\,\mathbf{k}_{p',q',r'},\,\mathbf{v}\right )  d^3 \mathbf{v}
\cr \iota \displaystyle \left \{ \mathbf{k} \right \}_\mathbf{r} \times  \widehat{\widehat{\delta\mathbf{E}}} \left ( \omega,\,\mathbf{k} \right )  &=& \mathbf{0}
\end{array} \right .
\end{equation}
where we have noted $\forall \mathbf{k} \in V_{\mathbf{k}_g},$
\begin{equation}
 {
\displaystyle \left \{ \mathbf{k} \right \}_\mathbf{r} = \displaystyle \left \{ \begin{array}{lcl}
     & {\displaystyle \left ( k_x,\, k_y,\,k_z\right ) }^t & \mathrm{ \, for \, the \, spectral \, solver }
\cr & {\displaystyle \left ( \displaystyle \frac{ \sin{\left ( k_x \Delta_x / 2\right )} }{ \Delta_x / 2 },\,  \displaystyle \frac{ \sin{\left ( k_y \Delta_y / 2\right )} }{ \Delta_y / 2 } ,\, \displaystyle \frac{ \sin{\left ( k_z \Delta_z / 2\right )} }{ \Delta_z / 2 } \right ) }^t & \mathrm{ \, for \, the \, 2nd \, order \,FDTD \, scheme}
\end{array} \right . 
}
\end{equation}
and $\forall p,\,q,\,r$, $\displaystyle \left \{ \mathbf{k}_{p,q,r} \right \}_\mathbf{r} = \displaystyle \left \{ \mathbf{k} \right \}_\mathbf{r} $.
By substituting $\widehat{\widehat{\delta f}}_{a_c} \displaystyle \left ( \omega,\, \mathbf{k}_{p',q',r'},\, \mathbf{v} \right )$ from (\ref{discretized_linearized_Klimontovich}) into the expression of $\delta \widehat{\widehat{ \mathbf{E}}} \left ( \omega,\, \mathbf{k} \right ) $ from (\ref{discretized_Maxwell3}), replacing $\mathbf{k}$ by $\mathbf{k}_{p,q,r}$, changing the indices $\left ( p',q',r' \right )$ into $\left (p'-p,q'-q,r'-r\right )$ in the resulting equation, one can impose then $\left ( p,q,r \right ) = \left ( -p',-q',-r' \right )$ and use the $\mathbf{k}_g $-periodicity of $\displaystyle \left \{ \mathbf{k} \right \}_\mathbf{r}$, $ \widehat{\widehat{\delta\mathbf{E}}}$ and $ \delta \widehat{f}_{a_c}$ to obtain the discrete Laplace-Fourier transform (\ref{discrete_Laplace_Fourier}) of electrostatic field statistical fluctuations \cite{Langdon1970b}. It reads 
\begin{equation}
\label{calculus_start}
 \widehat{\widehat{\delta\mathbf{E}}} \left ( \omega,\, \mathbf{k} \right ) = \displaystyle \frac{- 4 \pi  \left \{\mathbf{k}\right\}_\mathbf{r} }{ \varepsilon_L \displaystyle \left ( \omega, \mathbf{k} \right ) {\left \{\mathbf{k}\right\}_\mathbf{r}}^2} \displaystyle \sum_{p,q,r} \overset{\frown}{\mathrm{S}} \left ( \mathbf{k}_{p,q,r} \right ) \displaystyle \sum_a q_a \sum_m \displaystyle \int_{\Reals^3} \displaystyle \frac{ \delta \widehat{f}_{a_c} \left ( \mathbf{k},\,\mathbf{v},\,t_1 \right )  }{ \mathbf{k}_{p,q,r} \cdot \mathbf{v} - \omega_m } d^3 \mathbf{v}
\end{equation}
where the longitudinal permittivity of the PIC simulation plasma reads
\begin{equation}
\label{numerical_permitivity}
\begin{array}{lll}
 \varepsilon_L \displaystyle \left ( \omega, \mathbf{k} \right ) &=& 1 - \displaystyle \sum_a \displaystyle \frac{ { \omega_{p_a} }^2 }{ {\displaystyle \left \{\mathbf{k}\right\}_\mathbf{r}}^2 } \displaystyle \sum_{p,q,r}  {\overset{\frown}{\mathrm{S}} \left ( \mathbf{k}_{p,q,r} \right )}^2  \displaystyle \sum_{m} \displaystyle \int_{\Reals^3}  \displaystyle \frac{ \displaystyle \left \{\mathbf{k}\right\}_\mathbf{r} }{ \mathbf{k}_{p,q,r} \cdot \mathbf{v} - \omega_m } \cdot \displaystyle \frac{ d F_{a0} }{ d \mathbf{v} } d^3 \mathbf{v}
\end{array}
\end{equation}
with
\begin{equation}
{\omega_{p_a}} = \displaystyle \sqrt{ \displaystyle \frac{ 4 \pi \bar{n}_a {q_a}^2 }{ m_a } }
\end{equation}
the Langmuir angular frequencies of each plasma species. We can check that it tends to the longitudinal permittivity of collisionless plasmas in the limit $\forall \xi \in \left \{x,\,y,\,z,\,t \right \},\, \Delta_\xi \rightarrow 0$. The study of poles $\omega = \omega \left ( \mathbf{k} \right )$ given by the dispersion relation $\varepsilon_L \displaystyle \left ( \omega, \mathbf{k} \right ) = 0$ provides the stability criterion of the simulated plasma. We have assumed the simulated plasma remains stable, both physically and numerically. It means that all these poles have a negative imaginary part $\mathrm{Im} \left \{ \omega \left ( \mathbf{k} \right ) \right \} < 0$. In general, the stability criterion depends on the macroparticle shape and initial macroparticle velocity probability distributions $F_{a_0}$. 
For example, it has been shown that electrostatic PIC simulations of plasmas at equilibrium using the first or greater orders B-spline interpolating function (and same spatial spacings in all directions for simplicity) are stable if the condition 
\begin{equation}
\Delta_t \leq \Delta_x / v_{T_e} < 1 / \omega_p
\end{equation}
is respected \cite{Birdsall1991} where 
\begin{equation}
{\omega_p}^2 = {\omega_{p_e}}^2 + {\omega_{p_i}}^2 \approx  {\omega_{p_e}}^2 
\end{equation}
is the Langmuir plasma angular frequency in the limit $ Z m_e / m_i \ll 1 $. 

\begin{figure}
\centering
\includegraphics[height=7.cm]{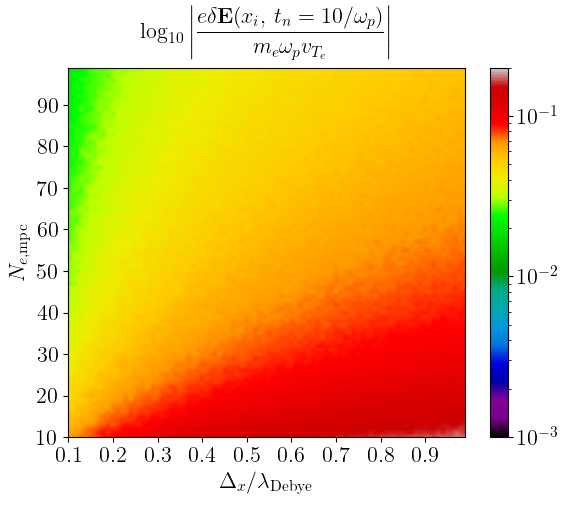}
\includegraphics[height=7.cm]{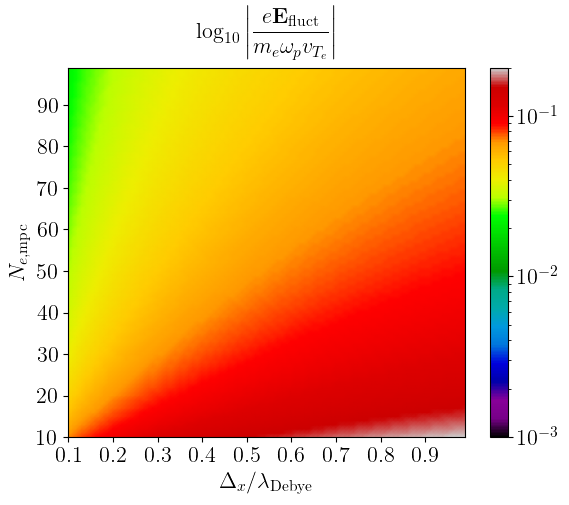}\\
\includegraphics[height=7.cm]{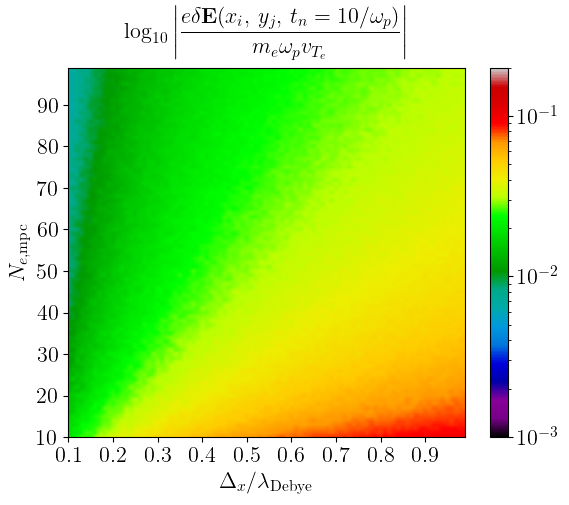}
\includegraphics[height=7.cm]{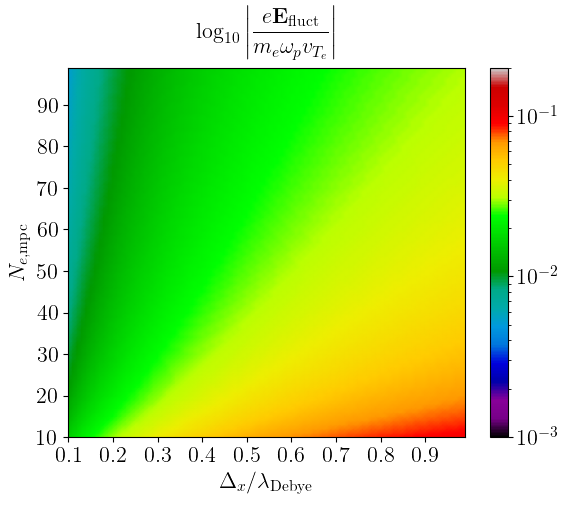}\\
\includegraphics[height=7.cm]{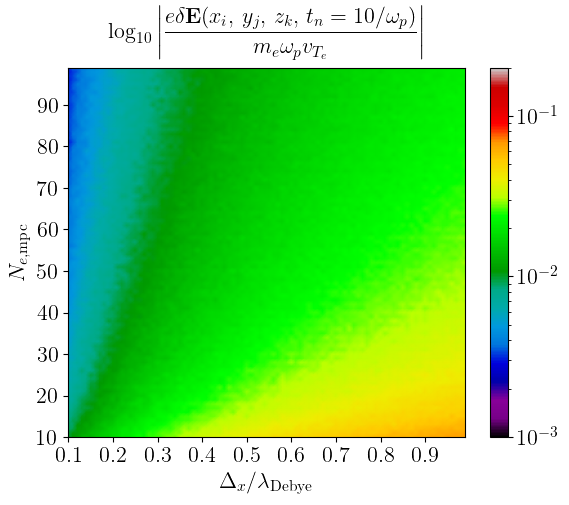}
\includegraphics[height=7.cm]{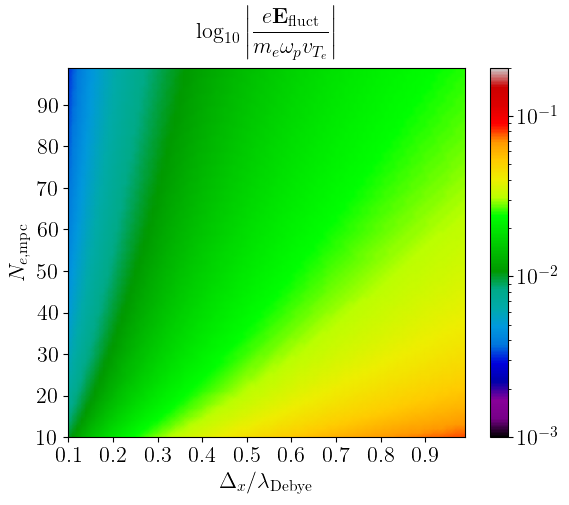}\\
\caption{Average over all spatial cells of electrostatic field fluctuations amplitudes at the discrete time $t_n = 10/\omega_p$ (Left) and their analytical estimates (\ref{electrostatic_field_fluctuations_estimate}) (Right) in 1D (Top), 2D (Middle) and 3D (Bottom) periodic electrostatic PIC plasmas at Maxwell-Boltzmann equilibrium simulated with the spectral electrostatic PIC codes from the UPIC framework \cite{Decyk1995} \cite{Decyk2007} \cite{Decyk2014}; each simulation out of the 30,000  uses a total of 4096 spatial cells and interpolation functions $\Pi^{(1)}_{\Delta_x}$ without filtering.}
\label{Figure2}
\end{figure}

One can now deduce the expression of electrostatic field statistical fluctuations autocorrelation from the discrete Laplace-Fourier transform of electrostatic field statistical fluctuations  (\ref{calculus_start}). It reads
\begin{equation}
\label{calculus_mm0}
\begin{array}{lll}
\mathbb{E} \displaystyle \left \{ {\delta \mathbf{E} \left ( \mathbf{r}_{i,j,k},\,t_n \right )}^2 \right \} &=& \displaystyle \int_{\iota \nu - \omega_g / 2}^{\iota \nu + \omega_g/2} \displaystyle \frac{d \omega}{ 2 \pi } \displaystyle \int_{V_{\mathbf{k}_g}} \displaystyle \frac{d^3 \mathbf{k}}{ {\left ( 2 \pi \right )}^3 } \displaystyle \int_{\iota \nu' - \omega_g / 2}^{\iota \nu' + \omega_g/2} \displaystyle \frac{d \omega'}{ 2 \pi } \displaystyle \int_{V_{\mathbf{k}_g}} \displaystyle \frac{d^3 \mathbf{k}'}{ {\left ( 2 \pi \right )}^3 }
\cr & & \mathbb{E} \displaystyle \left \{  \widehat{\widehat{\delta\mathbf{E}}} \left ( \omega, \mathbf{k} \right )  \widehat{\widehat{\delta\mathbf{E}}} \left ( \omega',\,\mathbf{k}' \right ) \right \} \exp{\displaystyle \left \{ - \iota \displaystyle \left [ \left (\omega + \omega' \right ) t_n - \left ( \mathbf{k} + \mathbf{k}' \right ) \cdot \mathbf{r}_{i,j,k} \right ] \right \}}
\end{array}
\end{equation}
where
\begin{equation}
\label{calculus_mm1}
\begin{array}{ll}
\mathbb{E} \displaystyle \left \{  \widehat{\widehat{\delta\mathbf{E}}} \left ( \omega, \mathbf{k} \right )  \widehat{\widehat{\delta\mathbf{E}}} \left ( \omega',\,\mathbf{k}' \right ) \right \} &=  \displaystyle \frac{ {\left ( 4 \pi \right )}^2 }{ \varepsilon_L \displaystyle \left ( \omega, \mathbf{k} \right ) \varepsilon_L \displaystyle \left ( \omega', \mathbf{k}' \right ) }  \displaystyle \frac{  {\left \{\mathbf{k}\right\}_\mathbf{r}} \cdot {\left \{\mathbf{k}'\right\}_\mathbf{r}} }{ {\left \{\mathbf{k}\right\}_\mathbf{r}}^2 {\left \{\mathbf{k}'\right\}_\mathbf{r}}^2 } \displaystyle \sum_{p,q,r} \displaystyle \sum_{p',q',r'} \overset{\frown}{\mathrm{S}} \left ( \mathbf{k}_{p,q,r} \right )  \overset{\frown}{\mathrm{S}} \left ( \mathbf{k}'_{p',q',r'} \right ) 
\cr & \displaystyle \sum_{a,b} {q_a} {q_b} \displaystyle \sum_{m,m'} \displaystyle \int_{\Reals^3} \displaystyle \int_{\Reals^3} \displaystyle \frac{ \mathbb{E} \displaystyle \left \{ \delta \widehat{f}_{a_c} \left ( \mathbf{k},\,\mathbf{v},\,t_1 \right )\delta \widehat{f}_{b_c} \left ( \mathbf{k}',\,\mathbf{v}',\,t_1 \right )  \right \}  }{ \displaystyle \left ( \mathbf{k}_{p,q,r} \cdot \mathbf{v} - \omega_m \right )  \displaystyle \left ( {\mathbf{k}'}_{p',q',r'} \cdot \mathbf{v}' - {\omega'}_{m'} \right ) } d^3 \mathbf{v} d^3 \mathbf{v}'.
\end{array}
\end{equation}
It depends on the weighted macroparticle center phase-space density fluctuations Fourier transforms correlation at the simulation start $t_1$. One can derive the exact analytical expression of the latter. One finds according to section 2.2
\begin{equation}
\label{correlation}
\begin{array}{l}
\mathbb{E} \displaystyle \left \{ \delta f_{a_c} \left ( \mathbf{r},\,\mathbf{v},\,t_1 \right ) \delta f_{b,c} \left ( \mathbf{r}',\,\mathbf{v}',\,t_1 \right ) \right \} 
= \delta_{ab}  \delta N_a \delta \left ( \mathbf{r} - \mathbf{r}' \right ) n_{a_c} \displaystyle \left ( \mathbf{r},\,t_1\right )  F_{a0} \displaystyle \left ( \mathbf{v} \right ) \displaystyle \left [ \delta \displaystyle \left ( \mathbf{v} - \mathbf{v}' \right ) - F_{a0} \displaystyle \left ( \mathbf{v}' \right )\right ].
\end{array}
\end{equation}
The first term in the square bracket corresponds to the case where there is only one macroparticle in the infinitezimal phase-space volume located between $\left ( \mathbf{r},\,\mathbf{v} \right )$ and $\left ( \mathbf{r} + d^3 \mathbf{r},\,\mathbf{v} + d^3 \mathbf{v} \right )$. Replacing consequently $\delta \left ( \mathbf{r} - \mathbf{r}' \right ) $ by its equivalent expression $ \delta_{i,i'} \delta_{j,j'} \delta_{k,k'} / \Delta_x \Delta_y \Delta_z$ in a discretized space, one can compute the Fourier transform (\ref{Fourier_transform}) of the initial weighted macroparticle center phase-space densities correlation (\ref{correlation}) expressed at the grid points $\mathbf{r}_{i,j,k}$ and $\mathbf{r}_{i',j',k'}$. It reads
\begin{equation}
\label{discretized_correlation}
\begin{array}{l}
\mathbb{E} \displaystyle \left \{ \delta \widehat{f}_{a_c} \left ( \mathbf{k},\,\mathbf{v},\,t_1 \right ) \delta \widehat{f}_{b,c} \left ( \mathbf{k}',\,\mathbf{v}',\,t_1 \right ) \right \} 
= \delta_{ab} \delta N_a {\displaystyle \left ( 2 \pi \right )}^3 \bar{n}_a \delta \left ( \mathbf{k} + \mathbf{k}',\,\mathbf{k}_g \right ) F_{a0} \displaystyle \left ( \mathbf{v} \right ) \displaystyle \left [ \delta \displaystyle \left ( \mathbf{v} - \mathbf{v}' \right ) - F_{a0} \displaystyle \left ( \mathbf{v}' \right )\right ]
\end{array}
\end{equation}
according to Assumption \ref{Assumption_homogeneous}. By performing successively the integration over $\mathbf{k}'$, $\mathbf{v}'$ and $\omega'$ in (\ref{calculus_mm0}) (see Appendix A), one finds
\begin{equation}
\label{electrostatic_field_fluctuations_spectrum_main_results}
\mathbb{E} \displaystyle \left \{  \widehat{\widehat{\delta\mathbf{E}}} \left ( \omega, \mathbf{k} \right )  \widehat{\widehat{\delta\mathbf{E}}} \left ( \omega',\,\mathbf{k}' \right ) \right \}  \underset{\omega_ p t_n \gg 1}{=} {\displaystyle \left ( 2 \pi \right )}^4 \delta \displaystyle \left ( \omega + \omega',\,\omega_g  \right ) \delta \displaystyle \left ( \mathbf{k} + \mathbf{k}',\,\mathbf{k}_g \right ) \widehat{\widehat{ {\delta\mathbf{E}}^2 }}   \left ( \omega, \mathbf{k} \right )
\end{equation}
with 
\begin{equation}
\label{electrostatic_field_fluctuations_spectrum}
\begin{array}{lcl}
\widehat{\widehat{ {\delta\mathbf{E}}^2 }}   \left ( \omega, \mathbf{k} \right )&=& \displaystyle \int_{\iota \nu' - \omega_g / 2}^{\iota \nu' + \omega_g/2} \displaystyle \frac{d \omega'}{ 2 \pi } \displaystyle \int_{V_{\mathbf{k}_g}} \displaystyle \frac{d^3 \mathbf{k}'}{ {\left ( 2 \pi \right )}^3 }
\cr & & \mathbb{E} \displaystyle \left \{  \widehat{\widehat{\delta\mathbf{E}}} \left ( \omega, \mathbf{k} \right )  \widehat{\widehat{\delta\mathbf{E}}} \left ( \omega',\,\mathbf{k}' \right ) \right \} \exp{\displaystyle \left \{ - \iota \displaystyle \left [ \left (\omega + \omega' \right ) t_n - \left ( \mathbf{k} + \mathbf{k}' \right ) \cdot \mathbf{r}_{i,j,k} \right ] \right \}}
\cr &\underset{\omega_ p t_n \gg 1}{=}& \displaystyle \frac{ 32 \pi^3  }{ {\varepsilon_L \displaystyle \left ( \omega, \mathbf{k} \right )}^2 } \displaystyle \frac{  1 }{ {\left \{\mathbf{k}\right\}_\mathbf{r}}^2 } \displaystyle \sum_{p,q,r} {\overset{\frown}{\mathrm{S}} \left ( \mathbf{k}_{p,q,r} \right )}^2  \displaystyle \sum_a \delta N_a \bar{n}_a {q_a}^2 \displaystyle \sum_{m} \displaystyle \int_{\Reals^3} d^3 \mathbf{v} F_{a0} \displaystyle \left ( \mathbf{v} \right ) \delta \displaystyle \left ( \omega_m - \mathbf{k}_{p,q,r} \cdot \mathbf{v} \right ) .
\end{array}
\end{equation}
in the expression of single-time electrostatic field statistical fluctuations autocorrelation (\ref{calculus_mm0}). For example, we find for PIC simulation plasmas at equilibrium the finite estimate (see Appendix B) 
\begin{equation}
\label{electrostatic_field_fluctuations_estimate}
\displaystyle \left | \mathbf{E}_\mathrm{fluct} \right |  \underset{\omega_ p t_n \gg 1}{\sim}   \displaystyle \sqrt{ \delta N k_B T \eta \left ( k_g \lambda_D \right ) } \,\mathrm{with}\,\eta \left ( k_g \lambda_D \right ) = \displaystyle \left \{ \begin{array}{cll}
    \displaystyle \frac{ 4 \arctan{\displaystyle \left ( k_g \lambda_D / 2 \right ) }       }{ {\lambda_D}  } &\, \mathrm{in} \, 1\mathrm{D}
\cr \displaystyle \frac{1}{2} \displaystyle \frac{ \ln{\left [ 1+ {(k_g \lambda_D/2)}^2 \right ]}                     }{   {\lambda_D}^2} &\, \mathrm{in} \, 2\mathrm{D}
\cr \displaystyle \frac{2}{3}  \displaystyle \frac{ k_g \lambda_D - 2 \arctan{\displaystyle \left ( k_g \lambda_D/2 \right ) } }{ \pi {\lambda_D}^3} &\, \mathrm{in} \, 3\mathrm{D}
\end{array}
\right . .
\end{equation}
These estimates are compared with the amplitude of electrostatic field statistical fluctuations in electrostatic PIC plasma simulations in Figure \ref{Figure2} considering $\Delta_x = \Delta_y = \Delta_z$ for simplicity and using $$ 
\delta N = \displaystyle \frac{ \bar{n}_e {\Delta_x}^\text{dim} }{ N_{e,\text{mpc}} }
$$ 
according to (\ref{density}) and (\ref{NMPC}). 

By performing the similar derivation for the single-time PIC simulation plasma spatial density statistical fluctuations autocorrelation for each species, one also finds
\begin{equation}
\begin{array}{lll}
\mathbb{E} \displaystyle \left \{ {\delta n_a \left ( \mathbf{r}_{i,j,k},\,t_n \right )}^2 \right \} &=& \displaystyle \int_{\iota \nu - \omega_g / 2}^{\iota \nu + \omega_g/2} \displaystyle \frac{d \omega}{ 2 \pi } \displaystyle \int_{V_{\mathbf{k}_g}} \displaystyle \frac{d^3 \mathbf{k}}{ {\left ( 2 \pi \right )}^3 } \displaystyle \int_{\iota \nu' - \omega_g / 2}^{\iota \nu' + \omega_g/2} \displaystyle \frac{d \omega'}{ 2 \pi } \displaystyle \int_{V_{\mathbf{k}_g}} \displaystyle \frac{d^3 \mathbf{k}'}{ {\left ( 2 \pi \right )}^3 }
\cr & & \mathbb{E} \displaystyle \left \{  \widehat{\widehat{\delta n_a}} \left ( \omega, \mathbf{k} \right ) \widehat{\widehat{\delta n_a}} \left ( \omega',\,\mathbf{k}' \right ) \right \} \exp{\displaystyle \left \{ - \iota \displaystyle \left [ \left (\omega + \omega' \right ) t_n - \left ( \mathbf{k} + \mathbf{k}' \right ) \cdot \mathbf{r}_{i,j,k} \right ] \right \}}
\end{array}
\end{equation}
with
\begin{equation}
\label{density_fluctuations_spectrum_main_results}
\mathbb{E} \displaystyle \left \{  \widehat{\widehat{\delta n_a}} \left ( \omega, \mathbf{k} \right ) \widehat{\widehat{\delta n_a}} \left ( \omega',\,\mathbf{k}' \right ) \right \}  \underset{\omega_ p t_n \gg 1}{=} {\displaystyle \left ( 2 \pi \right )}^4 \delta \displaystyle \left ( \omega + \omega',\,\omega_g  \right ) \delta \displaystyle \left ( \mathbf{k} + \mathbf{k}',\,\mathbf{k}_g \right ) \widehat{\widehat{ {\delta n_a}^2 }}  \left ( \omega, \mathbf{k} \right )
\end{equation}
and
\begin{equation}
\label{density_fluctuations_spectrum}
\widehat{\widehat{ {\delta n_a}^2 }}  \left ( \omega, \mathbf{k} \right ) \underset{\omega_ p t_n \gg 1}{=}  \displaystyle \frac{ 2 \pi \delta N_a \bar{n}_a }{ {\varepsilon_L \displaystyle \left ( \omega, \mathbf{k} \right )}^2 }  \displaystyle \sum_{p,q,r} {\overset{\frown}{\mathrm{S}} \left ( \mathbf{k}_{p,q,r} \right )}^2 \displaystyle \int_{\Reals^3} d^3 \mathbf{v} F_{a0} \displaystyle \left ( \mathbf{v} \right ) \delta \displaystyle \left ( \omega - \mathbf{k}_{p,q,r} \cdot \mathbf{v}, \omega_g \right ).
\end{equation}
Note that these lasts are the single-time autocorrelation of plasma particle densities statistical fluctuations 
\begin{equation}
\delta n_a \displaystyle \left ( \mathbf{r}_{i,j,k},\,t_n \right ) = \displaystyle \int_{\Reals^3} \delta f_a \displaystyle \left ( \mathbf{r}_{i,j,k},\,\mathbf{v},\,t_n\right ) d^3 \mathbf{v}
\end{equation}
and not the one of weighted macroparticle center spatial densities statistical fluctuations
\begin{equation}
\delta n_{a_c} \displaystyle \left ( \mathbf{r}_{i,j,k},\,t_n \right ) = \displaystyle \int_{\Reals^3} \delta f_{a_c} \displaystyle \left ( \mathbf{r}_{i,j,k},\,\mathbf{v},\,t_n\right ) d^3 \mathbf{v}.
\end{equation}
The single-time plasma electrical charge statistical fluctuations autocorrelation can be obtained  starting from (\ref{electrostatic_field_fluctuations_spectrum}) or from (\ref{density_fluctuations_spectrum}). One can note that initial single-time statistical fluctuations correlation of differing plasma species quantities necessarily cancels according to (\ref{correlation}).
Fluctuations spectra (\ref{electrostatic_field_fluctuations_spectrum}) and (\ref{density_fluctuations_spectrum}) are the same as the ones obtained by \cite{Langdon1970a} \cite{Langdon1979}, however generalizing them to simulations of multiple species plasmas using today's algorithms and arbitrary macroparticle weights factor $\delta N_a$. To derive these spectra, the authors used a test macroparticle formalism similar to the method used by \cite{Thompson1960} \cite{Hubbard1961} and later justified by \cite{Rostoker1964} from the classical kinetic theory of plasmas. It consists in perturbing a discretized Vlasovian plasma by one macroparticle propagating along a straight-line orbit and averaging the resulting linear PIC simulated plasma quantity over all possible macroparticle initial conditions. The validity of their method is ensured by the first term, corresponding to the case where there is only one macroparticle in the infinitesimal volume of phase-space, in our results (\ref{correlation}) of initial weighted macroparticle center phase-space density correlations. While providing the physical interpretation of the results up to the first order, it may lead to confusion between Vlasov simulations and PIC simulations. Therefore, it seems to us that our method relates with fewer ambiguity equations computed by an electrostatic PIC code with the kinetic theory of plasmas. In addition, our approach adds the information that the statistical fluctuation spectra are only valid on a time scale $\omega_p t_n \gg 1$. It means that one must let the simulation run during a time greater than $\sim 10 / \omega_p$ before obtaining comparable spectra by directly computing it from the PIC simulation results. 

\subsection{Kinetic equation of a PIC simulation electrostatic plasma}

Performing the same reasoning as in the previous section, one can also find the single-time correlation between interpolated electrostatic field statistical fluctuations and weighted macroparticle center phase-space density statistical fluctuations at the grid points $\mathbf{r}_{i,j,k}$ and time steps $t_n$ for each species. We find starting from (\ref{discretized_linearized_Klimontovich}), (\ref{calculus_start}) and (\ref{discretized_correlation}), assuming again Assumptions \ref{Assumption_consistency_errors}, \ref{Assumption_homogeneous},  \ref{Assumption_collisions}, \ref{Assumption_stationary}  and considering again only non-damped terms on time scales $\omega_p t_n \gg 1$
\begin{equation}
 \begin{array}{l}
\mathbb{E} \displaystyle \left \{  \widehat{\widehat{\delta\mathbf{E}}}_s \left ( \omega', \mathbf{k}' \right ) \widehat{\widehat{\delta f}}_{a_c} \left ( \omega,\,\mathbf{k},\,\mathbf{v}_a \right ) \right \} 
 \underset{\omega_ p t_n \gg 1}{=}  {\displaystyle \left ( 2 \pi \right )}^4 \delta \displaystyle \left ( \omega + \omega',\,\omega_g\right ) \delta \displaystyle \left ( \mathbf{k} + \mathbf{k}',\,\mathbf{k}_g \right )  \widehat{\widehat{ \delta \mathbf{E}_s \delta f}}_{a_c}  \left ( \omega, \mathbf{k},\,\mathbf{v}_a \right )
\end{array}
\end{equation}
where
\begin{equation}
\label{discrete_calculus_1}
\begin{array}{lcl}
\widehat{\widehat{ \delta \mathbf{E}_s \delta f}}_{a_c}  \left ( \omega, \mathbf{k},\,\mathbf{v}_a \right ) &\underset{\omega_ p t_n \gg 1}{=} & - \iota \displaystyle \sum_{p,q,r} \displaystyle \frac{ 8 \pi^2 {\overset{\frown}{\mathrm{S}} \left ( \mathbf{k}_{p,q,r} \right )}^2  }{ {\varepsilon_L \displaystyle \left ( \omega, \mathbf{k} \right )}  }  \displaystyle \frac{ \left \{ \mathbf{k}_{p,q,r} \right \}_\mathbf{r}}{ { \left \{ \mathbf{k}_{p,q,r}  \right \}_\mathbf{r}}^2 }  \delta N_a \bar{n}_a {q_a} \displaystyle \sum_{m} F_{a0} \displaystyle \left ( \mathbf{v} \right ) \delta \displaystyle \left ( \omega_m - \mathbf{k}_{p,q,r} \cdot \mathbf{v}_a \right )
\cr &+& \hspace{0.65em} \iota \displaystyle \sum_{p,q,r}  \bar{n}_a \displaystyle \frac{q_a}{m_a} \displaystyle \sum_{m}  \displaystyle \frac{ {\overset{\frown}{\mathrm{S}} \displaystyle \left ( \mathbf{k}_{p,q,r} \right )}^2   }{ \mathbf{k}_{p,q,r} \cdot \mathbf{v}_a - \omega_m   }  \left \{ \mathbf{k}_{p,q,r} \right \}_\mathbf{r} \cdot  \displaystyle \frac{d F_{a0}}{d \mathbf{v}_a} \widehat{\widehat{ {\delta\mathbf{E}}^2 }}  \left ( \omega, \mathbf{k} \right ) \displaystyle \frac{ \left \{ \mathbf{k}_{p,q,r} \right \}_\mathbf{r} }{ {\left \{ \mathbf{k}_{p,q,r} \right \}_\mathbf{r}}^2 }.
\end{array}
\end{equation}
It allows for estimating the discretized right-hand side of kinetic equations (\ref{averaged_Klimontovich}) depending on  
\begin{equation}
\label{discrete_calculus_2}
\begin{array}{lcl}
& & \mathbb{E} \displaystyle \left \{ {\delta \mathbf{E}_s \left ( \mathbf{r}_{i,j,k},\,t_n \right )} \delta f_{a_c} \displaystyle \left ( \mathbf{r}_{i,j,k},\,\mathbf{v}_a,\,t_n\right ) \right \} 
\cr &=& \displaystyle \int_{\iota \nu - \omega_g / 2}^{\iota \nu + \omega_g/2} \displaystyle \frac{d \omega}{ 2 \pi } \displaystyle \int_{V_{\mathbf{k}_g}} \displaystyle \frac{d^3 \mathbf{k}}{ {\left ( 2 \pi \right )}^3 } \displaystyle \int_{\iota \nu' - \omega_g / 2}^{\iota \nu' + \omega_g/2} \displaystyle \frac{d \omega'}{ 2 \pi } \displaystyle \int_{V_{\mathbf{k}_g}} \displaystyle \frac{d^3 \mathbf{k}'}{ {\left ( 2 \pi \right )}^3 }  
\cr & & \mathbb{E} \displaystyle \left \{  \widehat{\widehat{\delta\mathbf{E}}}_s \left ( \omega', \mathbf{k}' \right ) \widehat{\widehat{\delta f}}_{a_c} \left ( \omega,\,\mathbf{k},\,\mathbf{v}_a \right ) \right \} \exp{\displaystyle \left \{ - \iota \displaystyle \left [ \left (\omega + \omega' \right ) t_n - \left ( \mathbf{k} + \mathbf{k}' \right ) \cdot \mathbf{r}_{i,j,k} \right ] \right \}}
\cr &\underset{\omega_ p t_n \gg 1}{=}& \displaystyle \int_{\iota \nu - \omega_g / 2}^{\iota \nu + \omega_g/2}\displaystyle \frac{d \omega}{ 2 \pi } \displaystyle \int_{V_{\mathbf{k}_g}} \displaystyle \frac{d^3 \mathbf{k}}{ {\left ( 2 \pi \right )}^3 }\widehat{\widehat{ \delta \mathbf{E}_s \delta f}}_{a_c}  \left ( \omega, \mathbf{k},\,\mathbf{v} \right ).
\end{array}
\end{equation}
One finds after integration over $\omega$ (see Appendix C)
\begin{equation}
\label{collision_operator}
\begin{array}{lll}
&& - \displaystyle \frac{q_a}{m_a}  \displaystyle \frac{ \partial }{\partial \mathbf{v}_a} \cdot \mathbb{E} \displaystyle \left \{ {\delta \mathbf{E}_s \left ( \mathbf{r}_{i,j,k},\,t_n \right )} \delta f_{a_c} \displaystyle \left ( \mathbf{r}_{i,j,k},\,\mathbf{v},\,t_n\right ) \right \} 
\cr &\underset{\omega_ p t_n \gg 1}{=}& - \displaystyle \frac{1}{ m_a} \displaystyle \frac{ \partial }{\partial \mathbf{v}_a} \cdot \displaystyle \sum_b \bar{n}_a \bar{n}_b  \displaystyle \int_{\Reals^3}  \mathbf{Q} \displaystyle \left ( \mathbf{v}_a,\,\mathbf{v}_b\right ) \cdot \displaystyle \left [ \displaystyle \frac{ \delta N_b }{m_a} F_{b0} \left ( \mathbf{v}_b \right ) \displaystyle \frac{ d F_{a0} }{ d \mathbf{v}_a } - \displaystyle \frac{ \delta N_a }{m_b} F_{a0} \left ( \mathbf{v}_a \right ) \displaystyle \frac{ d F_{b0} }{ d \mathbf{v}_b } \right ] d^3 \mathbf{v}_b
\end{array}
\end{equation}
depending on the tensor
\begin{equation}
\label{collision_tensor}
\mathbf{Q} \displaystyle \left ( \mathbf{v}_a,\,\mathbf{v}_b\right ) 
= - 2 \,{q_a}^2 {q_b}^2 \displaystyle \int_{\Reals^3} \displaystyle \frac{ {\overset{\frown}{\mathrm{S}} \left ( \mathbf{k} \right ) }^2  { \left \{ \mathbf{k}\right \}_\mathbf{r} } \otimes { \left \{ \mathbf{k}\right \}_\mathbf{r} } }{  {\varepsilon_L \left ( \mathbf{k} \cdot \mathbf{v}_a,\,\mathbf{k}\right )}^2 { \left \{ \mathbf{k}\right \}_\mathbf{r} }^4 } \displaystyle \sum_{p,q,r} {\overset{\frown}{\mathrm{S}} \left ( \mathbf{k}_{p,q,r} \right )}^2 \displaystyle \sum_m \delta \displaystyle \left ( \mathbf{k} \cdot \mathbf{v}_a - \mathbf{k}_{p,q,r} \cdot \mathbf{v}_b - m \omega_g  \right )d^3 \mathbf{k}
\end{equation}
for 3D simulations ($d^3 \mathbf{k}$ and the entire 3D space of integration $\Reals^3$ should be replaced by $ {(2 \pi)}^2 d k_x$ and $\Reals$ or by $ 2 \pi \, d k_x d k_y $ and $\Reals^2$ for 1D or 2D simulations respectively). It tends to the Lenard-Balescu tensor for collisionless plasmas in the limit $\forall \xi \in \left \{x,\,y,\,z,\,t \right \},\, \Delta_\xi \rightarrow 0$. According to Assumptions \ref{Assumption_homogeneous}, \ref{Assumption_stationary} and the macroparticle shape function properties (\ref{particle_shape_property}), we have necessarily
\begin{equation}
\bar{n}_{a} \displaystyle \left ( \mathbf{r}_{i,j,k},\,t_n \right ) = \delta N_a \displaystyle \sum_{i,j,k} S \displaystyle \left ( \mathbf{r}_{i,j,k} - \mathbf{r}_{a,\ell} \left ( t_1 \right ) \right ) = \bar{n}_a.
\end{equation} 
and therefore 
\begin{equation}
\bar{f}_{a} \displaystyle \left ( \mathbf{r}_{i,j,k},\,\mathbf{v},\,t_n \right )  = \bar{n}_a F_{a_0} \displaystyle \left ( \mathbf{v} \right ) = \bar{f}_{a_c} \displaystyle \left ( \mathbf{r}_{i,j,k},\,\mathbf{v},\,t_n \right ) .
\end{equation}
Neglecting the expectation of the term depending on the internal tension forces of macroparticles (\ref{non_physical_term}), kinetic equations corresponding to an electrostatic PIC simulation follow immediately. They read
\begin{equation}
\label{kinetic_equation}
\begin{array}{l}
\hspace{2.em}{\displaystyle \left . \displaystyle \frac{ \partial {\bar{f}}_{a} }{\partial {t}} \right |}^{i,j,k,n} + {\displaystyle \left . \displaystyle \frac{ \partial  }{\partial {\mathbf{r}} } \cdot \displaystyle \left ( {\mathbf{v}}_a {\bar{f}}_{a} \right ) \right |}^{i,j,k,n} + \displaystyle \frac{ \partial  }{\partial {\mathbf{v}}_a} \cdot \displaystyle \left ( \displaystyle \frac{{q}_a }{ {m}_a } {\bar{\mathbf{E}}}_s \left ( {\mathbf{r}}_{i,j,k},\, {t}_n \right ) {\bar{f}}_{a}  \left ( {\mathbf{r}}_{i,j,k},\, {\mathbf{v}}_a,\,{t}_n \right ) \right ) 
\cr \underset{ \omega_{p} {t}_n \gg 1}{=}
  -  \displaystyle \frac{1}{ {m}_a} \displaystyle \frac{ \partial }{\partial {\mathbf{v}}_a} \cdot \displaystyle \sum_b  \displaystyle \int_{\Reals^3} {\mathbf{Q}} \displaystyle \left ( {\mathbf{v}}_a,\,{\mathbf{v}}_b\right ) \cdot {\displaystyle \left . \displaystyle \left ( \displaystyle \frac{ {\delta N}_b }{{m}_a } {\bar{f}}_{b} \left ( {\mathbf{v}}_b \right ) \displaystyle \frac{ \partial {\bar{f}}_{a} }{ \partial {\mathbf{v}}_a } - \displaystyle \frac{ {\delta N}_a }{{m}_b } {\bar{f}}_{a} \left ( {\mathbf{v}}_a \right ) \displaystyle \frac{ \partial {\bar{f}}_{b} }{ \partial {\mathbf{v}}_b } \right ) \right | }^{i,j,k,n} d^3 {\mathbf{v}}_b
\end{array}
\end{equation}
according to (\ref{averaged_Klimontovich}) and (\ref{collision_operator}), in agreement with \cite{Langdon1970a} \cite{Birdsall1991}. 
The component $\left ( m,\,p,\,q,\,r\right)  = \left (0,\,0,\,0,\,0\right ) $ in (\ref{collision_operator}) is physical. As shown by \cite{Dawson1983}, it describes the average friction and diffusion of macroparticles in the fluctuating component of the electrostatic force which are direct consequences from the emission \cite{Decyk1987} and absorption \cite{Landau1946} of plasma waves by the macroparticles in the simulation. However, if each macroparticle represent more than one real particle such that $\delta N_a= \delta N_b = \delta N \gg 1$, these physical processes are greatly overestimated by the weight factor $\delta N$ compared to reality according to the expression of the "collision operator" (\ref{collision_operator}). While the component $\left ( m,\,p,\,q,\,r\right)  = \left (0,\,0,\,0,\,0\right ) $ of the latter cancels for single species 1D plasmas \cite{Dawson1964} and for plasmas at equilibrium for which $\delta N_e = \delta N_i$ such that the diffusion of macroparticles compensates exactly their friction (in a plasma in thermal equilibrium, the rate of emission of the plasma waves is exactly balanced by their energy loss by Landau damping), this is never the case for its time aliases components $m \neq 0 $ and its spatial aliases $ p \neq 0 $, $ q \neq 0 $ or $ r \neq 0 $. Worst, they may even be at the origin of numerical instabilities \cite{Langdon1970a} \cite{Birdsall1991}. \cite{Langdon1970a} \cite{Birdsall1991} have shown that, in general, the "collision tensor" (\ref{collision_tensor}) has the following properties
\begin{equation}
\displaystyle \left \{ \begin{array}{l}
     \forall \mathbf{X},\, \mathbf{X}\cdot \mathbf{Q} \left ( \mathbf{v}_a,\,\mathbf{v}_b\right ) \cdot \mathbf{X} < 0
\cr \mathbf{Q} \left ( \mathbf{v}_a,\,\mathbf{v}_b\right ) = \mathbf{Q} \left ( \mathbf{v}_b,\,\mathbf{v}_a\right ) 
\end{array} \right . .
\end{equation} 
According to the kinetic equation (\ref{kinetic_equation}), a consequence is that the average PIC simulation plasma entropy
\begin{equation}
\label{macroscopic_entropy}
\bar{\mathrm{H}} \displaystyle \left ( \mathbf{r}_{i,j,k},\,t_n\right ) = - \displaystyle \sum_a \displaystyle \int_{\Reals^3} \bar{f}_a \displaystyle \left ( \mathbf{r}_{i,j,k},\,\mathbf{v}_a,\,t_n\right )  \ln{ \bar{f}_a \displaystyle \left ( \mathbf{r}_{i,j,k},\,\mathbf{v}_a,\,t_n\right ) }  d^3 \mathbf{v}_a,
\end{equation}
following average macroparticle straight line trajectories in space, necessarily increases
\begin{equation}
{\displaystyle \left .\displaystyle \frac{d \bar{\mathrm{H}} }{ d t } \displaystyle \right |} \displaystyle \left ( \mathbf{r}_{i,j,k},\,t_n\right ) > 0.
\end{equation}
We conclude that whatever the velocity distributions $F_{a0} \left ( \mathbf{v} \right )$ that are used to initialize macroparticle velocities using a random number generator, if one lets the simulation running during a sufficiently long time duration, the PIC simulation plasma will necessarily tend on average to a quasi equilibrium state that maximizes entropy. 

\section{Discussion}

\subsection{Neglect of numerical consistency errors compared to statistical fluctuations}

In this subsection, we discuss the simulation conditions for which Assumption \ref{Assumption_consistency_errors} is valid.  First, (\ref{electrostatic_field_fluctuations_estimate}) has to be compared with the numerical consistency errors associated with the leap-frog scheme (\ref{discretized_EoM}) and with the Maxwell solver.
For Maxwell spectral solvers, there is no numerical consistency error in space. The counterpart is the compulsory use of periodic boundary conditions. While it is not problematic here because we are considering an infinite and homogeneous plasma, it may become constraining under other circumstances. For the 2nd order FDTD scheme, the numerical consistency error in space is given by 
\begin{equation}
\label{space_discretization_error}
\mathbf{\boldsymbol{\varepsilon}}_\mathbf{r} \displaystyle \left ( \mathbf{r}_{i,j,k},\,t_n \right ) \, \mathrm{ with } \, \varepsilon_{\mathbf{r},\xi} \displaystyle \left ( \mathbf{r}_{i,j,k},\,t_n \right ) =  \displaystyle \frac{ {\Delta_\xi}^3 }{24}  { \displaystyle \left . \displaystyle \frac{ \partial^3 E_\xi }{ \partial \xi^3} \right |} \displaystyle \left ( \mathbf{r}_{i,j,k},\,t_n \right )  + O \left ( {\Delta_\xi}^5 \right ).
\end{equation}
The numerical consistency error on macroparticle locations due to the leap-frog scheme is
\begin{equation}
\displaystyle \frac{ {\Delta_t}^3 }{24}  { \displaystyle \left . \displaystyle \frac{ d^3 \mathbf{r}_{a,\ell} }{ d t^3} \right |} \displaystyle \left ( t_n \right )  + O \left ( {\Delta_t}^5 \right ).
\end{equation}
Therefore, injecting it in Maxwell equations, we can estimate the consequent numerical consistency error on the electrostatic field as
\begin{equation}
\label{time_discretization_error}
\mathbf{\boldsymbol{\varepsilon}}_t \displaystyle \left ( \mathbf{r}_{i,j,k},\,t_n \right ) = 4 \pi \displaystyle \frac{ {\Delta_t}^3 }{24}  \displaystyle \sum_a  Q_a \displaystyle \frac{q_a}{m_a}\displaystyle \sum_{\ell = 1}^{N_a} { \displaystyle \left . \displaystyle \frac{ \partial \mathbf{E}_s }{ \partial t} \right |} \displaystyle \left ( \mathbf{r}_{a,\ell} \displaystyle \left ( t_n \right ),\,t_n \right ) S  \displaystyle \left ( \mathbf{r}_{i,j,k} - \mathbf{r}_{a,\ell} \displaystyle \left ( t_n \right ) \right ) + O \left ( {\Delta_t}^5 \right ).
\end{equation}
By normalizing time, space, electrical charge and mass according to $\underline{t} = \omega_{p_e} t$, $\underline{\mathbf{r}} = \mathbf{r}  / \lambda_{D}$ where $\lambda_{D} = v_{T_e}/\omega_{p_e} $ is the electron Debye screening length, $\underline{q_a} = q_a / e$ and $\underline{m_a} = m_a / m_e$, respectively, and by assuming reasonably $Z m_e /m_i \ll 1$, we find the scalings 
\begin{equation}
\displaystyle \left | \underline{\mathbf{E}}_\mathrm{fluct} \right |  \underset{ \underline{t_n} \gg 1}{\sim} \displaystyle \sqrt{ \displaystyle \frac{ \delta N }{ 4 \pi \bar{n}_e {\lambda_D}^3 } \eta \left ( 2 \pi / \underline{\Delta_x} \right )},\,\displaystyle \left |  \underline{\mathbf{\boldsymbol{\varepsilon}}_\mathbf{r}} \right | \leq \alpha \displaystyle \frac{ {\underline{\Delta_x}}^3 }{ 24 } \mathrm{ and } \displaystyle \left |  \underline{\mathbf{\boldsymbol{\varepsilon}}_t} \right | \leq \beta  \displaystyle \frac{ {\underline{\Delta_t}}^3 }{ 24 } 
\end{equation}
where
$$
\alpha \left ( t_n \right ) = \displaystyle \max_{i,j,k}{  \displaystyle \left \{ \displaystyle \sqrt{   \displaystyle \sum_{\underline{\xi}=\underline{x},\underline{y},\underline{z}} {\displaystyle \left ({ \displaystyle \left . \displaystyle \frac{ \partial^3 \underline{E}_\xi }{ \partial \underline{\xi}^3} \right |} \displaystyle \left ( \underline{\mathbf{r}}_{i,j,k},\,\underline{t}_n \right ) \right )}^2  } \right \} }
$$
and
$$
\beta \left ( t_n \right ) = \displaystyle \max_{i,j,k}{  \displaystyle \left \{ \displaystyle \sqrt{   \displaystyle \sum_{\underline{\xi}=\underline{x},\underline{y},\underline{z}} {\displaystyle \left ({ \displaystyle \left . \displaystyle \frac{ \partial \underline{E}_{s,\xi} }{ \partial \underline{t}} \right |} \displaystyle \left ( \underline{\mathbf{r}}_{i,j,k},\,\underline{t}_n \right ) \right )}^2  } \right \} } \displaystyle \max_{i,j,k}{  \displaystyle \left \{ \displaystyle \frac{ n_e \displaystyle \left ( \underline{\mathbf{r}}_{i,j,k},\,\underline{t}_n \right ) }{ \bar{n}_e }\right \} }.
$$
Assumption \ref{Assumption_consistency_errors} is therefore valid if $\displaystyle \left | \underline{\mathbf{E}}_\mathrm{fluct} \right | \gg \displaystyle \left |  \underline{\mathbf{\boldsymbol{\varepsilon}}_\mathbf{r}} \right |$ and $ \displaystyle \left | \underline{\mathbf{E}}_\mathrm{fluct} \right | \gg \displaystyle \left |  \underline{\mathbf{\boldsymbol{\varepsilon}}_t} \right |$.
Considering roughly $\alpha \sim \beta  \sim 1$ and $\underline{\Delta_t} \sim \underline{\Delta_x} $, this is the case whenever 
\begin{equation}
 \displaystyle \frac{ {\underline{\Delta_x}}^{6} }{\eta \left ( 2 \pi / \underline{\Delta_x} \right ) 24^2 }  \ll \displaystyle \frac{ \delta N }{ 4 \pi \bar{n}_e {\lambda_D}^3 }
\end{equation}
which is not constraining in many cases. It means that statistical fluctuations plays a crucial role in PIC simulations using a random number generator.

\subsection{Stationarity of the plasma distribution function in PIC simulations}

In this subsection, we discuss the simulation conditions for which Assumption \ref{Assumption_stationary} is valid. It seems paradoxical  that the right-hand side of kinetic equations (\ref{kinetic_equation}) that we have obtained according to Assumptions \ref{Assumption_consistency_errors}, \ref{Assumption_homogeneous},  \ref{Assumption_collisions} and  \ref{Assumption_stationary} never cancels while their left-hand side cancels according to Assumptions \ref{Assumption_consistency_errors}, \ref{Assumption_homogeneous} and \ref{Assumption_stationary}. The paradox is resolved if Assumption \ref{Assumption_stationary} remains valid only on a time scale $\omega_p t_n$ during which the right-hand side of kinetic equations (\ref{kinetic_equation}) is sufficiently small thus ensuring the distribution functions (\ref{distribution_function}) remain homogeneous and stationary. Above a certain simulation time duration much greater than $\omega_p t_n$, the distribution functions begin to evolve with time. To estimate this simulation time threshold, let us consider the 3D electrostatic spectral PIC simulation of a plasma consisting of immobile ions and thermal electrons that uses equal weights $\delta N_e = \delta N_i = \delta N$ and a smoothing function filtering all spatial frequency aliases. We also neglect time aliasing effects assuming a sufficiently small time step $\Delta_t$ and we assume $\Delta_x = \Delta_y = \Delta_z = \Delta$ and $a_x = a_y = a_z = a \leq \Delta$ such that $\overset{\frown}{\mathrm{S}} \left ( \mathbf{k} \right ) \approx \exp{ \left ( - a^2 \mathbf{k}^2 / 2 \right )} $. In this case, the PIC Coulomb logarithm, defined according to  
\begin{equation}
\ln{ \Lambda \left ( \mathbf{v}_e \right ) } = \displaystyle \int_{k_\mathrm{min}}^{k_\mathrm{max}} \displaystyle \frac{ {\overset{\frown}{\mathrm{S}} \left ( \mathbf{k} \right )}^4  d k}{k \, {\varepsilon_L \left ( \mathbf{k} \cdot \mathbf{v}_e,\,\mathbf{k} \right )}^2  },
\end{equation}
only depends on macroelectron velocities $\mathbf{v}_e$ if they have a velocity much greater than the thermal velocity $\left | \mathbf{v}_e\right | \gg v_{T_e}$, implying the PIC simulation plasma permittivity is close to unity $ \varepsilon_L \left ( \mathbf{k} \cdot \mathbf{v}_e,\,\mathbf{k} \right ) \approx 1$. Here, we have noted $ k = \left | \mathbf{k} \right |$ and $\mathbf{\boldsymbol{\kappa}} = \mathbf{k} / k = {\left ( \cos{\theta},\,\sin{\theta} \cos{\varphi},\,\sin{\theta} \sin{\varphi}\right )}^t$. We have also introduced arbitrary cutoffs $ k_\mathrm{min}$ and $ k_\mathrm{max} $ that are not necessary for the convergence of the Coulomb logarithm contrary to the one obtained for point particles. In this particular case, the Lenard-Balescu-like tensor (\ref{collision_tensor}) reduces to the Landau collision tensor
\begin{equation}
\begin{array}{lll}
\mathbf{Q} \left ( \mathbf{v}_e,\,\mathbf{v}_b\right ) &=& - 2 \displaystyle \frac{ {e}^2 {q_b}^2 \ln{ \Lambda \left ( \mathbf{v}_e \right ) } }{ \left | \mathbf{v}_e - \mathbf{v}_b \right | } \displaystyle \int_{0}^{2\pi} d \varphi \displaystyle \int_0^{\pi} \sin{\theta} d \theta\,  \mathbf{\boldsymbol{\kappa}} \otimes \mathbf{\boldsymbol{\kappa}} \,\delta \displaystyle \left ( \mathbf{\boldsymbol{\kappa}} \cdot \displaystyle \frac{ \mathbf{v}_e - \mathbf{v}_b }{ \left | \mathbf{v}_e - \mathbf{v}_b \right | } \right )
\cr &=& - 2 \displaystyle \frac{ {e}^2 {q_b}^2 \ln{ \Lambda \left ( \mathbf{v}_e \right ) } }{ \left | \mathbf{v}_e - \mathbf{v}_b \right | } \pi \displaystyle \left [ \mathbf{I} - \displaystyle \frac{ \left ( \mathbf{v}_e - \mathbf{v}_b \right ) \otimes \left ( \mathbf{v}_e - \mathbf{v}_b \right )    }{ {\left | \mathbf{v}_e - \mathbf{v}_b \right |}^2  } \right ]
\end{array}
\end{equation}
that describes small angle binary Coulomb collisions between point particles. It seems very strange to obtain such a collision integral since we have neglected close-encounter Coulomb collisions between macroparticles and consequently between real particles. In the textbook \cite{Landau1981}, they explain "it is sufficient to consider fluctuations in a collisionless plasma to calculate the collision integral because the important Fourier components of the electric field in collisions in plasmas are those with $ \left |\mathbf{k} \right | \gtrapprox 1 / \lambda_D \gg 1 / \lambda_\mathrm{mfp}$, so that collisions may be neglected" where $\lambda_\mathrm{mfp}$ denotes real particle mean free paths due to close-encounter binary Coulomb collisions. As a consequence, "collisions" are naturally taken into account in electrostatic PIC simulations of collisionless plasmas using a random number generator to initialize macroparticle velocities. Knowing from the kinetic theory of plasmas that the smallest relaxation rate is given by the electron-ion collision frequency, let us neglect in a first attempt the electron-electron fluctuations correlation term in the kinetic equation (\ref{kinetic_equation}). Considering then the limit $Z m_e / m_i \ll 1$, one can neglect the friction experienced by such electrons compared to their diffusion so that the right hand-side of the kinetic equation (\ref{kinetic_equation}) reduces to a Lorentz-like collision operator whose time scale is governed by the overestimated electron-ion collision frequency
\begin{equation}
\nu_{ei} \left ( \mathbf{v}_e \right ) = \delta N \displaystyle \frac{ 4 \pi \bar{n}_i Z^2 e^4  }{ {m_e}^2 { \left | \mathbf{v}_e \right |  }^3 } \ln{ \Lambda } \left ( \mathbf{v}_e \right ) \lesssim \bar{\nu}_{ei} = \delta N \displaystyle \frac{ 4 \pi \bar{n}_e Z e^4 }{ {m_e}^2 { v_{T_e}  }^3 } \ln{ \Lambda } \left ( \mathbf{v}_{T_e} \right ).
\end{equation}
We conclude that Assumption \ref{Assumption_stationary} remains valid on a time scale $t_n$ such that 
\begin{equation}
\label{time_scale}
1 \ll  \omega_{p} t_n \ll \displaystyle \frac{ \omega_p }{ \bar{\nu}_{ei} } = \displaystyle \frac{4 \pi \bar{n}_e {\lambda_D}^3}{ Z \delta N \ln{ \Lambda } \left ( \mathbf{v}_{T_e} \right )} \approx \displaystyle \frac{ \lambda_\mathrm{mfp} }{ \delta N \lambda_D }
\end{equation}
that is to say for not too long electrostatic PIC simulations of "not too dense" collisionless plasmas; "not too dense" referring to the value of macroparticle weight factors $\delta N$ constrained by today's computer technology.  Although overestimating them by the weight factor $\delta N$, the term $ \left ( m,\,p,\,q,\,r\right ) = \left ( 0,\,0,\,0,\,0\right ) $ of (\ref{collision_operator}) describes single-time correlations between the positions of real plasma particles due to their long-range Coulomb interactions. It is consequently not a numerical noise contrary to what is usually considered. Contrary to "not too dense" collisionless plasmas, the distribution functions evolve faster than the statistical fluctuations correlation in electrostatic PIC simulations of "dense" collisionless plasmas or collisional plasmas for which $ \omega_p / \bar{\nu}_{ei} \lessapprox 1 $. Assumption 4. is therefore not valid in this case. In addition, the physics of such fast evolution is not taken into account by electrostatic PIC codes because they underestimate close-encounter Coulomb collisions in this regime where $1 / \lambda_D \lessapprox 1 / \delta N \lambda_\mathrm{mfp} < 1 / \lambda_\mathrm{mfp}$. To model correctly this fast evolution of distribution functions, a Langevin force based on a Monte-Carlo model of multiple binary Coulomb collisions between real particles per time step $\Delta_t$  should be added into (\ref{EoM}) such as the one proposed by \cite{Nanbu1997}. Such a Monte-Carlo algorithm emulates statistical fluctuations due to binary Coulomb collisions between particles and ensures Bogoliubov hypothesis. However, these PIC simulations are out of the scope of this paper. 

In order to give orders of magnitude, let us consider a laser-generated Helium plasma ($Z=2$), heated to temperatures up to $T_e \approx 1 \mathrm{ keV}$ and where the electron density is about $\bar{n}_e \approx 5 \times 10^{19} \mathrm{ cm}^{-3}$ ($\omega_p \approx 4\times 10^{14} \mathrm{ rad.s}^{-1}$), the Debye length is about $\lambda_D \approx 0.03 \,\mu\mathrm{m}$. The resulting number of electrons in a Debye cube is about $ \bar{n}_e {\lambda_D}^3 \approx 1776 $. If we choose reasonably $\Delta = 0.01\,\mu\mathrm{m} \approx \lambda_D / 3$, we have necessarily $N_{e,\mathrm{mpc}} {\left ( \lambda_D / \Delta \right )}^\mathrm{3} \gtrapprox 266$ for the assumption 2. to be valid. It leads to a maximum possible value of macroelectron weights $\delta N_e \lessapprox 7$ in 3D simulations for matching the correct plasma electron density. We deduce Assumption 4. remains valid on a time scale $\omega_p t_n \ll 1670$. Under such simulation conditions, assumption 1. is valid if one chooses $\omega_p \Delta_t \lessapprox \Delta / \lambda_D$ to ensure the numerical stability of the simulation. Indeed, $\underline{\Delta}^6/ \eta \left ( 2 \pi / \underline{\Delta}\right )24^2 \approx 1 \times 10^{-15} \eta \left ( 2 \pi / \underline{\Delta}\right ) \ll \delta N_e / 4 \pi \bar{n}_e {\lambda_D}^3 \approx 3\times 10^{-5}$ in this case. Considering a reasonable number of macroparticles  simulated with today's computer technology $N_e \approx 1 \times 10^{11}$ using a highly parallelized PIC code and several $10^5$ CPU $\times$ hours with a computational cost typical of $\approx 100\,\mathrm{ns}/\mathrm{time step}/\mathrm{macroparticle}$ and a simulation time duration $L_t = 1000/ \omega_p$, we conclude that we are limited to plasma simulation boxes of $L_x=L_y=L_z\approx 40\,\mu\mathrm{m}$ by using such a macroelectron weight $\delta N_e \approx 7$. 

Performing the same reasoning on the main term considering the same assumptions for 2D electrostatic PIC simulations but defining a Coulomb logarithm 
\begin{equation}
\displaystyle \frac{ \ln{ \Lambda^{\mathrm{2D}} \left ( \mathbf{v}_e \right ) } }{ \lambda_D }= \displaystyle \int_{k_\mathrm{min}}^{k_\mathrm{max}} \displaystyle \frac{ {\overset{\frown}{\mathrm{S}} \left ( \mathbf{k} \right )}^4  d k}{{\varepsilon_L \left ( \mathbf{k} \cdot \mathbf{v}_e,\,\mathbf{k} \right )}^2  }
\end{equation}
instead, we find Assumption \ref{Assumption_stationary} remains valid on a time scale $t_n$ such that
\begin{equation}
\label{time_scale_2D}
1 \ll  \omega_{p} t_n \ll \displaystyle \frac{ \omega_p }{ \bar{\nu}_{ei}^{\mathrm{2D}} } = \displaystyle \frac{2 \pi \bar{n}_e {\lambda_D}^2}{ Z \delta N \ln{ \Lambda^{\mathrm{2D}} \left ( \mathbf{v}_{T_e} \right ) }}
\end{equation}
in agreement with \cite{Montgomery1970}, \cite{Hockney1971} and \cite{Hockney1981}. For 1D electrostatic PIC simulations, one finds the main term $ \left ( m,\,p\right ) = \left ( 0,\,0\right ) $ is identically zero owing to the Dirac delta function similarly as for the 1D Lenard-Balescu or the 1D Landau collision terms in agreement with \cite{Dawson1964}. Indeed, according to the latter, the observed relaxation time 
\begin{equation}
\label{time_scale_1D_a}
\displaystyle \frac{\omega_p }{ \bar{\nu}^{\mathrm{1D}}_\mathrm{Dawson} } = \displaystyle \frac{ 1 }{  10} \, { \displaystyle\left ( \displaystyle \frac{ \bar{n}_e {\lambda_D} }{ \delta N   } \right )}^2
\end{equation} 
of 1D electrostatic PIC simulation plasmas is due to the 3-body collisions between macroparticles and we should have considered the fluctuations equations (\ref{linearized_Klimontovich}) and (\ref{Maxwell3}) without neglecting the quadratic terms (Assumption 3) to obtain such a result. However, the aliasing terms do not cancel in agreement with \cite{Birdsall1991} and \cite{Montgomery1970}. One finds for example a relaxation time 
\begin{equation}
\label{time_scale_1D_b}
\displaystyle \frac{ \omega_p }{ \bar{\nu}_{(0,1)}^{\mathrm{1D}} } = \displaystyle \frac{2 {(\mathbf{k}_g \lambda_D)}^2 V }{ v_{T_e} } \displaystyle \frac{\bar{n}_e {\lambda_D}}{ \delta N }
\end{equation}
where
$$
\displaystyle \frac{1}{V} = \displaystyle \frac{1}{V_e} + \displaystyle \frac{Z m_e / m_i }{V_i} \,\mathrm{and}\,
\displaystyle \frac{1}{V_b} = \displaystyle \int_{-\infty}^{\infty}  {\displaystyle \left ( \displaystyle \frac{ {\overset{\frown}{\mathrm{S}} \left ( \displaystyle \frac{ - k_g v'_{b,x} }{v_{e,x} - v'_{b,x}} \right )} {\overset{\frown}{\mathrm{S}} \left ( \displaystyle \frac{ - k_g v_{e,x} }{v_{e,x} - v'_{b,x}} \right )}  }{ \varepsilon_L \displaystyle \left ( \displaystyle \frac{ -k_g v_{e,x} v'_{b,x} }{v_{e,x} - v'_{b,x}},\, \displaystyle \frac{ - k_g v'_{b,x} }{v_{e,x} - v'_{b,x}} \right ) } \right )}^2 {\displaystyle \left ( 1 - \displaystyle \frac{ v_{e,x} }{ v'_{b,x} } \right )}^2  \displaystyle \frac{ F_{b_0} ( v'_{b,x} ) }{v'_{b,x}} d v'_{b,x}
$$
considering only the spatial aliasing term $ \left ( m,\,p\right ) = \left ( 0,\,1\right ) $ when deriving the collision operator  (\ref{collision_operator}) that reads 
% \begin{equation}
% \begin{array}{lll}
% && - \displaystyle \frac{q_e}{m_e}  \displaystyle \frac{ \partial }{\partial \mathbf{v}_e} \cdot \mathbb{E} \displaystyle \left \{ {\delta \mathbf{E}_s \left ( \mathbf{r}_{i,j,k},\,t_n \right )} \delta f_{e_c} \displaystyle \left ( \mathbf{r}_{i,j,k},\,\mathbf{v},\,t_n\right ) \right \} 
% \cr &\underset{\omega_ p t_n \gg 1}{\approx }& -  \displaystyle \frac{ \delta N }{ {m_e}^2 } \displaystyle \frac{ \partial }{\partial {v}_{e,x}} \cdot \displaystyle \sum_{b=e,i} \bar{n}_b  \displaystyle \int_{-\infty}^\infty  Q \displaystyle \left ( {v}_{e,x},\, {v'}_{b,x} \right ) \cdot \displaystyle \left [ F_{b0} \left ( {v'}_{b,x} \right ) \displaystyle \frac{ \partial \bar{f_{e_c}} }{ \partial {v}_{e,x} } - \displaystyle \frac{ m_e }{m_b} \bar{f_{e_c}} \left ( {v}_{e,x} \right ) \displaystyle \frac{ d F_{b0} }{ d {v'}_{b,x} } \right ] d {v'}_{b,x}
% \cr &\underset{\omega_ p t_n \gg 1}{\approx }& \displaystyle \frac{ \partial }{\partial v_{e,x}} \left [ {\nu}_{(0,1)}^{\mathrm{1D}} {v_{T_e}}^2 \displaystyle \frac{  \partial  \bar{f_{e_c}} }{  \partial v_{e,x} } \right ]
% \end{array}
% \end{equation}
\begin{equation}
- \displaystyle \frac{q_e}{m_e}  \displaystyle \frac{ \partial }{\partial \mathbf{v}_e} \cdot \mathbb{E} \displaystyle \left \{ {\delta \mathbf{E}_s \left ( \mathbf{r}_{i,j,k},\,t_n \right )} \delta f_{e_c} \displaystyle \left ( \mathbf{r}_{i,j,k},\,\mathbf{v},\,t_n\right ) \right \} 
\underset{\omega_ p t_n \gg 1}{\approx } \displaystyle \frac{ \partial }{\partial v_{e,x}} \displaystyle \left ( {\nu}_{(0,1)}^{\mathrm{1D}} {v_{T_e}}^2 \displaystyle \frac{  \partial  \bar{f_{e_c}} }{  \partial v_{e,x} } \right )
\end{equation}
in agreement with \cite{Birdsall1991} and the empirical diffusion term proposed by \cite{Virtamo1979}. As a conclusion for 1D electrostatic PIC simulations, we conclude Assumption \ref{Assumption_stationary} remains valid on a time scale $t_n$ such that 
\begin{equation}
\label{time_scale_1D}
1 \ll  \omega_{p} t_n \ll \displaystyle \frac{ \omega_p }{ \bar{\nu}_{ei}^{\mathrm{1D}} }
\end{equation} 
with $\bar{\nu}^{\mathrm{1D}}$ scaling as (\ref{time_scale_1D_b}) for a large number of macroparticle centers per Debye length and as (\ref{time_scale_1D_a}) for a small number of macroparticle centers per Debye length as already suggested by \cite{Birdsall1991}.

\subsection{Evolution of the empirical entropy in electrostatic PIC simulation plasmas}

Landau damping of electrostatic waves is a time-reversible process \cite{Landau1946}. \cite{Decyk1987} have shown that the emission of electrostatic wakefield is also a time-reversible process. Therefore, it may seem paradoxical that we concluded that whatever velocity distributions $F_{a0} \left ( \mathbf{v} \right )$ that are used to initialize macroparticle velocities using a random number generator, if one lets the simulation running during a sufficiently long time duration, the PIC-simulated plasma will necessarily tend on average to a quasi-equilibrium state that maximizes the entropy. The physical reason for this entropy increase is the average distribution function entropy increase: any deterministic and time-reversible statistical fluctuations around the average solution makes tend the average solution to be the one that maximizes the entropy. Consequently, even if they are time-reversible, the statistical realizations also tend to the solution that maximizes entropy since they are supposed to be close to the average result in the limit $N_{a,\mathrm{mpc}}\gg1$. Let us emphasize here that the macroscopic PIC-simulated plasma entropy (\ref{macroscopic_entropy}) is not the empirical PIC-simulated plasma entropy 
\begin{equation}
\label{experimental_entropy}
\mathrm{H}_{\Pi} \displaystyle \left ( \mathbf{r}_{i,j,k},\,t_n\right ) = - \displaystyle \sum_a \displaystyle \sum_{\mathrm{l,m,n}} f_{a,\Pi} \displaystyle \left ( \mathbf{r}_{i,j,k},\,\mathbf{v}_{\mathrm{l,m,n}},\,t_n\right ) \ln{ f_{a,\Pi} \displaystyle \left ( \mathbf{r}_{i,j,k},\,\mathbf{v}_{\mathrm{l,m,n}},\,t_n\right ) }  {\Delta_{v}}^3
\end{equation}
that one can compute directly from the PIC simulation by discretizing the velocity space according to bins of volume ${\Delta_{v}}^3$ located between velocity grid points $\mathbf{v}_{l,m,n} = \mathbf{I}_{l,m,n} \cdot \mathbf{\boldsymbol{\Delta}}_v $ where $ \mathbf{\boldsymbol{\Delta}}_v = {\displaystyle \left ( \Delta_{v} ,\,   \Delta_{v} ,\, \Delta_{v} \right )}^t $ and by counting the number of macroparticles contained inside each bins 
\begin{equation}
\label{experimental_distribution_function}
\begin{array}{lll}
f_{a,\Pi} \displaystyle \left ( \mathbf{r}_{i,j,k},\mathbf{v}_{\mathrm{l,m,n}},t_n\right ) &=&  \displaystyle \int_{\Reals^3} d^3 \mathbf{v} \displaystyle \int_{\Reals^3} d^3 \mathbf{r} f_{a,c} \displaystyle \left ( \mathbf{r},\,\mathbf{v},\,t_n\right ) S \left ( \mathbf{r} - \mathbf{r}_{i,j,k} \right ) \Pi_{\mathbf{\boldsymbol{\Delta}}_v}^{(0)} \left ( \mathbf{v} - \mathbf{v}_{\mathrm{l,m,n}} \right )
\cr &=& \displaystyle \int_{\Reals^3} d^3 \mathbf{v} \hspace{0.5em}f_{a} \displaystyle \left ( \mathbf{r}_{i,j,k},\,\mathbf{v},\,t_n\right ) \Pi_{\mathbf{\boldsymbol{\Delta}}_v}^{(0)} \left ( \mathbf{v} - \mathbf{v}_{\mathrm{l,m,n}} \right )
\cr &=& \delta N_a \displaystyle \sum_{\ell=1}^{N_a} S \displaystyle \left ( \mathbf{r}_{i,j,k} - \mathbf{r}_{a,\,\ell} \displaystyle \left ( t_n \right ) \right )  \Pi_{\mathbf{\boldsymbol{\Delta}}_v}^{(0)} \left ( \mathbf{v}_{\mathrm{l,m,n}} - \mathbf{v}_{a,\ell} \left ( t_n \right )\right ).
\end{array}
\end{equation}
Indeed, by differentiating the latter empirical distribution functions $f_{a,\Pi} \displaystyle \left ( \mathbf{r},\mathbf{v},t\right )$ with time, one obtains the kinetic equation
\begin{equation}
\label{experimental_distribution_function_kinetic_equation}
\displaystyle \frac{ \partial f_{a,\Pi} }{\partial t} +  \displaystyle \frac{ \partial  }{\partial \mathbf{r}} \cdot \displaystyle \left ( \mathbf{v} f_{a,\Pi} \right ) +  \displaystyle \frac{ \partial  }{\partial \mathbf{v}} \cdot \displaystyle \left ( \displaystyle \frac{q_a}{m_a} \mathbf{E}_s \left ( \mathbf{r},\,t \right ) f_{a,\Pi}  \right ) = C_a' \left ( \mathbf{r},\,\mathbf{v},\,t\right )
\end{equation}
where
\begin{equation}
\label{experimental_entropy_source}
C_a' \left ( \mathbf{r},\,\mathbf{v},\,t\right ) =  C_a \left ( \mathbf{r},\,\mathbf{v},\,t\right ) - \displaystyle \frac{\partial }{ \partial \mathbf{r} } \cdot \displaystyle \left ( \delta N_a \displaystyle \sum_{\ell=1}^{N_a} \displaystyle \left [ \mathbf{v}_{a,\ell} \displaystyle \left (t \right ) - \mathbf{v} \right ] S \displaystyle \left ( \mathbf{r} - \mathbf{r}_{a,\,\ell} \displaystyle \left ( t \right ) \right ) \Pi_{\mathbf{\boldsymbol{\Delta}}_v}^{(0)} \left ( \mathbf{v} - \mathbf{v}_{a,\ell} \left ( t \right )\right ) \right)
\end{equation}
according to the continuous macroparticle equations of motion.
If $\Delta_{v_x} = \Delta_{v_y} = \Delta_{v_z} = \Delta_v $ is chosen sufficiently small such that there is a maximum of one macroparticle per bin volume, meaning $ \Pi_{\mathbf{\boldsymbol{\Delta}}_v}^{(0)} \left ( \mathbf{v} - \mathbf{v}_{l,m,n} \right ) \underset{\Delta_v \rightarrow 0}{\rightarrow} \delta  \left ( \mathbf{v} - \mathbf{v}_{l,m,n} \right )$ and $f_{a,\Pi} \displaystyle \left ( \mathbf{r}_{i,j,k},\mathbf{v}_{l,m,n},t_n\right )  \underset{\Delta_v \rightarrow 0}{\rightarrow} f_a \displaystyle \left ( \mathbf{r}_{i,j,k},\mathbf{v}_{l,m,n},t_n\right )$, the discretized empirical entropy (\ref{experimental_entropy}) is only determined by the right-hand side (\ref{non_physical_term}) of Klimontovich-like equations (\ref{Klimontovich_like_equation}) describing the effects of macroparticles internal tension forces. However, in practice, we do not have sufficient computer memory and there is always more than one macroparticle per velocity bin volume. As a consequence, we lose necessarily information about the simulation results and the second term in the right-hand side (\ref{experimental_entropy_source}) of equation (\ref{experimental_distribution_function_kinetic_equation}) cannot be neglected. Finally, let us stress here that, if one does not use a random number generator to initialize macroparticle velocities $\mathbf{V}_{a,\ell}$ at the simulation start $t_1$ by initializing, for example, a cold plasma where $\forall a,\,\ell,\,\mathbf{V}_{a,\ell} = \mathbf{0}$ instead, all equations are fully deterministic and there is no statistical fluctuations. For all PIC-simulated plasma  quantities $F$, $\mathbb{E} { \left \{ F \right \} } = F$ and the kinetic equations (\ref{averaged_PIC_Klimontovich}) become meaningless. Only the deterministic description (\ref{Klimontovich_like_equation}) applies in this case.

\section{The ensemble averaging technique}

\subsection{Reduction of statistical fluctuations via ensemble averaging}

Let us consider $N_\ess$ PIC simulations of the same stationary, homogeneous, infinite and fully ionized plasma that we have studied in the previous section 3, the simulations differing only by the randomly generated numbers used to initialize macroparticle velocities. We note for all PIC simulation plasma quantity $F \in \{ \delta \left ( \mathbf{v} - \mathbf{v}_{a,\ell} \left ( t_n\right ) \right )$, $\bar{f}_{a_c}$, $f_{a_c}$, $\delta f_{a_c}$, $\bar{n}_{a_c}$, $n_{a_c}$, $\delta n_{a_c}$, $\bar{f}_{a}$, $f_{a}$, $\delta f_{a}$, $\bar{n}_{a}$, $n_{a}$, $\delta n_{a}$, $\bar{\mathbf{E}}$, $\mathbf{E}$, $\delta \mathbf{E}$, $\bar{\mathbf{E}}_s$, $\mathbf{E}_s$ , $\delta \mathbf{E}_s  \}$,
\begin{equation}
\label{finite_ensemble_average}
\displaystyle \langle F \rangle = \displaystyle \frac{1}{ N_\ess } \displaystyle \sum_{\ess = 1}^{N_\ess} F^{(\ess)} \underset{ N_\ess \rightarrow \infty }{ \rightarrow } \mathbb{E} \displaystyle \left \{ F \right \}
\end{equation}
its finite ensemble average obtained from all PIC simulation results, each one denoted by the superscript $(\ess),\,\ess \in \left [ 1,\,N_\ess \right ]$.
By linearity, the expected value of the ensemble average of PIC simulation plasma quantities is also their expected value
\begin{equation}
\mathbb{E} \displaystyle \left \{ \displaystyle \langle F \rangle \right \} =  \mathbb{E} \displaystyle \left \{ F \right \}
\end{equation}
during one single PIC simulation. We thus have in particular
\begin{equation}
\mathbb{E} \displaystyle \left \{ \displaystyle \langle f_a \left ( \mathbf{r}_{i,j,k},\,\mathbf{v},\,t_n \right )  \rangle \right \} = \bar{f}_a \left ( \mathbf{r}_{i,j,k},\,\mathbf{v},\,t_n \right )
\end{equation}
and
\begin{equation}
\label{mean}
\mathbb{E} \displaystyle \left \{ \displaystyle \langle \mathbf{E} \left ( \mathbf{r}_{i,j,k},\,t_n \right )  \rangle \right \} = \bar{\mathbf{E}} \left ( \mathbf{r}_{i,j,k},\,t_n \right ).
\end{equation}
As a consequence, the deterministic description of the finite ensemble average of PIC-simulations consists in the kinetic equation 
\begin{equation}
\label{ensemble_Klimontovich_like_equation}
\begin{array}{lll}
& & {\displaystyle \left . \displaystyle \frac{ \partial \langle  f_{a} \rangle }{\partial t} \right |} \left ( \mathbf{r}_{i,j,k},\, \mathbf{v},\,t_n \right ) + {\displaystyle \left . \displaystyle \frac{ \partial  }{\partial \mathbf{r}} \cdot \displaystyle \left ( \mathbf{v} \langle f_{a} \rangle \right ) \right |} \left ( \mathbf{r}_{i,j,k},\, \mathbf{v},\,t_n \right ) + \displaystyle \frac{ \partial  }{\partial \mathbf{v}} \cdot \displaystyle \left ( \displaystyle \frac{q_a}{m_a} \langle \mathbf{E}_s \left ( \mathbf{r}_{i,j,k},\,t_n\right ) \rangle \langle f_{a} \left ( \mathbf{r}_{i,j,k},\,\mathbf{v},\, t_n \right ) \rangle \right )  
\cr &=& - \displaystyle \frac{q_a}{m_a}  \displaystyle \frac{ \partial }{\partial \mathbf{v}} \cdot \displaystyle \left [ \displaystyle \langle  {\delta \mathbf{E}_s } \left ( \mathbf{r}_{i,j,k},\,t_n\right ) \delta f_{a}  \left ( \mathbf{r}_{i,j,k},\, \mathbf{v},\,t_n  \right ) \rangle - \langle  {\delta \mathbf{E}_s } \left ( \mathbf{r}_{i,j,k},\,t_n\right ) \rangle \langle  \delta f_{a}  \left ( \mathbf{r}_{i,j,k},\, \mathbf{v},\,t_n  \right ) \rangle \right ] 
\cr && +  \displaystyle \langle C \displaystyle \left ( \mathbf{r}_{i,j,k},\,\mathbf{v},\,t_n \right ) \rangle 
\end{array}
\end{equation}
self-consistently coupled with the finite ensemble average of discretized Maxwell equations
\begin{equation}
\label{ensemble_PIC_Maxwell}
\displaystyle \left \{\begin{array}{lllll}
      {\displaystyle \left . \displaystyle \frac{\partial}{\partial \mathbf{r}} \,\, \cdot \,\langle{\mathbf{E}\rangle} \hspace{0.5em}\right |}^{i,j,k,n}&=& 4 \pi \displaystyle \sum_a q_a \displaystyle \int_{\Reals^3} \langle{f_{a} \displaystyle \left ( \mathbf{r}_{i,j,k},\,\mathbf{v},\,t_n \right ) \rangle} d^3 \mathbf{v}
\cr {\displaystyle \left . \displaystyle \frac{\partial}{\partial \mathbf{r}}  \times \langle{\mathbf{E}\rangle} \hspace{0.5em} \right |}^{i,j,k,n} &=& \mathbf{0}
\end{array} \right . .
\end{equation}
according to (\ref{Klimontovich_like_equation}), (\ref{PIC_Maxwell}) and (\ref{finite_ensemble_average}). Because all PIC simulations use different randomly generated numbers to initialize macroparticle velocities, they are all statistically independent. We thus deduce the relationship between the finite ensemble average of PIC simulation plasma quantity variances with their own variances from one single PIC simulation. For all PIC simulation plasma quantity  $G \in \left \{ \bar{n}_{a_c},\,n_{a_c},\,\delta n_{a_c},\,\bar{n}_a,\, \,n_{a},\,\delta n_{a},\, \bar{\mathbf{E}},\, \mathbf{E},\,\delta \mathbf{E},\,\bar{\mathbf{E}}_s,\,\mathbf{E}_s ,\, \delta \mathbf{E}_s\right \}$, we have
\begin{equation}
\mathbb{E} \Big \{ {\Big ( \displaystyle \langle G \rangle - \mathbb{E} \displaystyle \left \{  \displaystyle \langle G \rangle  \right \} \Big ) }^2 \Big \} =  \displaystyle \frac{ 1 }{ N_\ess}  \mathbb{E} \Big \{ {\Big(  G - \mathbb{E} \displaystyle \left \{  G  \right \} \Big ) }^2  \Big \}.
\end{equation}
We thus obtain in particular 
\begin{equation}
\label{variance}
\mathbb{E} \Big \{ { \displaystyle \langle \delta \mathbf{E} \left ( \mathbf{r}_{i,j,k},\,t_n \right ) \rangle }^2 \Big \} = \displaystyle \frac{ \mathbb{E} \Big \{ {  \delta \mathbf{E} \left ( \mathbf{r}_{i,j,k},\,t_n \right ) }^2 \Big \} }{N_\ess}.
\end{equation}
If $N_\ess$ is sufficiently large, the finite ensemble average of electrostatic field statistical fluctuation $\langle \delta \mathbf{E} \displaystyle \left ( \mathbf{r}_{i,j,k},\,,t_n \right ) \rangle$ at grid points $\mathbf{r}_{i,j,k}$ and time steps $t_n$ follows a normal distribution law with mean  (\ref{mean}) and variance (\ref{variance}) according to the central limit theorem. One deduces 
\begin{equation}
 \displaystyle \left | \displaystyle \langle \delta \mathbf{E} \displaystyle \left ( \mathbf{r}_{i,j,k},\,,t_n \right ) \rangle \rangle \right | \underset{N_\ess \gg 1}{\leq} \displaystyle \sqrt{ 2 }\,\text{erf}^{-1} \displaystyle \left ( \displaystyle \frac{p}{100} \right ) \displaystyle \left | \displaystyle \langle \mathbf{E}_\text{fluct} \rangle \right |
\end{equation}
with a $p \%$-confidence where we have noted
\begin{equation}
\label{ensemble_averaged_electrostatic_field_fluctuations_estimate}
\displaystyle \left | \displaystyle \langle \mathbf{E}_\text{fluct} \rangle \right | = \displaystyle \frac{ \displaystyle \left | \displaystyle \mathbf{E}_\text{fluct} \right | }{ \sqrt{N_\ess}  }
\end{equation}
assuming again Assumptions \ref{Assumption_consistency_errors}, \ref{Assumption_homogeneous},  \ref{Assumption_collisions} and  \ref{Assumption_stationary} as in section 3.2. Considering again the particular case of PIC simulations of a collisionless plasma at equilibrium using equal weights $\delta N_e = \delta N_i = \delta N$ and $\Delta_t < \Delta_x / v_T = \Delta_y / v_T = \Delta_z / v_T < 1 / \omega_p $, it reads 
\begin{equation}
\label{ensemble_averaged_electrostatic_field_fluctuations_estimate_example}
\displaystyle \left | \displaystyle \langle \mathbf{E}_\text{fluct} \rangle \right | \underset{\omega_ p t_n \gg 1}{\sim} \displaystyle \sqrt{\displaystyle \frac{ \delta N }{ N_\ess } \displaystyle \frac{ k_B T }{ {\lambda_D}^3 } \eta \left ( k_g \lambda_D \right ) }.
\end{equation}
A finite ensemble average of $N_\ess$ PIC simulations thus reduces statistical fluctuation amplitudes of electrostatic fields and densities by a factor $\sqrt{N_\ess}$ compared to one single electrostatic PIC simulation. Also, since 
$ \delta N \propto 1 / N_{e,\text{mpc}} $ according to its definition (\ref{density}), we deduce there is an equivalence between running one PIC simulation using $N_{e,\text{mpc}}$ macroparticles and a finite ensemble average of $N_{e,\text{mpc}} / N_{e,\text{mpc}}'$ PIC simulations using each one $N_{e,\text{mpc}}'$ macroelectrons per cell. We can thus apply the central limit theorem to estimate the statistical fluctuations in one single PIC simulations using a sufficiently large number of macroparticle per cell $N_{e,\text{mpc}} \gg N_{e,\text{mpc}}' \geq 2$ in 1D, 4 in 2D and 8 in 3D as we did in section 3.2.

\subsection{Kinetic equation for an ensemble average of electrostatic PIC simulations}

Similarly, if we note $F'$ another PIC simulation plasma quantity distinct from $F$, the statistical independence of PIC simulations makes the finite ensemble average of PIC simulation plasma quantity covariance between $F$ and $F'$ decrease by a factor $\sqrt{N_\ess}$ compared to their covariance from one single PIC simulation
\begin{equation}
\mathbb{E} \Big \{ {\Big ( \displaystyle \langle F \rangle - \mathbb{E} \displaystyle \left \{  \displaystyle \langle F \rangle  \right \} \Big ) } {\Big ( \displaystyle \langle F' \rangle - \mathbb{E} \displaystyle \left \{  \displaystyle \langle F' \rangle  \right \} \Big ) } \Big \} =  \displaystyle \frac{ 1 }{ N_\ess}  \mathbb{E} \Big \{ {\Big(  F - \mathbb{E} \displaystyle \left \{  F  \right \} \Big ) } {\Big (  F' - \mathbb{E} \displaystyle \left \{  F'  \right \} \Big ) } \Big \}.
\end{equation}
We thus obtain in particular
\begin{equation}
\label{discretized_covariance2}
\mathbb{E} \displaystyle \left \{ \displaystyle \langle \delta \widehat{f}_{a_c} \left ( \mathbf{k},\,\mathbf{v},\,t_1 \right ) \rangle \displaystyle \langle \delta \widehat{f}_{b,c} \left ( \mathbf{k}',\,\mathbf{v}',\,t_1 \right ) \rangle  \right \} 
= \delta_{ab} \displaystyle \frac{\delta N_a}{ N_\ess } {\displaystyle \left ( 2 \pi \right )}^3 \bar{n}_a \delta \left ( \mathbf{k} + \mathbf{k}',\,\mathbf{k}_g \right ) F_{a0} \displaystyle \left ( \mathbf{v} \right ) \displaystyle \left [ \delta \displaystyle \left ( \mathbf{v} - \mathbf{v}' \right ) - F_{a0} \displaystyle \left ( \mathbf{v}' \right )\right ].
\end{equation}
Therefore, by defining statistical fluctuations
\begin{equation}
\displaystyle \langle \delta f_a  \left ( \mathbf{r}_{i,j,k},\, \mathbf{v},\,t_n \right ) \rangle =  \displaystyle \langle f_a  \left ( \mathbf{r}_{i,j,k},\, \mathbf{v},\,t_n \right ) \rangle - \displaystyle \langle \bar{f}_a  \left ( \mathbf{r}_{i,j,k},\, \mathbf{v},\,t_n \right ) \rangle
\end{equation}
and
\begin{equation}
\displaystyle \langle \delta \mathbf{E} \left ( \mathbf{r}_{i,j,k},\,t_n \right ) \rangle =  \displaystyle \langle \mathbf{E} \left ( \mathbf{r}_{i,j,k},\,t_n \right ) \rangle - \displaystyle \langle \bar{\mathbf{E}} \left ( \mathbf{r}_{i,j,k},\,t_n \right ) \rangle
\end{equation}
between the finite ensemble average of PIC simulation results and the expected ones, by performing the same derivation as in section 3.3 using (\ref{discretized_covariance2}) instead of (\ref{discretized_correlation}) and assuming again Assumptions \ref{Assumption_consistency_errors}, \ref{Assumption_homogeneous},  \ref{Assumption_collisions} and  \ref{Assumption_stationary}, we obtain the probabilistic description of the finite ensemble average of PIC simulations. The kinetic equation reads
\begin{equation}
\label{ensemble_average_kinetic_equation}
\begin{array}{l}
\hspace{1.5em} {\displaystyle \left . \displaystyle \frac{ \partial \langle {\bar{f}}_{a} \rangle }{\partial {t}} \right |} \left ( {\mathbf{r}}_{i,j,k},\, {\mathbf{v}}_a,\,{t}_n \right ) + {\displaystyle \left . \displaystyle \frac{ \partial  }{\partial {\mathbf{r}} } \cdot \displaystyle \left ( {\mathbf{v}}_a \langle {\bar{f}}_{a} \rangle \right ) \right |} \left ( {\mathbf{r}}_{i,j,k},\, {\mathbf{v}}_a,\,{t}_n \right )  + \displaystyle \frac{ \partial  }{\partial {\mathbf{v}}_a} \cdot \displaystyle \left ( \displaystyle \frac{{q}_a }{ {m}_a }  \langle {\bar{\mathbf{E}}}_s \left ( {\mathbf{r}}_{i,j,k},\, {t}_n \right ) \rangle \langle  {\bar{f}}_{a}  \left ( {\mathbf{r}}_{i,j,k},\, {\mathbf{v}}_a,\,{t}_n \right ) \rangle \right ) 
\cr \underset{ \omega_{p} {t}_n \gg 1}{=}
  -  \displaystyle \frac{1}{ {m}_a} \displaystyle \frac{ \partial }{\partial {\mathbf{v}}_a} \cdot \displaystyle \sum_b  \displaystyle \int_{\Reals^3} {\mathbf{Q}} \displaystyle \left ( {\mathbf{v}}_a,\,{\mathbf{v}}_b\right ) \cdot {\displaystyle \left . \displaystyle \left ( \displaystyle \frac{ {\delta N}_a }{ N_\ess } \displaystyle \frac{ \langle {\bar{f}}_{a} \left ( {\mathbf{v}}_a \right ) \rangle }{ m_b }\displaystyle \frac{ \partial \langle {\bar{f}}_{b} \rangle }{ \partial {\mathbf{v}}_b } - \displaystyle \frac{ {\delta N}_b }{ N_\ess } \displaystyle \frac{ \langle {\bar{f}}_{b} \left ( {\mathbf{v}}_b \right ) \rangle }{ m_a}\displaystyle \frac{ \partial \langle {\bar{f}}_{a}\rangle  }{ \partial {\mathbf{v}}_a } \right ) \right | } \left ( {\mathbf{r}}_{i,j,k},\, {t}_n \right ) d^3 {\mathbf{v}}_b.
\end{array}
\end{equation}
It is self-consistently coupled with the ensemble average of expected discretized Maxwell equations
\begin{equation}
\label{ensemble_average_expected_Maxwell}
\displaystyle \left \{\begin{array}{lllll}
      {\displaystyle \left . \displaystyle \frac{\partial}{\partial \mathbf{r}} \,\, \cdot \,\langle{\bar{\mathbf{E}}\rangle} \hspace{0.5em}\right |}^{i,j,k,n}&=& 4 \pi \displaystyle \sum_a q_a \displaystyle \int_{\Reals^3} \langle{\bar{f}_{a} \displaystyle \left ( \mathbf{r}_{i,j,k},\,\mathbf{v},\,t_n \right ) \rangle} d^3 \mathbf{v}
\cr {\displaystyle \left . \displaystyle \frac{\partial}{\partial \mathbf{r}}  \times \langle{\bar{\mathbf{E}}\rangle} \hspace{0.5em} \right |}^{i,j,k,n} &=& \mathbf{0}
\end{array} \right . .
\end{equation}
via the ensemble average of expected electrostatic field interpolated at macroparticle center locations
\begin{equation}
\langle \bar{\mathbf{E}}_s \displaystyle \left ( \mathbf{r}_{a,\ell} \left ( t_n \right ),\,t_n \right ) \rangle =  \displaystyle \sum_{i,j,k} \langle \bar{\mathbf{E}} \left ( \mathbf{r}_{i,j,k},\, t_n \right ) \rangle S \displaystyle \left ( \mathbf{r}_{a,\ell} \left ( t_n \right ) - \mathbf{r}_{i,j,k} \right ) \Delta_x \Delta_y \Delta_z.
\end{equation}
Consequently, the kinetic equations (\ref{ensemble_average_kinetic_equation}) of the ensemble average of distribution functions tend to the Vlasov equation in the limit of an infinite number of simulations $N_\ess / \delta N \rightarrow \infty$. More realistically, the ensemble average of $N_\ess = \delta N$ independent electrostatic PIC simulations modeling the same macroscopic plasma with macroparticle weights $\delta N = \delta N_e = \delta N_i$ is strictly equivalent to the numerical resolution of the Vlasov-Lenard-Balescu/Maxwell set of equations discretized in space and time only. Let us mitigate here these two important results (\ref{ensemble_averaged_electrostatic_field_fluctuations_estimate}) and (\ref{ensemble_average_kinetic_equation}) by reminding the reader that they are only valid assuming Assumptions \ref{Assumption_consistency_errors}, \ref{Assumption_homogeneous},  \ref{Assumption_collisions} and  \ref{Assumption_stationary}, that is, only for simulation time durations $L_t$ of collisionless plasmas using sufficiently small cell sizes and time steps such that
\begin{equation}
 \displaystyle \frac{ {\left ( \omega_{p_e} \Delta_t \right)}^{6} }{ { \eta \left ( k_g \lambda_D \right )} 24^2 }  \ll \displaystyle \frac{ \delta N / N_\ess}{ 4 \pi \bar{n}_e {\lambda_D}^3 } \ll \displaystyle \frac{1}{Z \omega_{p_e} L_t}
\end{equation}
similarly as we obtained in the previous section 4.

Precautions have been made in the previous paragraphs to distinguish between PIC simulation plasma quantities $F$, $F'$ and $G$. Indeed, the variance of plasma particle phase-space densities at the grid point $\mathbf{r}_{i,j,k}$ and velocity $\mathbf{v}$ naturally diverges due to the contribution of the term expressing the case where there is only one macroparticle in the phase-space volume located between $\left ( \mathbf{r}_{i,j,k},\,\mathbf{v}\right )$  and $\left ( \mathbf{r}_{i,j,k}+ \mathbf{\boldsymbol{\Delta}} ,\,\mathbf{v} + d^3 \mathbf{v}\right )$. To the best of our knowledge, there is no unbiased estimator of distribution function statistical realizations. However, to troubleshoot this issue, one may use the biased statistical estimator introduced in section 3.3 for the empirical distribution function (\ref{experimental_distribution_function}) to express
\begin{equation}
\label{statistical_estimator}
f_{a,\Pi} \displaystyle \left ( \mathbf{r}_{i,j,k},\mathbf{v}_{l,m,n},t_1\right ) = \displaystyle \frac{\bar{n}_a}{ N_{a,mpc} } \displaystyle \sum_{\ell_{i,j,k}=1}^{N_{a,mpc}} \Pi_{\mathbf{\boldsymbol{\Delta}}_v}^{(0)} \left ( \mathbf{v}_{l,m,n} - \mathbf{V}_{a,\ell_{i,j,k}} \right )
\end{equation}
according to Assumption 2.
Assuming $\Delta_{v_x} = \Delta_{v_y} = \Delta_{v_z} = \Delta_v \ll v_{T_a}$ for simplicity, one obtains
\begin{equation}
\label{mean2}
\begin{array}{lcl}
\bar{f}_{a,\Pi} \displaystyle \left ( \mathbf{r}_{i,j,k},\mathbf{v}_{l,m,n},t_1\right ) &=& \mathbb{E}{ \displaystyle \left \{ f_{a,\Pi} \displaystyle \left ( \mathbf{r}_{i,j,k},\mathbf{v}_{l,m,n},t_1\right ) \right \} } 
\cr &\underset{ \Delta_v \ll v_{T_a}}{ = }& \bar{n}_a F_{a_0} \left ( \mathbf{v}_{l,m,n} \right ) + \bar{n}_a \displaystyle \frac{  {\Delta_v}^2 }{ 24 } \displaystyle \sum_{\xi} \displaystyle \frac{d^2 F_{a_0}}{ d v_{\xi}^2 } \left ( \mathbf{v}_{l,m,n} \right ) + O \left [ { \displaystyle \left ( \displaystyle \frac{ \Delta_v }{ v_{T_a} } \right ) }^4\right ]
\end{array}
\end{equation}
and
\begin{equation}
\label{variance2}
\mathbb{E} \displaystyle \left \{ \delta f_{a,\Pi} \left ( \mathbf{r}_{i,j,k},\,\mathbf{v}_{l,m,n},\,t_1 \right )^2 \right \} \underset{ \Delta_v \ll v_{T_a}}{ = } \displaystyle \frac{ {\left . \bar{n}_a \right .}^2 }{ N_{a,\text{mpc}} } F_{a0} \displaystyle \left ( \mathbf{v}_{l,m,n} \right ) \displaystyle \left [ \displaystyle \frac{1}{ {\Delta_v}^3 } - F_{a0} \displaystyle \left ( \mathbf{v}_{l,m,n} \right )\right ].
\end{equation}
Therefore, in the limit of large number of macroparticles per cell, the statistical estimator of particle phase-space densities (\ref{statistical_estimator}) follows a normal distribution law with mean (\ref{mean2}) and variance (\ref{variance2}) according to the central limit theorem. Assuming the PIC simulation plasma is stable, we deduce in the limits $\omega_p t_n \gg 1$, $\Delta_v \ll v_{T_a}$, $N_{a,\text{mpc}} \gg 1$ and $N_\ess \gg 1$
\begin{equation}
\label{statistical_phase_space_density_fluctuations_estimate}
 \displaystyle \left | \displaystyle  \delta f_{a,\Pi} \left ( \mathbf{r}_{i,j,k},\,\mathbf{v}_{l,m,n},\,t_n \right )  \right | \leq \bar{n}_a \displaystyle \frac{  {\Delta_v}^2 }{ 24 } \displaystyle \sum_{\xi} \displaystyle \left | \displaystyle \frac{d^2 F_{a_0}}{ d v_{\xi}^2 } \left ( \mathbf{v}_{l,m,n} \right ) \right |
+ \displaystyle \sqrt{ \displaystyle \frac{ 2 }{ N_{a,\text{mpc}} } }\,\text{erf}^{-1} \displaystyle \left ( \displaystyle \frac{p}{100} \right ) \bar{n}_a \displaystyle \sqrt{ \displaystyle \frac{ F_{a0} \displaystyle \left ( \mathbf{v}_{l,m,n} \right ) }{ {\Delta_v}^3 } }
\end{equation}
and
\begin{equation}
 \displaystyle \left | \displaystyle \langle \delta f_{a,\Pi} \left ( \mathbf{r}_{i,j,k},\,\mathbf{v}_{l,m,n},\,t_n \right ) \rangle  \right | \leq \bar{n}_a \displaystyle \frac{  {\Delta_v}^2 }{ 24 } \displaystyle \sum_{\xi} \displaystyle \left | \displaystyle \frac{d^2 F_{a_0}}{ d v_{\xi}^2 } \left ( \mathbf{v}_{l,m,n} \right ) \right |
+ \displaystyle \sqrt{ \displaystyle \frac{ 2 }{ N_{a,\text{mpc}} N_\ess} }\,\text{erf}^{-1} \displaystyle \left ( \displaystyle \frac{p}{100} \right ) \bar{n}_a \displaystyle \sqrt{ \displaystyle \frac{ F_{a0} \displaystyle \left ( \mathbf{v}_{l,m,n} \right ) }{ {\Delta_v}^3 } }
\end{equation}
with a $p \%$-confidence. Statistical fluctuations of phase-space densities depends consequently on a scaling different from $1/\sqrt{ N_{a,\text{mpc}} }$ in single PIC simulations. Indeed, by using velocity bin sizes $  \Delta_v \propto v_{T_a} / {N_{a,\text{mpc}}}^n $ that decrease with decreasing number of macroparticles per cell, the optimal choice minimizing the bias is $n = 1 / \left ( 4 + \text{dim} \right )$ where "dim" denotes here the PIC simulation velocity space dimension. It provides instead a scaling $1/ \sqrt{{ N_{a,\text{mpc}}}^{4 /  \left ( 4 + \text{dim} \right ) } }$ for single PIC simulations and $1/ \sqrt{{ \left ( N_{a,\text{mpc}} \times N_\ess \right ) }^{4 /  \left ( 4 + \text{dim} \right ) } }$ for a finite ensemble average of $N_\ess$ PIC simulations.

\subsection{Application to linear Landau damping}

\begin{figure}
\centering
\includegraphics[height=6.cm]{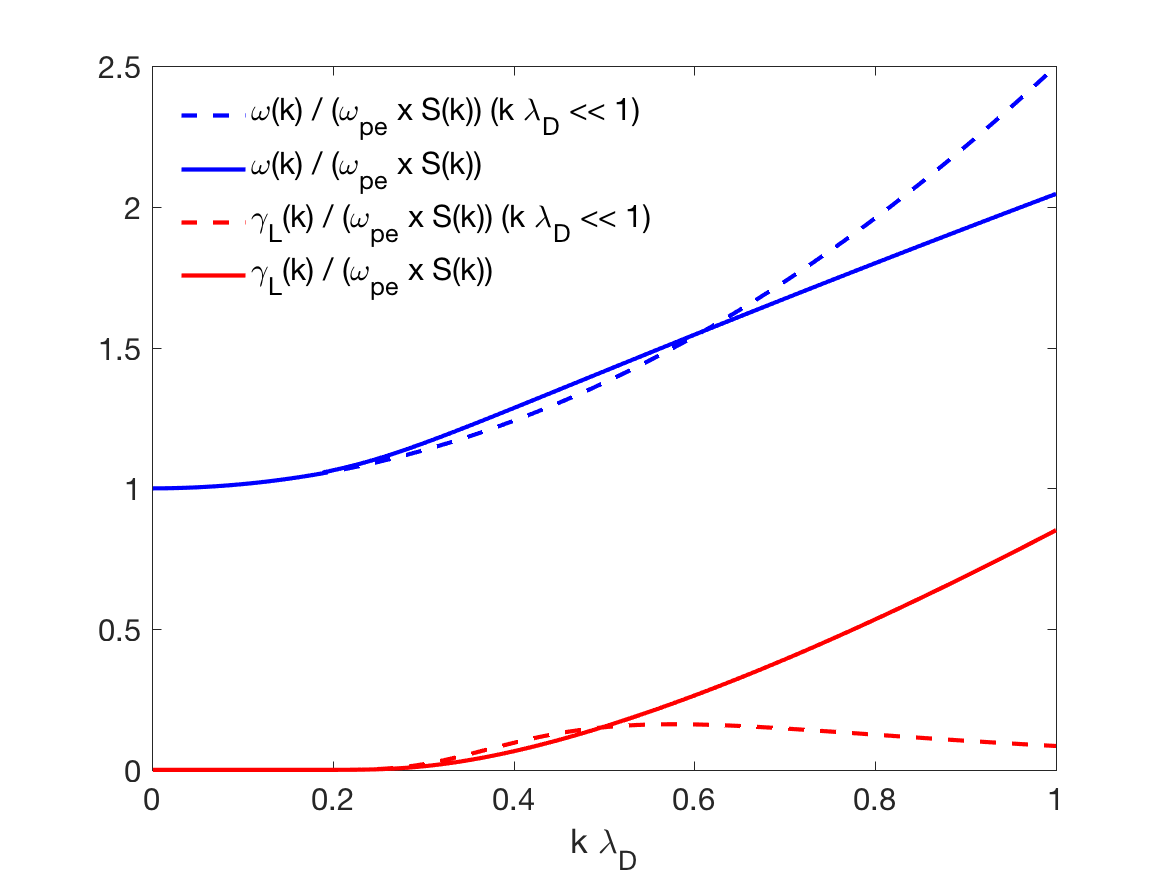}
\includegraphics[height=6.cm]{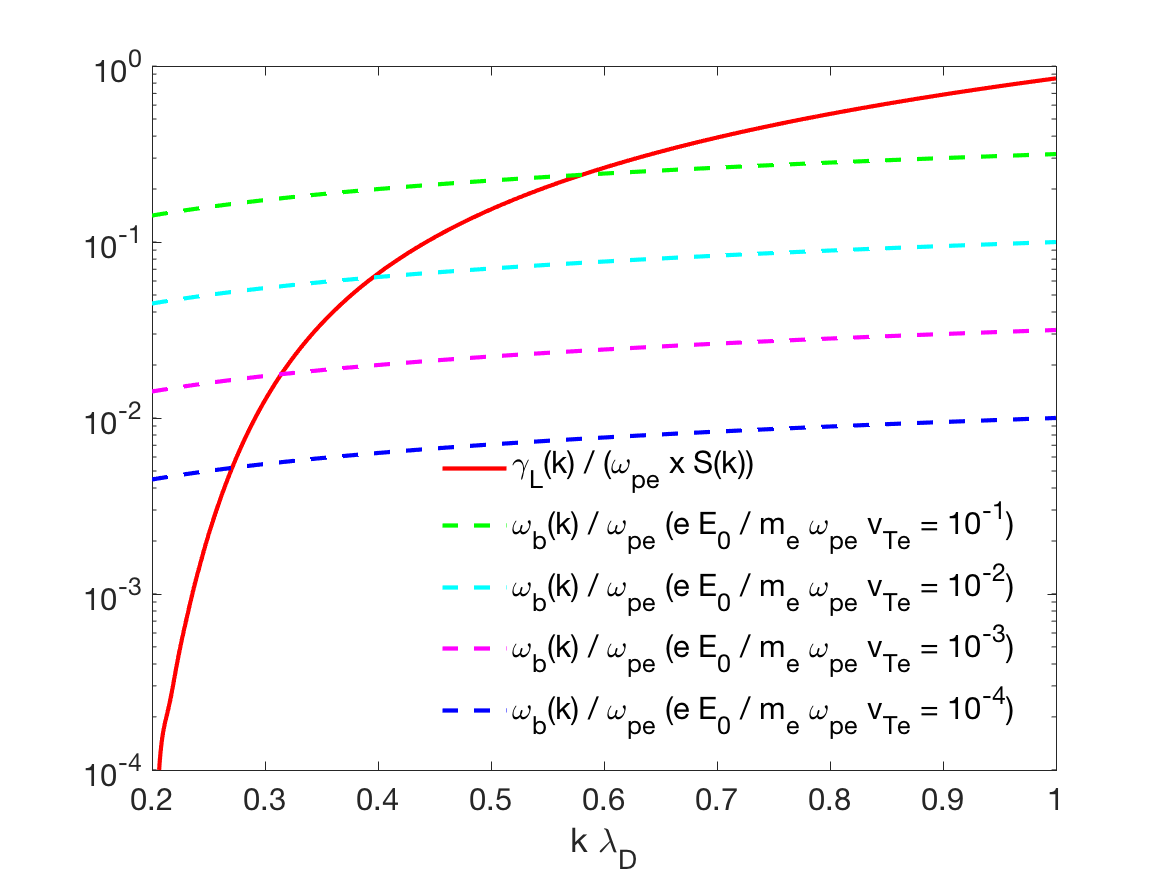}
\caption{(Left) Comparison between electrostatic wave angular frequency $\omega (k)$ and its Landau damping rate $\gamma_L (k)$ versus its  spatial angular frequency $k \lambda_D$ obtained using Taylor expansions in the limit $k \lambda_D \ll 1$ (dashed curves) and exact theoretical ones (full curves). (Right) Comparisons between Landau damping rates $\gamma_L$ and bounce frequencies $\omega_b$ versus spatial angular frequency of electrostatic waves $k \lambda_D$ in the interesting regime for different electrostatic wave amplitudes $E_0$.}
\label{Figure3}
\end{figure}

To illustrate how to decrease the amplitude of statistical fluctuations in PIC simulations by ensemble averaging, we consider here the linear Landau damping of a small amplitude electrostatic wave propagating in the 1D-3V PIC simulation plasma described in section 2. In order to generate the electrostatic wave, we perturb the PIC simulation plasma by driving an electrostatic field $ \mathbf{E}_\text{d} \left ( x_i,\, t_n \leq  0 \right )= E_\text{d} \cos{\displaystyle \left [  k_0 x_i - \omega_0 ( t_n + \tau )  \right ]} \mathbf{e}_x $ from $\omega_{p_e} t_n = - \omega_{p_e} \tau = - 3.6 $ to $\omega_{p_e} t_n=0$ at a spatial angular frequency $k_0 \lambda_D = 0.319$.
\cite{Landau1946} predicted that if particles move in a collisionless plasma at a velocity $v_x$ close to the wave phase velocity $v_{\varphi} = \omega  / k $ of an electrostatic wave, they see an almost constant electrostatic field and can thus interact with the wave. If $v_x \lessapprox v_{\varphi}$, trapped particles earn energy from the wave and if $v_x \gtrapprox v_{\varphi}$, they lose energy to it. Consequently, if one consider a PIC simulation plasma consisting of immobile ions and thermal electrons with an initial Maxwellian distribution of velocities (\ref{Maxwellian}), more macroelectrons will earn energy from the wave than the electrostatic wave from macroelectrons since $\partial F_{e_0} / \partial v_x \left (v_x > 0 \right )< 0$ in this case. As a result, the driven electrostatic wave is linearly damped at the Landau damping rate $\gamma_L$ in a first-order approximation. If one uses a Gaussian smoothing function (\ref{smoothing_function}) and chooses a sufficiently small time step $ \omega_{p_e} \Delta_t = 0.2$, aliasing effects are small in the expression of the PIC simulation plasma permittivity (\ref{numerical_permitivity}) and one may deduce from the resulting dispersion relation $\varepsilon_L \left ( \omega - \iota \gamma_L,\,k \right ) = 0 $ 
\begin{equation}
\displaystyle \frac{\omega \left ( k \right )}{\omega_{p_e}} \underset{ k \lambda_D \ll 1}{=}  \overset{\frown}{\text{S}}\left ( k \right ) \displaystyle \left [ 1 + \displaystyle \frac{3}{2} {\left ( k \lambda_D\right )}^2 \right ]
\end{equation} 
and
\begin{equation}
\label{Landau_damping_rate}
\displaystyle \frac{ \gamma_L \left ( k \right ) }{  \omega_{p_e} }\underset{ k \lambda_D \ll 1}{=} \displaystyle \sqrt{ \displaystyle \frac{ \pi }{ 8 } } \displaystyle \frac{ \overset{\frown}{\text{S}}\left ( k \right ) }{ { \left (k \lambda_D \right ) }^3 } \exp{ \displaystyle \left [ - \displaystyle \frac{ 1 }{ 2 { \left (k \lambda_D \right ) }^2 } - \displaystyle \frac{3}{2} \right ] }
\end{equation}
in the limit $k \lambda_D \ll 1$. Actually, $ \omega_0 = \omega \left ( k_0 \right ) \approx 1.18 \omega_{p_e} \overset{\frown}{\text{S}}\left ( k_0 \right )  $ and $\gamma_{L_0}  = \gamma_L \left ( k_0 \right ) \approx 1.96 \times 10^{-2} \omega_{p_e} \overset{\frown}{\text{S}}\left ( k_0 \right ) $ at $k_0 \lambda_D = 0.319$ if one solves numerically the dispersion relation; cf. the left panel of Figure \ref{Figure3}.
By using spatial grid spacings $\Delta_x = \lambda_D$, a quadratic interpolating function $\Pi_{\Delta_x}^{(2)}$ and a smoothing parameter $a_x = 0.866667 \lambda_D$ such that we lose the less physical information from spatial angular frequency spectra as possible, the macroparticle shape (\ref{particle_shape}) is $\overset{\frown}{\text{S}}\left ( k_0 \right ) \approx 0.903$ at the considered spatial angular frequency $k_0 \lambda_D = 0.319$ and it doesn't affect too much the theoretical results.

\begin{figure}
\centering
\includegraphics[height=6.cm]{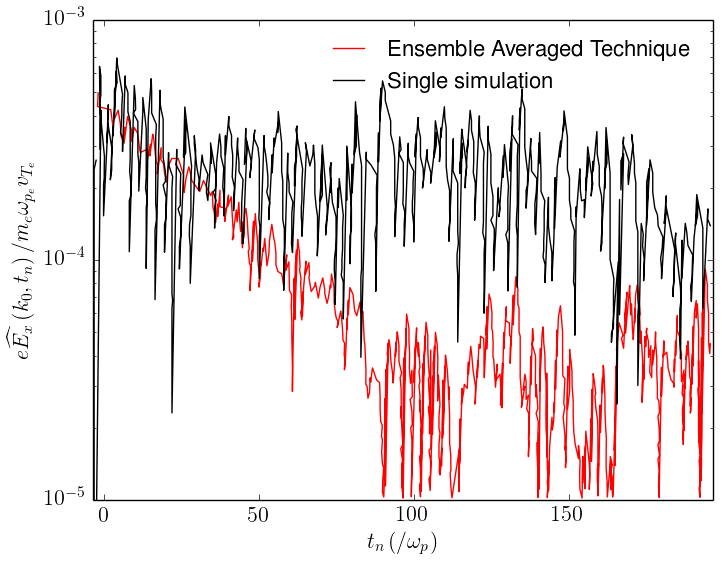}
\includegraphics[height=6.cm]{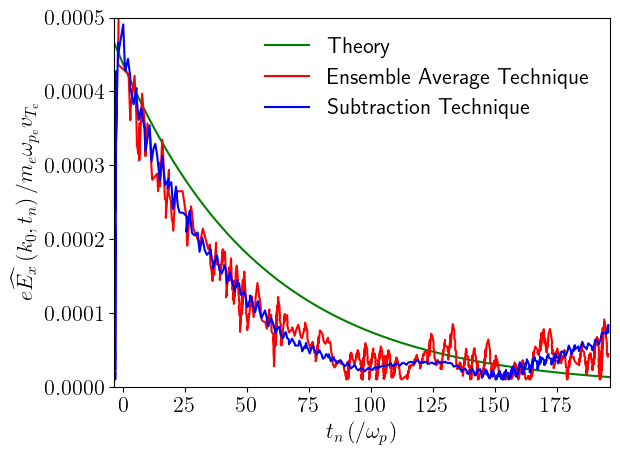}
\caption{(Left) Comparison of linearly Landau-damped electrostatic wave spatial Fourier mode $p_0$ from one single electrostatic 1D-3V PIC simulation (black curve) and an average of 30 simulations (red curve). (Right) Comparison of linearly Landau-damped electrostatic wave spatial Fourier mode $p_0$ from Landau theory (green curve), from an average of 30 electrostatic 1D-3V PIC simulations (red curve) and from the subtraction of one nominal simulation without perturbation from one PIC simulation with perturbation (blue curve).}
\label{Figure4}
\end{figure}

As macroelectrons damp the wave, they also earn energy. As a result, at the end of this linear stage more macroelectrons finish with a velocity $v_x \gtrapprox v_\phi$ compared to those who finish with a velocity $v_x \lessapprox v_\phi$ so that the sign of the resulting distribution of electron velocities changes and macroelectrons give back the received energy to the wave. By studying the energy conservation between the electrostatic wave and one trapped macroelectron, one deduces the non-linear Landau damping occurs at the bounce angular frequency
\begin{equation}
\label{bounce_frequency}
\omega_{b} \left ( k \right )= \displaystyle \sqrt{\displaystyle \frac{ e E_0 k}{ m_e } }
\end{equation}
where $E_0$ is the electrostatic wave amplitude. Consequently, the condition for observing the linear Landau damping of an electrostatic wave is given by the scaling $\omega \gg \gamma_{L} \gg \omega_{b} $ and the electrostatic wave amplitude is consequently constrained according to 
\begin{equation}
\label{amplitude_level}
\displaystyle \frac{ e E_0 }{ m_e \omega_{p_e} v_{T_e} } \ll  \displaystyle \frac{ { \left ( \gamma_L / \omega_{p_e}\right ) }^2 }{ k \lambda_D } \ll \displaystyle \frac{ { \left ( \omega / \omega_{p_e}\right ) }^2 }{ k \lambda_D }.
\end{equation}
Plots in Figure \ref{Figure3} stop at the spatial angular frequency of $k \lambda_D = 1$. Indeed, above this value, the linear Landau damping rate has the same order of magnitude compared to the angular frequency of electrostatic wave oscillations. Therefore, the concept of electrostatic waves is meaningless in this regime. In the opposite regime where angular spatial frequencies $k \lambda_D \lessapprox 0.2$, the linear Landau damping rate decreases drastically with decreasing spatial angular frequencies by many order of magnitudes according to equation (\ref{Landau_damping_rate}). In this regime, Landau damping is negligible and electrostatic waves may consequently propagate almost freely in the PIC simulation plasma. In the right panel of Figure \ref{Figure3}, we compare the Landau damping rate $\gamma_L ( k )$ and bounce angular frequencies (\ref{bounce_frequency}) for different values of initial perturbing electrostatic field amplitude $E_0$ in the interesting regime of Landau damping $0.2 \lessapprox k \lambda_D \leq 1$. While the observation of non-linear Landau damping of a perturbing electrostatic wave is relatively easy, the observation of linear Landau damping is much more challenging due to the smaller initial amplitude of the perturbing electrostatic field that is needed and that is usually below the PIC simulation plasma electrostatic field statistical fluctuation amplitudes (\ref{statistical_fluctuation_amplitude_0}). In our case, $k_0 \lambda_D = 0.319$ leads to a value of $ { \left ( \gamma_L / \omega_{p_e}\right ) }^2 / k_0 \lambda_D  \approx 1.2 \times 10^{-3} $.  
In a first approximation and neglecting statistical fluctuations, the self-consistent PIC simulation driven plasma electrostatic wave increases linearly with time $ \mathbf{E}_x \left ( x_i,\,t_n \leq 0\right ) \approx E_\text{d} \alpha ( t_n + \tau ) \sin{\displaystyle \left [  k_0 x_i - \omega_0 ( t_n + \tau )\right ]} \mathbf{e}_x$ under the action of the driver where $\alpha$ is the growing rate of the field. 
In order to give orders of magnitude, let us consider a laser-generated Hydrogen plasma ($Z = 1$) at equilibrium with an electron density $\bar{n}_e \approx 5.4\times 10^{19} \text{ cm}^{-3}$ consisting of immobile ions ($m_e / m_i \ll 1 \Rightarrow v_{T_i} \ll v_{T_e}$) and thermal electrons at a temperature $T_e \approx 7$ keV. For such a 1D collisionless plasma, the number of electrons in a Debye cube is about $\bar{n}_e {\lambda_D}^3 \approx 32,768$ and the Debye length about $\lambda_D \approx 0.085\,\mu$m. We perform the simulations by using the open source spectral electrostatic PIC code BEPS developed from the UPIC framework \cite{Decyk2007} available at \url{https://github.com/UCLA-Plasma-Simulation-Group}. We use a macroelectron weight $\delta N_e = 1$, a number of macroelectrons $N_e = 8,388,608$ and a simulation box $L_x = 256 \lambda_D$ leading to a number $N_{e,\text{mpc}} = 32,768$ of macroelectrons per cell and $k_0 \lambda_D = 2 \pi p_0 / L_x $ with $p_0 = 13$. The choice of a macroelectron weight $\delta N_e = 1$ allows us to emphasize again that physical statistical fluctuations are also present in dedicated experiments and they are not a numerical noise. The input deck for single simulations can be found in Appendix D. Given all these simulation parameters, we have 
\begin{equation}
  \displaystyle \frac{ 4 \pi {\left ( \omega_{p_e} \Delta_t \right)}^{6} }{\arctan{ \left ( k_g \lambda_D \right )} 24^2 } \approx 9.8\times10^{-7} \leq \displaystyle \frac{ 1 }{ 4 \pi \bar{n}_e {\lambda_D}^3 } \approx 2.4 \times 10^{-6}\ll \displaystyle \frac{1}{ \omega_{p_e} L_t} = 5\times10^{-3}
\end{equation}
with $L_t = 200 / \omega_{p_e}$ ($N_t = 1000$ time iterations). It means according to the previous section 2 that all our estimates obtained in sections 3, 4 and 5 are valid for these simulations. In particular, consistency errors due to the leap-frog scheme are smaller than statistical fluctuation amplitudes 
\begin{equation}
\label{fluctuations_level}
 \displaystyle \left | \displaystyle \delta \mathbf{E} \displaystyle \left (x_i,\,,t_n \right ) \right | \underset{N_{e,\text{mpc}} \gg 1}{\leq} \displaystyle \sqrt{ 2 }\,\text{erf}^{-1} \displaystyle \left ( \displaystyle \frac{p}{100} \right ) \displaystyle \left | \mathbf{E}_\text{fluct} \right |
\end{equation}
with $e \displaystyle \left | \mathbf{E}_\text{fluct} \right | / m_e \omega_{p_e} v_{T_e} \approx 
5 \times 10^{-4}$ and a $p \%$-confidence. We choose $e E_\text{d} / m_e \omega_{p_e} v_{T_e} = 10^{-3}$. According to simulations, we measure a growing rate of the field $\alpha \approx 0.243\, \omega_{p_e}$; cf. Figure \ref{Figure4}. 
At $t_n = 0$, the driver is stopped and we expect the driven traveling electrostatic wave is Landau-damped according to
\begin{equation}
\mathbf{E}_x \left ( x_i,\, t_n \geq 0 \right ) \underset{ \omega_b t \ll 1 }{ \approx } E_0   \sin{\displaystyle \left [  k_0 x_i - \omega_0 ( t_n + \tau ) \right ]} \exp{ \displaystyle \left ( - \gamma_{L_0}  t_n \right ) }\mathbf{e}_x
\end{equation}
with $ e E_0 /  m_e \omega_{p_e} v_{T_e} =  e E_\text{d} \alpha \tau /  m_e \omega_{p_e} v_{T_e} \approx 8.75 \times 10^{-4}$. It is just slightly above the statistical fluctuations level and leads to a bounce frequency $\omega_{b_0} = \omega_b (k_0) = 1.67 \times 10^{-2} \omega_{p_e} $ slightly lower than the linear Landau damping rate $\gamma_{L_0}$ in agreement with the simulations; cf Figure \ref{Figure3}.  The observation of linear Landau damping is therefore difficult to observe in such electrostatic PIC simulations or dedicated experiments. A method proposed by \cite{Decyk1987} has already been applied by \cite{Grismayer2011} for resolving this problem. The subtraction technique consists in running two identical PIC simulations that use the same randomly generated numbers to initialize macroparticle velocities at the simulation start. Then, the nominal simulation without perturbation is subtracted from the one initialized with the small amplitude perturbing electrostatic field. Here, we use the ensemble average of $N_\ess = 30$ simulations and we compare the different results. In the left panel of Figure \ref{Figure4}, we have plotted the spatial mode $p_0$ of electrostatic field discrete Fourier transform 
\begin{equation}
\widehat{E}_x \left ( k_0,\, t_n  \right ) = \displaystyle \frac{1}{N_x} \displaystyle \sum_{i=1}^{N_x} E_x \left ( x_i,\,t_n \right ) \exp{ \displaystyle \left ( - \iota  2 \pi p_0 x_i / L_x \right ) }
\end{equation}
from one single simulation and from their ensemble average. We can see statistical fluctuation amplitudes (\ref{fluctuations_level}) are effectively reduced by a factor $\sqrt{30} \approx 5.5$ in agreement with our estimate  (\ref{ensemble_averaged_electrostatic_field_fluctuations_estimate}). After $\omega_{p_e} t_n \approx 20$, one single PIC simulation doesn't allow for the observation of the tiny electrostatic field amplitude (\ref{amplitude_level}) below statistical fluctuations (\ref{fluctuations_level}) while the electrostatic wave amplitude continue to decrease until $\omega_{p_e} t_n \approx 100$ where it reaches the reduced statistical fluctuation amplitudes $\displaystyle \left | \displaystyle \langle \mathbf{E}_\text{fluct} \rangle \right | \approx 9 \times 10^{-5}$. An advantage of the ensemble averaging technique is that simulations do not need to be parallelized compared to an equivalent but more-computationally-extensive simulation using $N_\ess$ times more macroparticles and macroparticle weights $\delta N_e / N_\ess$.
In the right panel of Figure \ref{Figure4}, the ensemble average results is compared with the substraction technique and the theoretical result
\begin{equation}
 \widehat{E}_x \left ( k_0,\, t_n > 0 \right )  \underset{ \omega_b t \ll 1 }{ \approx } \displaystyle \frac{E_0}{ 2 } \exp{ \left ( - \gamma_{L_0} t_n \right ) }.
\end{equation}
We find good agreements between the two different methods and the Vlasov/Maxwell theory.

\section{Conclusions}

PIC simulations are one of the most widely used and most powerful tools in plasma physics, being a core component and methodology of plasma physics as a scientific discipline. However, the statistical physics principles of a PIC code (and what is actually computed) are still a source of confusion among many users and developers. Conflicting views assume that a PIC code evolves the macroscopic Vlasov-Maxwell set of equations, others consider PIC simulations describe the microscopic Klimontovich/Maxwell set of equations, while others claim that PIC simulations cannot be compared to (or used with) the kinetic theory of plasmas. In this paper, we clarify the kinetic properties of PIC simulations. 
We have derived the deterministic description of a PIC simulation electrostatic plasma, or PIC plasma. This description consists of a Klimontovich-like equation discretized in space and time, but not in velocity space, with a source term (in the right-hand side of the Klimontovich-like equation) due to the internal tension force of macroparticles. In the left-hand side of the Klimontovich-like equation, the collective electrostatic force interpolated at macroparticle locations is self-consistently coupled with the discretized Maxwell equations describing the evolution of the (approximated) microscopic fields. The source terms of the discretized Maxwell's equations are obtained by depositing the electrical charge and current of macroparticles onto the spatial grid such that the Coulomb electrostatic fields generated by macroparticles are reduced at distances smaller than the spatial grid spacing. If the minimum possible distance between particles due to their binary electrostatic interaction (the Landau length) is not resolved, then close-encounter binary Coulomb collisions between macroparticles are underestimated in PIC codes.  To understand the statistical properties of PIC plasmas, we have applied the theory of fluctuations in collisionless plasmas derived by \cite{Rostoker1961,Klimontovich1962,Dupree1963} starting from the statistical fluctuations typically implemented by the random sampling used to initialize the velocities of macroparticles. By neglecting numerical consistency errors from the numerical schemes, we used the exact discrete Laplace-Fourier transform of the PIC kinetic equations coupled with the exact discrete Laplace-Fourier transform of self-consistent discretized Maxwell equations. We derived the single-time autocorrelation of the electrostatic field and plasma densities fluctuations, as well as the the single-time correlation between the fluctuations of weighted macroparticle center phase-space density and electrostatic field. This has allowed us to estimate the deviations from the expected results of the simulation by applying the central limit theorem to the statistical fluctuations in the limit of a large number of macroparticles per cell as well as deriving the kinetic equations for each PIC simulation plasma species. The former is useful to determine the spatial spectrum seeding an eventual physical (or numerical) instability in a PIC simulation plasma or to study a physical process of small amplitude that can be hidden in the "numerical noise" (fluctuations), and the latter allowed us to emphasize that "collisions" are indeed taken into account in a PIC plasma, where "collisions" mean here the the friction and diffusion of macroparticles in the fluctuating component of the electrostatic field due to Landau damping and plasma waves emission by macroparticles. The kinetic equations of the PIC plasma are indeed described by a Vlasov-Lenard-Balescu-like equation with non-physical spatial aliases in the "collision" integral (that can be mitigated by using a smoothing function with the cost of eventually losing a physical part of the plasma spatial spectrum) and with time aliases (that can be mitigated by using small time steps). Our theory recovers all results previously obtained \cite{Langdon1970a}, confirming their approach of perturbing a discretized Vlasovian plasma with a macroparticle test and then averaging the obtained physical quantity over the initial macroparticle velocity distribution. Our theory also extends their results to multiple-species multiple-weights PIC plasmas. The underlying hypothesis and assumptions for the validity of the theory have been systematically justified and/or estimated, closely connecting this with the PIC plasma stability and the simulation time duration, with estimates for these bounds provided as a function of the dimensionality of the PIC plasma, the number of real electrons in the plasma Debye sphere and the macroparticle weight factors. It is important to observe that the physical statistical fluctuations are overestimated when weight factors $\delta N > 1 $ are used i.e. when a much-smaller-than-in-reality number of macroparticles is chosen to represent the real plasma particles phase-space.  We have used our theoretical framework to explore the ”ensemble averaging technique” as a possible technique to recover lower/physical statistical fluctuations. In this method, a series of PIC simulations, representing identical statistical realizations of the same macroscopic initial conditions, are averaged. The equivalent statistical realizations, corresponding to different initial microscopic conditions of the PIC plasma, are obtained from the random sampling from the same distribution function of initial macroparticle velocities. It is shown that the ensemble average of $N_\ess = \delta N$ electrostatic PIC simulations is equivalent to the Vlasov-Lenard-Balescu/Maxwell set of equations discretized in space and time. Even for macroparticle weights of unity, the statistical fluctuations may hide theoretical features predicted in the Vlasov limit of an infinite number of electrons in the Debye sphere. In this case, the ensemble averaging technique can also be used to study in detail such microphysics processes beyond the limits imposed by statistical fluctuations. In order to illustrate the ”ensemble averaging technique”, we have applied it to the study of linear Landau damping of one electrostatic wave perturbing a PIC plasma in equilibrium by averaging the results obtained from $N_\ess = 30 $ simulations.
%, by averaging the results obtained from $N_\ess = 30 $ simulations that uses each one $N_{e,\text{mpc}} = 32,768$ macroelectrons per cell and a macroelectron weight $\delta N_e = 1$. 
The amplitude of the statistical fluctuations is reduced by a factor $\sqrt{N_\ess}$, thus allowing the ensemble averaging technique to identify short-lived very small amplitude perturbations that would be undetected in a single simulation. The method shows good agreement with the theory and the subtraction technique. 
This technique is also particularly suited for modern computer architectures, 
% Finally, except for some cases such as 3D-3V simulations with extremely large spatial simulation box and using a spectral Maxwell solver, the simulation time and RAM memory needed by a PIC simulation usually depend linearly on the total number of macroparticles used to sample all plasma species phase-spaces. 
since all $N_\ess$ simulations that will be averaged can run in parallel, allowing for speed ups by a factor $N_\ess$ the global simulation (including the individual simulations over which the ensemble averaging is going to be performed), as compared with an equivalent simulation that uses $N_\ess \times N_{e,\text{mpc}}$ macroelectrons per cell and a lower macroelectron weight $\delta N_e / N_\ess$. All $N_\ess$ simulations can be run in a sequential way too. The ensemble averaging technique can thus also be used if the equivalent simulation needs more RAM memory than what is available. The flexibility of the technique, the theoretical understanding presented in this paper, and the outstanding computational resources now available, open the way to a broader and wider use of this technique in all areas of plasma physics using PIC simulations.

\section*{Appendix}

\subsection*{Appendix A\\Derivation of the Laplace-Fourier transform of electrostatic field fluctuations single-time autocorrelation}

\begin{equation}
\label{useful_property}
\mathrm{For \, all \, functions \,g,} \displaystyle \left \{ \begin{array}{lll}
&\displaystyle \int_{V_{\mathbf{k}_g}} \displaystyle \frac{ d^3 \mathbf{k} }{ {\displaystyle \left ( 2 \pi \right )}^3 } \displaystyle \sum_{p,q,r} g \left ( \mathbf{k}_{p,q,r} \right ) &= \displaystyle \int_{\Reals^3} \displaystyle \frac{ d^3 \mathbf{k} }{ {\displaystyle \left ( 2 \pi \right )}^3 } g \left ( \mathbf{k} \right )
\cr &\displaystyle \int_{\iota \nu - \omega_g / 2}^{\iota \nu + \omega_g / 2} \displaystyle \frac{ d \omega }{  2 \pi   } \displaystyle \sum_{m} g \left ( \omega_{m} \right ) &= \displaystyle \int_{\iota \nu - \infty}^{\iota \nu + \infty}  \displaystyle \frac{ d \omega }{  2 \pi   } g \left ( \omega \right )
\end{array} \right . .
\end{equation}
Therefore, by substituting (\ref{discretized_correlation}) into (\ref{calculus_mm0}), one can perform the integration over $\mathbf{k}'$ by using the $\mathbf{k}_g$-periodicity property of $\left \{\mathbf{k}\right\}_\mathbf{r}$, $\varepsilon_L $ and $\delta \widehat{f}_{a_c}$ and the changes of indices $\left ( p',q',r' \right )$ into $\left (p'-p,q'-q,r'-r\right )$.
One gets
\begin{equation}
\mathbb{E} \displaystyle \left \{ \displaystyle \frac{ {\delta \mathbf{E} \left ( \mathbf{r}_{i,j,k},\,t_n \right )}^2 }{8 \pi} \right \} = \displaystyle \int_{V_{\mathbf{k}_g}} \displaystyle \frac{d^3 \mathbf{k}}{ {\left ( 2 \pi \right )}^3 } \displaystyle \int_{\iota \nu - \omega_g / 2}^{\iota \nu + \omega_g/2} \displaystyle \frac{d \omega}{ 2 \pi } W_{\omega,\mathbf{k}} \displaystyle \left ( \omega,\, \mathbf{k} \right ) 
\end{equation}
where we have introduced the spectral energy density of the electrostatic fluctuations
\begin{equation}
\label{calculus_mm2}
 {
\begin{array}{lcl}
W_{\omega,\mathbf{k}} \displaystyle \left ( \omega,\, \mathbf{k} \right ) &=& \displaystyle \int_{\iota \nu' - \omega_g / 2}^{\iota \nu' + \omega_g/2} \displaystyle \frac{d \omega'}{ 2 \pi }  \displaystyle \frac{  2 \pi  }{ \varepsilon_L \displaystyle \left ( \omega, \mathbf{k} \right ) \varepsilon_L \displaystyle \left ( \omega', -\mathbf{k} \right ) }  \displaystyle \frac{  1 }{ {\left \{\mathbf{k}\right\}_\mathbf{r}}^2 } \displaystyle \sum_{p,q,r} \overset{\frown}{\mathrm{S}} \left ( \mathbf{k}_{p,q,r} \right )  \overset{\frown}{\mathrm{S}} \left ( - \mathbf{k}_{p,q,r} \right ) \displaystyle \sum_a \delta N_a \bar{n}_a {q_a}^2
\cr &\times& \displaystyle \sum_{m} \displaystyle \sum_{m'} \displaystyle \int_{\Reals^3} d^3 \mathbf{v} \displaystyle \int_{\Reals^3} d^3 \mathbf{v}' \displaystyle \frac{ F_{a_0} \displaystyle \left ( \mathbf{v}\right ) \displaystyle \left [ \delta \displaystyle \left ( \mathbf{v} - \mathbf{v}' \right ) - F_{a0} \displaystyle \left ( \mathbf{v}' \right )\right ] }{ \displaystyle \left ( \mathbf{k}_{p,q,r} \cdot \mathbf{v} - \omega_m \right )  \displaystyle \left ( {\mathbf{k}}_{p,q,r} \cdot \mathbf{v}' + {\omega'}_{m'} \right ) }  \exp{\displaystyle \left [ - \iota \displaystyle \left (\omega + \omega' \right ) t_n  \right ]}.
\end{array}
}
\end{equation}
Since we have assumed that the PIC simulation plasma is stable which means that all poles of the longitudinal PIC simulation plasma permittivity (\ref{numerical_permitivity}) have a negative imaginary part, the second term in (\ref{calculus_mm2}) $\propto F_{a0} \left ( \mathbf{v} \right ) F_{a0} \left ( \mathbf{v}' \right ) $ has similarly poles $\omega'$ in the lower half of the $\omega'$-planes, only. Thus, the undamped part of the electrostatic field statistical fluctuations autocorrelation arises only from the contribution of the first term on large time scale $\omega_p t_n \gg 1 $.  We thus find after integration over $\mathbf{v}'$ and by using again the integral-sum property (\ref{useful_property})
\begin{equation}
\label{calculus_mm3}
 {
\begin{array}{lcl}
W_{\omega,\mathbf{k}} \displaystyle \left ( \omega,\, \mathbf{k} \right ) &\underset{\omega_p t_n \gg 1}{=}& \displaystyle \int_{\iota \nu' - \infty}^{\iota \nu' + \infty} \displaystyle \frac{d \omega'}{ 2 \pi }  \displaystyle \frac{  2 \pi  }{ \varepsilon_L \displaystyle \left ( \omega, \mathbf{k} \right ) \varepsilon_L \displaystyle \left ( \omega', -\mathbf{k} \right ) }  \displaystyle \frac{  1 }{ {\left \{\mathbf{k}\right\}_\mathbf{r}}^2 } \displaystyle \sum_{p,q,r} \overset{\frown}{\mathrm{S}} \left ( \mathbf{k}_{p,q,r} \right )  \overset{\frown}{\mathrm{S}} \left ( - \mathbf{k}_{p,q,r} \right ) \displaystyle \sum_a \delta N_a \bar{n}_a {q_a}^2
\cr &\times& \displaystyle \sum_{m} \displaystyle \int_{\Reals^3} d^3 \mathbf{v} \displaystyle \frac{ F_{a_0} \displaystyle \left ( \mathbf{v}\right ) }{ \displaystyle \left ( \mathbf{k}_{p,q,r} \cdot \mathbf{v} - \omega_m \right )  \displaystyle \left ( {\mathbf{k}}_{p,q,r} \cdot \mathbf{v} + {\omega'} \right ) }  \exp{\displaystyle \left [ - \iota \displaystyle \left (\omega + \omega' \right ) t_n  \right ]}.
\end{array}
}
\end{equation}
Then, since
\begin{equation}
\displaystyle \frac{1}{ \left ( \mathbf{k}_{p,q,r} \cdot \mathbf{v} - \omega_m \right ) \left ( \mathbf{k}_{p,q,r} \cdot \mathbf{v} + \omega' \right ) }  = \displaystyle \frac{1}{ \omega_m + {\omega'} } \displaystyle \left [  \displaystyle \frac{1}{\mathbf{k}_{p,q,r} \cdot \mathbf{v} - \omega_m} -  \displaystyle \frac{1}{\mathbf{k}_{p,q,r} \cdot \mathbf{v} + {\omega'} } \right ],
\end{equation}
the only one contributions that are not damped comes from the residue at poles ${\omega'} = - \omega_m$. In this sense, the factor $1/\left ( \omega_m + \omega' \right )$ is to be interpreted as $- \iota 2 \pi \delta \left ( \omega_m + \omega' \right )$ leading to
\begin{equation}
\label{calculusmm3}
 {
\begin{array}{lcl}
W_{\omega,\mathbf{k}} \displaystyle \left ( \omega,\, \mathbf{k} \right ) &\underset{\omega_ p t_n \gg 1}{=}& -  \iota  \displaystyle \frac{ 2 \pi  }{ {\varepsilon_L \displaystyle \left ( \omega, \mathbf{k} \right )}^2 }   \displaystyle \frac{  1 }{ {\left \{\mathbf{k}\right\}_\mathbf{r}}^2 } \displaystyle \sum_{p,q,r} {\overset{\frown}{\mathrm{S}} \left ( \mathbf{k}_{p,q,r} \right )}^2  \displaystyle \sum_a \delta N_a \bar{n}_a {q_a}^2 
\cr &\times&  \displaystyle \sum_{m} \displaystyle \lim_{\nu' \rightarrow 0^+} \displaystyle \int_{\Reals^3} d^3 \mathbf{v} F_{a0} \displaystyle \left ( \mathbf{v} \right ) \displaystyle \left [  \displaystyle \frac{1}{\mathbf{k}_{p,q,r} \cdot \mathbf{v} - \omega_{m} + \iota \nu'} -  \displaystyle \frac{1}{\mathbf{k}_{p,q,r} \cdot \mathbf{v} - \omega_{m} - \iota \nu'} \right ] 
\cr &\underset{\omega_ p t_n \gg 1}{=}& \displaystyle \frac{ 4 \pi^2  }{ {\varepsilon_L \displaystyle \left ( \omega, \mathbf{k} \right )}^2 } \displaystyle \frac{  1 }{ {\left \{\mathbf{k}\right\}_\mathbf{r}}^2 } \displaystyle \sum_{p,q,r} {\overset{\frown}{\mathrm{S}} \left ( \mathbf{k}_{p,q,r} \right )}^2  \displaystyle \sum_a \delta N_a \bar{n}_a {q_a}^2 
\cr &\times& \displaystyle \sum_{m} \displaystyle \int_{\Reals^3} d^3 \mathbf{v} F_{a0} \displaystyle \left ( \mathbf{v} \right ) \delta \displaystyle \left ( \omega_m - \mathbf{k}_{p,q,r} \cdot \mathbf{v} \right ) .
\end{array}
}
\end{equation}
Here, we have used the particle shape parity property $\overset{\frown}{\mathrm{S}} \left ( \mathbf{k} \right ) = \overset{\frown}{\mathrm{S}} \left ( -\mathbf{k} \right )$ and the $\left ( \omega_g,\,\mathbf{k}_g \right )$-periodicity property of the longitudinal PIC simulation plasma permittivity to simplify the equation. As a conclusion, one may express 
\begin{equation}
\mathbb{E} \displaystyle \left \{  \widehat{\widehat{\delta\mathbf{E}}} \left ( \omega, \mathbf{k} \right )  \widehat{\widehat{\delta\mathbf{E}}} \left ( \omega',\,\mathbf{k}' \right ) \right \}  \underset{\omega_ p t_n \gg 1}{=} {\displaystyle \left ( 2 \pi \right )}^4 \delta \displaystyle \left ( \omega + \omega',\,\omega_g  \right ) \delta \displaystyle \left ( \mathbf{k} + \mathbf{k}',\,\mathbf{k}_g \right ) \widehat{\widehat{ {\delta\mathbf{E}}^2 }}   \left ( \omega, \mathbf{k} \right )
\end{equation}
in the expression of electrostatic field statistical fluctuations autocorrelation (\ref{calculus_mm0}) where
\begin{equation}
\widehat{\widehat{ {\delta\mathbf{E}}^2 }}  \left ( \omega, \mathbf{k} \right ) = 8 \pi \, W_{\omega,\mathbf{k}} \displaystyle \left ( \omega,\, \mathbf{k} \right ) 
\end{equation}
reads (\ref{electrostatic_field_fluctuations_spectrum}).

\subsection*{Appendix B\\Derivation of the single-time autocorrelation of electrostatic field fluctuations in PIC simulation plasmas at equilibrium}

According to (\ref{permittivity_imaginary_part}), one finds back the fluctuation-dissipation theorem derived by \cite{Callen1951}
\begin{equation}
\label{fluctuation_dissipation_theorem}
W_{\omega,\mathbf{k}} \displaystyle \left ( \omega,\, \mathbf{k} \right ) = \displaystyle \frac{ \widehat{\widehat{ {\delta\mathbf{E}}^2 }}   \left ( \omega, \mathbf{k} \right ) }{8 \pi}\underset{\omega_ p t_n \gg 1}{=} - \delta N \displaystyle \frac{ k_B T }{ \omega } \displaystyle \frac{ \mathrm{Im}\displaystyle \left \{ \varepsilon_L \displaystyle \left ( \omega,\,\mathbf{k} \right )\right \} }{ {\varepsilon_L \displaystyle \left ( \omega, \mathbf{k} \right )}^2 }
\end{equation}
only in the particular case of PIC simulations of a plasma at equilibrium for which $\delta N_e = \delta N_i = \delta N$ and using a spectral solver with a smoothing function filtering all spatial frequency aliases $p,q,r \neq 0,0,0$. Indeed, only in this particular case, $\delta N$ can be extracted from the sum over all species in (\ref{electrostatic_field_fluctuations_spectrum}) simplifying the electrostatic fluctuations energy spectrum into  
\begin{equation}
\label{fluctuation_dissipation_theorem2}
W_{\omega,\mathbf{k}} \displaystyle \left ( \omega,\, \mathbf{k} \right ) \underset{\omega_ p t_n \gg 1}{=}  \displaystyle \frac{ 4 \pi^2 \delta N }{ {\varepsilon_L \displaystyle \left ( \omega, \mathbf{k} \right )}^2 } \displaystyle \frac{  {\overset{\frown}{\mathrm{S}} \left ( \mathbf{k} \right )}^2 }{ {\mathbf{k}}^2 }  \displaystyle \sum_a  \bar{n}_a {q_a}^2 \displaystyle \int_{\Reals^3} d^3 \mathbf{v} F_{a0} \displaystyle \left ( \mathbf{v} \right ) \delta \displaystyle \left ( \omega - \mathbf{k} \cdot \mathbf{v}, \omega_g \right )
\end{equation}
while the imaginary part of PIC simulation electrostatic plasmas longitudinal permittivity (\ref{permittivity_imaginary_part}) reduces to
\begin{equation}
\label{numerator}
\mathrm{Im}\displaystyle \left \{ \varepsilon_L \displaystyle \left ( \omega,\,\mathbf{k} \right )\right \} = - \displaystyle \frac{ 4 \pi^2 \omega}{k_B T} \displaystyle \frac{ { \overset{\frown}{\mathrm{S}} \left ( \mathbf{k} \right )}^2 }{ \mathbf{k}^2 } \sum_a \bar{n}_a {q_a}^2 \displaystyle \int_{\Reals^3} F_{a0} \displaystyle \left ( \mathbf{v}\right ) \delta \displaystyle \left ( \omega -  \mathbf{k} \cdot \mathbf{v}, \omega_g \right ) d^3 \mathbf{v}
\end{equation}
thus providing the fluctuation-dissipation theorem (\ref{fluctuation_dissipation_theorem}) by recognizing (\ref{numerator}) in the numerator of (\ref{fluctuation_dissipation_theorem2}).
Still in this particular case, one can get a simple estimate of the electrostatic field fluctuations energy spatial spectrum
\begin{equation}
W_\mathbf{k} \left ( \mathbf{k} \right ) = \displaystyle \int_{\iota \nu - \omega_g / 2}^{\iota \nu + \omega_g/2} \displaystyle \frac{d \omega}{ 2 \pi } W_{\omega,\mathbf{k}} \displaystyle \left ( \omega,\,\mathbf{k} \right ).
\end{equation} 
It reads
\begin{equation}
W_\mathbf{k} \left ( \mathbf{k} \right ) \underset{\omega_ p t_n \gg 1}{=} \delta N  \displaystyle \frac{k_B T}{2} \displaystyle \frac{ {\overset{\frown}{\mathrm{S}} \left ( \mathbf{k} \right )}^2  }{ {\overset{\frown}{\mathrm{S}} \left ( \mathbf{k} \right )}^2 + { \displaystyle \left ( \mathbf{k} \lambda_D \right )}^2  }
\end{equation}
by performing the $\omega$-integration using the residue theorem, assuming $\Delta_t \leq \Delta_x / v_{T_e}$ and where we have introduced the Debye screening length $\lambda_D = v_{T_e} / \omega_{p_e}$. Assuming then $a_x,\, a_y\,\mathrm{ and } \,a_z \ll \lambda_D $ and $\Delta_x = \Delta_y = \Delta_z \ll \lambda_D$ for simplicity, one may consider $\overset{\frown}{\mathrm{S}} \left ( \mathbf{k} \right ) \sim 1$ and estimate
\begin{equation}
\mathbb{E} \displaystyle \left \{ \displaystyle \frac{ {\delta \mathbf{E} \left ( \mathbf{r}_{i,j,k},\, t_n \right )}^2 }{8 \pi} \right \} = \displaystyle \int_{V_{\mathbf{k}_g}} \displaystyle \frac{d^3 \mathbf{k}}{ {\left ( 2 \pi \right )}^3 } W_\mathbf{k} \left ( \mathbf{k} \right ).
\end{equation}
While diverging in 2D and 3D collisionless plasmas for which $\delta N = 1$, $N_e \rightarrow \infty$ and $\Delta_x = \Delta_y = \Delta_z = \Delta_t \rightarrow 0$, we find (\ref{electrostatic_field_fluctuations_estimate}) for PIC simulation plasmas at equilibrium by replacing $d^3 \mathbf{k} / {\left ( 2 \pi \right )}^3$ with $d k_x / 2 \pi $ for 1D simulations and with $dk_x d k_y /  {\left ( 2 \pi \right )}^2$ for 2D simulations as well as the domain of integration $V_{\mathbf{k}_g}$ with the closed disk $ \{ (k_x,\,k_y) \, \mathrm{such} \, \mathrm{that} \,{k_x}^2 + {k_y}^2 \leq {(k_g / 2)}^2 \}$ for 2D simulations and with the closed sphere $ \{ (k_x,\,k_y,\,k_z) \, \mathrm{such} \, \mathrm{that} \, {k_x}^2 + {k_y}^2 + {k_y}^2 \leq {(k_g / 2)}^2 \}$ for 3D simulations instead of $V_{\mathbf{k}_g} = [-k_g/2,\,k_g/2]$, the closed square $[-k_g/2,\,k_g/2]^2$ and the closed cube $[-k_g/2,\,k_g/2]^3$ for 1D, 2D and 3D simulations respectively. An empirical factor $1/2$ for 2D simulations and $2/3$ for 3D simulations has been chosen to account for this latter approximation.

\subsection*{Appendix C\\Derivation of the "collision operator"}

According to (\ref{numerical_permitivity}) and the Plemelj formula, one may express the imaginary part of PIC simulation electrostatic plasmas longitudinal permittivity according to
\begin{equation}
\label{permittivity_imaginary_part}
 {
\begin{array}{l}
\mathrm{Im} \displaystyle \left \{ \varepsilon_L \displaystyle \left ( \omega, \mathbf{k} \right ) \right \} =  \displaystyle \frac{ \pi }{ {\displaystyle \left \{\mathbf{k}\right\}_\mathbf{r}}^2 } \displaystyle \sum_{p,q,r}  {\overset{\frown}{\mathrm{S}} \left ( \mathbf{k}_{p,q,r} \right )}^2  \displaystyle \sum_a { \omega_{p_a} }^2 \displaystyle \int_{\Reals^3} d^3 \mathbf{v}  \displaystyle \left \{\mathbf{k}\right\}_\mathbf{r}  \cdot \displaystyle \frac{ d F_{a0} }{ d \mathbf{v} } \delta \displaystyle \left ( \omega - \mathbf{k}_{p,q,r} \cdot \mathbf{v} ,\,\omega_g\right ).
\end{array}
}
\end{equation}
Therefore, since the collision operator (\ref{discrete_calculus_2}) is necessarily real, the first term in (\ref{discrete_calculus_1}) gives
\begin{equation}
 {
\begin{array}{lll}
&&  \displaystyle \int_{\iota \nu - \omega_g / 2}^{\iota \nu + \omega_g/2} \displaystyle \frac{d \omega}{ 2 \pi } \displaystyle \int_{V_{\mathbf{k}_g}} \displaystyle \frac{d^3 \mathbf{k}}{ {\left ( 2 \pi \right )}^3 } \displaystyle \sum_{p,q,r} { \left \{ \mathbf{k}_{p,q,r} \right \}_\mathbf{r} } \displaystyle \frac{ 8 \pi^2 \bar{n}_a q_a  {\overset{\frown}{\mathrm{S}} \left ( \mathbf{k}_{p,q,r} \right ) }^2 }{ {\varepsilon_L \left ( \omega,\,\mathbf{k}\right )}^2 { \left \{ \mathbf{k}_{p,q,r}\right \}_\mathbf{r}}^2 }  \delta N_a  
\cr &&  \hspace{12.em} \times \mathrm{Im} \displaystyle \left \{ \varepsilon_L \left ( \omega,\,\mathbf{k}\right )   \right\}  F_{a0} \left ( \mathbf{v}_a \right ) \delta \left ( \omega - \mathbf{k}_{p,q,r} \cdot \mathbf{v}_a, \omega_g \right )
\cr &=&  \displaystyle \int_{\iota \nu - \omega_g / 2}^{\iota \nu + \omega_g/2} \displaystyle \frac{d \omega}{ 2 \pi } \displaystyle \int_{V_{\mathbf{k}_g}} \displaystyle \frac{d^3 \mathbf{k}}{ {\left ( 2 \pi \right )}^3 } \displaystyle \sum_{p,q,r} { \left \{ \mathbf{k}_{p,q,r}\right \}_\mathbf{r} } \displaystyle \frac{ 32 \pi^4 \bar{n}_a q_a  {\overset{\frown}{\mathrm{S}} \left ( \mathbf{k}_{p,q,r} \right ) }^2 }{ {\varepsilon_L \left ( \omega,\,\mathbf{k}\right )}^2 { \left \{ \mathbf{k}_{p,q,r}\right \}_\mathbf{r} }^4 } \delta N_a  
\cr && \hspace{12.em} \times \displaystyle \sum_{p',q',r'}  {\overset{\frown}{\mathrm{S}} \left ( \mathbf{k}_{p',q',r'} \right )}^2 \displaystyle \sum_b \displaystyle \frac{ \bar{n}_b {q_b}^2 }{ m_b }  \displaystyle \int_{\Reals^3} { \left \{ \mathbf{k}_{p',q',r'} \right \}_\mathbf{r} } \cdot \displaystyle \frac{ d F_{b0} }{ d \mathbf{v}_b }
\cr && \hspace{12.em} \times \; \delta  \left ( \omega - \mathbf{k}_{p',q',r'} \cdot \mathbf{v}_b,\,\omega_g \right ) d^3 \mathbf{v}_b F_{a0} \left ( \mathbf{v}_a \right ) \delta \left ( \omega - \mathbf{k}_{p,q,r} \cdot \mathbf{v}_a,\,\omega_g \right ) .
\end{array}
}
\end{equation}
For the same reason, the contribution of the second term in (\ref{discrete_calculus_1}) comes from its imaginary part that can be simplified into 
\begin{equation}
 {
\begin{array}{l}
 \hspace{1.em} - \displaystyle \int_{V_{\mathbf{k}_g}} \displaystyle \frac{d^3 \mathbf{k}}{ {\left ( 2 \pi \right )}^3 } \displaystyle \int_{\iota \nu - \omega_g / 2}^{\iota \nu + \omega_g/2} \displaystyle \frac{d \omega}{ 2 \pi }  \mathrm{Im} \displaystyle \Big \{ \displaystyle \sum_{p,q,r} { \left \{ \mathbf{k}_{p,q,r}\right \}_\mathbf{r} } \displaystyle \frac{ 32 \pi^2 \bar{n}_a q_a  {\overset{\frown}{\mathrm{S}} \left ( \mathbf{k}_{p,q,r} \right ) }^2 }{ {\varepsilon_L \left ( \omega,\,\mathbf{k}\right )}^2 { \left \{ \mathbf{k}_{p,q,r}\right \}_\mathbf{r}}^4 } \displaystyle \frac{\pi}{m_a}  \displaystyle \sum_{p',q',r'}  {\overset{\frown}{\mathrm{S}} \left ( \mathbf{k}_{p',q',r'} \right )}^2
\cr \hspace{16.em} \times \displaystyle \sum_{m}  \displaystyle \frac{ { \left \{ \mathbf{k}_{p,q,r}\right \}_\mathbf{r}} }{ \omega_m - \mathbf{k}_{p,q,r} \cdot \mathbf{v}_a } \cdot  \displaystyle \frac{d F_{a0}}{d \mathbf{v}_a}  \displaystyle \sum_b \delta N_b \bar{n}_b {q_b}^2 
\cr \hspace{16em} \times  \displaystyle \int_{\Reals^3} F_{b0} \displaystyle \left ( \mathbf{v}_b \right ) \delta \displaystyle \left ( \omega - \mathbf{k}_{p',q',r'} \cdot \mathbf{v}_b, \, \omega_g \right ) d^3 \mathbf{v}_b \Big \}
\cr = - \displaystyle \int_{V_{\mathbf{k}_g}} \displaystyle \frac{d^3 \mathbf{k}}{ {\left ( 2 \pi \right )}^3 } \displaystyle \int_{\iota \nu - \omega_g / 2}^{\iota \nu + \omega_g/2} \displaystyle \frac{d \omega}{ 2 \pi } \displaystyle \sum_{p,q,r} { \left \{ \mathbf{k}_{p,q,r}\right \}_\mathbf{r} } \displaystyle \frac{ 32 \pi^4 \bar{n}_a q_a  {\overset{\frown}{\mathrm{S}} \left ( \mathbf{k}_{p,q,r} \right ) }^2 }{ {\varepsilon_L \left ( \omega,\,\mathbf{k}\right )}^2 { \left \{ \mathbf{k}_{p,q,r}\right \}_\mathbf{r}}^4 } \displaystyle \frac{1}{m_a}  \displaystyle \sum_{p',q',r'}  {\overset{\frown}{\mathrm{S}} \left ( \mathbf{k}_{p',q',r'} \right )}^2   
\cr \hspace{14.em} \times \, \delta \left ( \omega - \mathbf{k}_{p,q,r} \cdot \mathbf{v}_a,\omega_g \right ) { \left \{ \mathbf{k}_{p,q,r}\right \}_\mathbf{r}} \cdot  \displaystyle \frac{d F_{a0}}{d \mathbf{v}_a} \displaystyle \sum_b \delta N_b \bar{n}_b {q_b}^2 
\cr \hspace{14.em}\times \displaystyle \int_{\Reals^3} F_{b0} \displaystyle \left ( \mathbf{v}_b \right ) \delta \displaystyle \left ( \omega - \mathbf{k}_{p',q',r'} \cdot \mathbf{v}_b, \, \omega_g \right ) d^3 \mathbf{v}_b
\end{array}
}
\end{equation}
according to the Plemelj formula. By performing the integration over $\omega$ using the integral-sum properties (\ref{useful_property}), one finally deduces (\ref{collision_operator}).

\subsection*{Appendix D - Landau damping BEPS input deck}

pinput2\\
IDRUN = 1\\
INDX =   8, INDY =   1, NPX = 8388608, NPY = 2\\
NPXB =   0, NPYB =   0, INORDER = 2,\\
POPT = 2, DOPT = 2, DJOPT = 2\\
NTW = 1, NTP = 0, NTD = 10, NTA = 0, NTV = 0, NTS = 0\\
NUSTRT = 1, NTR = 0, PSOLVE = 1, RELATIVITY = 0\\
TEND =  500.0, DT =   0.200, CI = 0.1\\
QME =  -1.000, VTX =   1.000, VTY =   1.000, VTZ =   1.000\\
VX0 =   0.000, VY0 =   0.000, VZ0 =   0.000\\
VDX =   0.000, VDY =   0.000, VDZ =   0.000\\
VTDX =   0.000, VTDY =   0.000, VTDZ =   0.000\\
AX =    0.866667, AY =   0.866667\\
NSRAND = 0, NDPROF = 0
AMPDX = 0.0, SCALEDX = 0.0, SHIFTDX = 0.0\\
AMPDY = 0.0, SCALEDY = 0.0, SHIFTDY = 0.0\\
MOVION = 0, NPXI =  8192, NPYI =  2048\\
NPXBI =  0, NPYBI = 0\\
QMI =  1.000, RMASS = 100.0\\
RTEMPXI =   20.000, RTEMPYI =   20.000, RTEMPZI =   20.000\\
VXI0 =   0.000, VYI0 =   0.000, VZI0 =   0.000\\
VDXI =   0.000, VDYI =   0.000, VDZI =   0.000\\
RTEMPDXI =   1.000, RTEMPDYI =   1.000, RTEMPDZI =   1.000\\
NSRANDI = 0, NDPROFI = 0\\
AMPDXI = 0.0, SCALEDXI = 0.0, SHIFTDXI = 0.0\\
AMPDYI = 0.0, SCALEDYI = 0.0, SHIFTDYI = 0.0\\
SORTIME = 100, SORTIMI = 250\\
NPLOT = 0, IDPAL = 1, NDSTYLE = 1, SNTASKS=1,\\
ntw=1,ntp=50\\
$\,$\\
pinput2 jf\\
driver select= 1\\
amp=0.005,wavemode=13,wavew=1.18\\
rise = 0.,flat = 0.,fall=0.,yrise fall=1000.0\\
timerise=1.2,timeflat=1.2,timefall=1.2,\\
center1 = 200, center2 = 0, phase offset = 4\\
bow = 0., bow power = 0.\\
superGauss = 2\\
$\,$\\
fvxmax=12., fvymax=12.\\
nphbx=150,nphby=150\\
nphxx=20,nphyx=20\\
nphxy=20,nphyy=2

\section*{Acknowledgments}
This work was partially sponsored by NNSA, NSF under Grant ACI-1339893 and the European Research Council through the project ERC AdG InPairs no. 695088.

\bibliographystyle{apalike}
\bibliography{refs}

\end{document}